\documentclass{pasj02}
\usepackage{color}

\usepackage[switch,mathlines]{lineno}
\newcommand{\oneE}{1E 161348$-$5055}    
\newcommand{\AXS}{AXS~J161730$-$505505}   
\newcommand{\RCW}{RCW 103}

\newcommand{\zz}{Z_4^2}
\newcommand{\Ntrial}{N_{\rm trial}}
\newcommand{\Mseg}{M_{\rm seg}}
\newcommand{\EL}{E_{\rm L}}
\newcommand{\EU}{E_{\rm U}}

\newcommand{\Ppr}{P_{\rm pr}}
\newcommand{\Prot}{P_{\rm rot}}
\newcommand{\Pch}{{\cal P}_{\rm ch}}
\newcommand{\Pchone}{{\cal P}_{\rm ch}^{(1)}}
\begin{document}
\SetRunningHead{Makishima, K.}{Pulsation in RCW 103}
\Received{2025/10/31}%{yyyy/mm/dd}
\Accepted{2026/01/17}%{yyyy/mm/dd}
\title{Evidence for 1.01 s Pulsations of the Central Compact Object 
in the Supernova Remnant \RCW\  with ASCA, XMM-Newton, and NuSTAR}
\author{Kazuo  \textsc{Makishima}\altaffilmark{1,2}}
\email{maxima@phys.s.u-tokyo.ac.jp}
\altaffiltext{1}{
Department of Physics, The University of Tokyo,
7-3-1 Hongo, Bunkyo-ku, Tokyo 113-0033}
\altaffiltext{2}{
Kavli Institute for the Physics and Mathematics of the Universe (WPI),
The University of Tokyo,
5-1-5 Kashiwa-no-ha, Kashiwa, Chiba, 
277-8683}
\author{Nagomi Uchida\altaffilmark{3}}
\altaffiltext{3}{Institute of Space and Astronautical Science,
3-1-1 Yoshinodai, Chuo-ku, Sagamihara, Kanagawa, 252-5210, Japan}
\author{Teruaki Enoto\altaffilmark{4,5}}
\altaffiltext{4}{Department of Physics, Kyoto University,
Kitashirakawa Oiwake-cho, Sakyo-ku, Kyoto, 606-8502, Japan }
\altaffiltext{5}{Extreme Natural Phenomena RIKEN Hakubi Research Team,
Cluster for Pioneering Research, RIKEN, \\
2-1 Hirosawa, Wako, Saitama, 351-0198, Japan }
%
% Don't change these lines
%\label{firstpage}
\maketitle
\begin{abstract}
The neutron-star X-ray source \oneE,  associated with the supernova remnant RCW 103,
exhibits clear intensity variations with a period of 6.67 hr.
To clarify the nature of this object and its long periodicity,  
detailed timing studies were applied to its archival X-ray data,
taken with  ASCA (in 1993), XMM-Newton (in 2001, 2005, and 2016),
and NuSTAR (2016 and 2017).
It was assumed that the 6.67 hr period arises due to the beat between 
the rotation and free precession periods of the star that is slightly aspherical.
By removing  timing perturbations to be caused by this long periodicity,
the six data sets consistently yielded evidence for 
pulsations at periods of $P\approx 1.01$ s,
to be interpreted as the objects' spin period,
although the optimum energy range differed among the data sets.
The measured six periods accurately line up  on a linear spin-down trend of
$\dot P = 1.097 \times 10^{-12}~{\rm s~s}^{-1}$.
The object is implied to have a characteristic age of 14.7 kyr,
a spin-down luminosity of $4.2\times 10^{34}$ erg s$^{-1}$
which is insufficient to power the X-ray luminosity,
a dipole magnetic field of  $ \sim 4.6\times 10^{13}$ G,
and a toroidal field of $\sim 7 \times 10^{15}$ G.
Its similarity and dissimilarity to magnetars are discussed.
An emission geometry, which crudely explain these results, is presented. 
\end{abstract}
\email{maxima@phys.s.u-tokyo.ac.jp}

%%%%%% 1 %%%%%%
\section{Introduction}
\label{sec:intro}
%%%%%%%%%%%
The X-ray source \oneE, hereafter abbreviated as 1E~1613,
is a radio-quiet neutron star (NS)
located at the projected center of the  supernova remnant (SNR) RCW 103,
and once  classified as a Central Compact Object (CCO).
Although recent studies \citep{DeLuca17} indicate
that it may not be a typical object of this class,
we conventionally regard  it as a CCO in the present paper.
The object was discovered with the Einstein Observatory in 1979 \citep{1Ediscovery},
but it then became undetectable. Using the ASCA observatory,
\citet{GPH97}, hereafter GPH97, reconfirmed it in 1993 in an X-ray  bright state,
and  in 1997 at a 10\% flux level \citep{GPV99}.
On long time scales,1E 1613 is thus highly variable
by up to 2.5 orders of magnitude (e.g., DL06),
sometimes emitting long flares \citep{Esposito19}
and short bursts \citep{Rea16}.
The object thus resembles  magnetars
(e.g., \cite{Rea16, RCW_magnetar}).

The search for a rapid  NS rotation  in 1E~1613 was
performed repeatedly  (e.g,  GPH97, \cite{GPV99}, \cite{DeLuca07}),
but none was successful. 
Instead, with  an XMM-Newton observation in 2005,
a periodic X-ray intensity variation  was found 
by \citet{DeLuca06}, hereafter DL06, 
at  an extremely long period of
 %\vspace*{-2mm}
\begin{equation}
T= 6.67 \pm 0.03~{\rm hr} = 24.01 \pm 0.11~{\rm ks}.
\label{eq:6.67hr}
\end{equation}
This periodicity had been suggested before \citep{Garmire00},
and reconfirmed in later observations \citep{Rea16, Esposito19}.
Although a straightforward interpretation of this $T$ 
is the NS rotation (DL06, \cite{Rea16}),
it appears to be way too long,
because  the implied rotational frequency is 
$\sim 3 \times 10^{-8}$ of the break-up frequency of a typical NS.
It would be difficult  to extract the NS's angular momentum to such a nearly zero level.
\citet{Pizzolato08}  proposed that 1E~1613 forms a binary 
with a low-mass companion in a 6.67 hr orbit,
and its spin is magnetically locked to the orbital revolution.
However, such a  binary that survived the supernova explosion
is not known.

Before the report of equation~(\ref{eq:6.67hr}) by DL06, 
\citet{HeylHernquist02}, hereafter HH02,
tried to explain the earlier suggestion of the long periodicity \citep{Garmire00}
in the following way.
Namely, due to an intense ($\sim 10^{16}$ G) toroidal magnetic field,
this NS is deformed to asphericity of $\epsilon  \equiv \Delta I /I \sim 10^{-4}$,
where $I$ is the moment of inertia and $\Delta I$ is its difference
between the parallel and perpendicular directions to the symmetry axis;
the NS hence undergoes free precession with a  period $P$,
which differs only slightly from the  rotation period at $P/(1+\epsilon)$;
the two periods produce a beat effect 
at a long period of $T=P/\epsilon$ (often called the slip period),
\footnote{``Precession" of an axisymmetric rigid body means the motion of 
its symmetry axis  around the  angular momentum vector \citep{LandauLifshitz}.
It occurs with a period  close to the rotational period,
and their beat  appears at 6.67 hr in the present case.
Although the beat period itself is sometimes called the precession period, as in HH02,
we do not follow this terminology.}
which should be identified with  equation~(\ref{eq:6.67hr});
the pulsation at $P$ must lurk around $1$ s;
the X-ray emission  comes from a displaced hot spot on the NS surface;
and some appropriate  geometry can explain
the large flux variation at $T$,
while keeping the pulses  at $P$ difficult to detect directly.
A similar idea was invoked  by \citet{ZhuXu06},
to explain the suggested  $\sim 77$ min periodicity
in the enigmatic bursting radio transient GCRT J1745$-$3009.

So far, we have accumulated plenty of observational evidence 
supporting the HH02 scenario.
Namely, altogether seven magnetars we studied were all
found to exhibit periodical pulse-phase modulation (PPM),
in such a way that the arrival times of their hard X-ray pulses at a period $P$ 
are modulated periodically with a long  period $T$ 
at several tens ks (\cite{Makishima14}, 2016, 2019, 2021ab, 2024ab).
Just as HH02 assume,
we interpret that these NSs are magnetically deformed to $\epsilon \sim 10^{-4}$,
and perform free precession, to produce the PPM 
at  $T \approx P/\epsilon$.
This effect is observed exclusively in their spectral hard component,
often making their hard X-ray pulsation difficult to detect.
Nevertheless, their soft component, being free from the PPM,
exhibits a regular pulsation at $P$.
With this knowledge, 
we can correct (``demodulate") the photon arrival times of hard X-ray photons 
for the PPM effect, and securely detect the hard X-ray pulsation.

In the present study, we analyze  archival data  of 1E~1613,
from ASCA, XMM-Newton, and NuSTAR.
Guided by our magnetar results, 
we try to detect the predicted fast ($P \sim 1$ s) pulsation (HH02),
by correcting the photon arrival tames for  the 6.67 hr PPM.
In contrast to our magnetar studies where $P$ is known and $T$ is unknown,
here $T$ is known and $P$ is unknown,

%%%%%% 2 %%%%%%
\section{Observations}
\label{sec:obs}
%%%%%%%%%%%%%

%==========2.1========
\subsection{Data selection}
\label{subsec:obs_data_selection}
%%%%%%%%%%%%

Although 1E~1613 has been observed by many X-ray satellites,
not all the available archival data sets are necessarily
suited to the present first-cut attempt,
of which the aim is described as above.
We hence select data sets that satisfy all the following six criteria;
these are considered to provide necessary conditions
for a reliable detection of a PPM-affected fast pulsation. 
\begin{itemize}
\item[(i)] 
The observation was made with adequate imaging information,
so that we can reduce the background.
Those taken with the Suzaku HXD are hence excluded,
because the signal intensity from 1E~1613 (even when it was rather bright)
will be at most a few percent  of the HXD background,
which is of the level of $\sim 7$ mCrab at 15 keV.
Although the HXD afforded the detection of periodic 
pulse-phase modulations from several magnetars, 
they were considerably brighter than 1E~1613.
\item[(ii)] 
The data have a modest ($\lesssim 10\%$ at 5 keV) energy resolution,
because the PPM  in magnetars often show 
complex energy dependence \citep{Makishima24a},
which requires us to carefully choose the energy range to be analyzed.
%The Chandra HRI data are hence excluded.
%
\item[(iii)] The data have a time resolution better than $\sim 25$ ms,
so that the 4th harmonic, which is of vital importance \citep{Makishima24a},
of the putative $P \sim 1$ s pulsation is fully resolved.
Data from the ASCA GIS and NuSTAR,
with time resolutions of $61~\mu$ s and $<65~\mu$ s,
satisfy this criterion,
while  some data taken with CCD detectors are excluded
depending on the data read-out mode.
\item[(iv)] The total data span is $S \gtrsim 20$ ks,
so as to cover a significant ($\gtrsim 75\%$) fraction of $T=24.0$ ks.
Unless this condition is fulfilled, errors in the PPM correction procedure
becomes very large, even if $T$ is fixed.
The errors would propagate to the uncertainty in $P$,
and make its identification difficult.
\item [(v)] The data provide a sufficient number of signal photons,
which is crudely proportional to 
$Q \equiv f_{\rm x} \/ \eta_{\rm net} \/ A_{100}$,
a quantity to be called a data quality measure.
Here, $f_{\rm x}$ is the 0.5--2 keV flux (in $10^{-12}$ erg s$^{-1}$ cm$^{-1}$),
$\eta_{\rm net}$ is the net exposure in ks, 
and $A_{100}$ is the instrumental effective area at 5 keV
\footnote{We assume $A_{100}=0.24, 0.95, 0.70$, and 0.30
for  ASCA GIS, XMM-Newton EPIC,  NuSTAR FPMA+B, 
and Chandra ACIS, respectively.}
in units of 100 cm$^{2}$.
We then require $Q \geq 40$, because this will give $\gtrsim 3000$ signal photons
(under an appropriate spectral assumption),
which are adequate based on our experience with the seven magnetars.
\item[(vi)] The data are free from major instrumental artifacts,
e.g., event pile ups or dead times.
This is because the present analysis is more subject to such artifacts
than a simple pulsation search.
\end{itemize}

The application of these criteria to the available archives have left us 
the six data sets listed in table~\ref{tbl:datasets}.
The first of them, named ASCA93, was taken  in 1993
with the Gas Imaging Spectrometer (GIS) onboard ASCA,
while the next three (XMM01, XMM05 and XMM16) with the EPIC-pn camera 
onboard XMM-Newton in 2001, 2005 and 2016, respectively.
The remaining two, NuS16 and NuS17, were 
taken with NuSTAR in 2016 and 2017, respectively.
The six data sets span  24 yrs in total.

Of the six data sets, our primary emphasis is put on ASCA93
which led  GPH97 to the 1E~1613 rediscovery,
because the object was then very bright (though with $Q \sim 50$; see table~\ref{tbl:datasets}),
and the data have a moderate gross exposure
which  covers about 2.6 cycles of $T$.
In addition, the GIS \citep{GIS1,GIS2} is suited to this kind of timing studies,
thanks to  its  high time resolution ($61~\mu$s),
low and stable background, 
high hard X-ray  efficiency,
and insignificant dead time or pile-up effects.
In fact, a few  novel timing results have been derived 
from the GIS archive (\cite{Makishima23}, 2024b).
The object was observed with ASCA  again in 1997 \citep{GPV99},
but it was then fainter  by an order of magnitude,
with $Q \sim 7$ which does not meet Criterion  (v).
The 1997 ASCA data are not analyzed here.

Once a candidate period is found in ASCA93,
we use the three XMM-Newton data sets to reconfirm it,
and constrain the period change rate $\dot{P}$ with progressive accuracies.
Among them, XMM01 was taken when 1E~1613 was
half as bright as in ASCA93 (table~\ref{tbl:datasets}), yielding $Q \sim 45$.
In  XMM05, the object was rather faint,
but the exposure was long enough to give $Q \sim 43$.
In addition, this data set is worth analyzing, 
because it led  DL06 to arrive at equation~(\ref{eq:6.67hr}).
The XMM16 data were obtained in 2016 August \citep{Esposito19},
when 1E~1613 was as bright  as in ASCA93.
Although another XMM-Newton observation was made in 2018,
we do not analyze this data set,
because the employed time resolution of 73.4 ms 
does not satisfy Criterion (iii).

For sky regions near1E~1613, 
the NuSTAR archive provides three data sets,
taken in 2016, 2017,  and  2018. 
However, the 2018 data are not useable,
because the object was just outside the detector field of view.
We hence analyze the remaining two, NuS16 and NuS17.
As reported by \citet{Esposito19},
the object was very bright in 2016 $(Q\sim 270$),
whereas its flux decreased to a quarter in 2017 $(Q\sim 68$).
Obviously, the hard X-ray information from NuSTAR is very important,
because the PPM  in the seven magnetars 
was seen only in their spectral hard component
which appears  at $\gtrsim 10$ keV.
Nevertheless, the ASCA and XMM-Newton data
(both limited to $\lesssim 10$ keV)  still remain valuable,
because the previous pulse non-detection from 1E~1613 
in these soft energies suggests 
that the PPM disturbance continues way down to $<10$ keV.
In any case,  we may need to carefully adjust,
for each data set,
the lower energy limit $\EL$ of the analysis.

The source has been observed  24 times with Chandra, from 1999 to 2024.
Among them, 20 data sets are short, $S \le 18$ ks, violating Criterion (iv).
Most of these data sets also fail to satisfy Criterion (v).
Of the remaining four, one (ID 7619) does not meet Criterion (ii)
because the focal plan instrument was the HRI.
Other two (ID 970 and 501340) were taken with the ACIS 
but in the TE mode with insufficient time resolution (3.2 s), against Criterion (iii).
We are hence left with the only one data set, ID 500209,
taken on 2002 March 3 with  the ACIS in the CC mode.
The source was moderately bright ($f_{\rm x} \approx 1.1$),
but when combined with the limited  exposure ($\eta_{\rm net} \approx 48.6$)
and $A_{100}=0.3$ for Chandra, we derive $Q \sim 16$
which falls  short of Criterion (v).
In addition, the effects of event pile up, Criterion (vi),
 is at present unknown.
Thus, no Chandra archive can be used in the present study.
%%%%%%%%%%%%

%%T%T%T%T%T%T%T%T%T%T%T%T%T%T%T%T%T%T%T%T%T%T%T%T%T%T%T%T%T%T%T%T%T%T%T%
\begin{table*}
\caption{Archival data sets utilized in the present study.}
\label{tbl:datasets}
\vspace*{-3mm}
\begin{center}
\begin{tabular}{lccccrrccccccc}
\hline %-----------------------------------------------------
Data set & Instrument  &Obs ID &   \multicolumn{2}{c}{Observation} 
                 &   \multicolumn{2}{c}{Exposure (ks)$^{*}$} &   \multicolumn{2}{c}{Number of photons$^{\dagger}$}& Flux$^{\ddagger}$&$Q^{\#}$\\
          &         &           & Date (y/m/d)& MJD  & Gross  &  Net  & Soft& Hard   \\
\hline %------------------------------------------------------
\hline %------------------------------------------------------
ASCA93 & ASCA GIS2+3    &50035000  &1993/08/17  & 49216    & 63.4&  39.5  & 79954 & 5249  & 5.2& 50\\
%\hline %------------------------------------------------------
%ASCA97 & ASCA GIS     &55041000 & 1997/09/04  &50695   & 131.3& 63.8   & 138974 & 5661& 0.5\\
\hline %------------------------------------------------------
XMM01& XMM-Newton pn &0113050601&2001/09/03&52155   & 19.6 & 18.5 & 17246  & 7446 & 2.6 & 45\\
\hline %------------------------------------------------------
XMM05&XMM-Newton pn&0302390101& 2005/08/23 & 53605  & 87.9 & 82.6  & 21556 & 5752 & 0.6& 43\\
\hline %------------------------------------------------------
XMM16&XMM-Newton pn&0743750201&2016/08/19 &57619 &81.0 & 81.0 & 159706 & 75940&4.8& 420\\
\hline %------------------------------------------------------
NuS16&NuSTAR FPMA+B&90201028002&2016/06/25 &57565 &129.6 & 65.9 & 37910&2808& 5.9& 270\\
\hline %------------------------------------------------------
NuS17&NuSTAR FPMA+B&30301017002
     &2017/06/02 &57906 &130.5& 65.6& 9588&1131&1.6& 68\\
\hline %------------------------------------------------------
\end{tabular}
\end{center}
\begin{itemize}
\setlength{\itemsep}{0mm}
  \item[$^{*}$] The gross exposure means the total data span.
  The net exposure for XMM-Newton is after removing background flares. 
  \item[$^{\dagger}$] Background-inclusive counts from the CCO. 
 For ASCA and XMM-Newton,  ``Soft'' and ``Hard" mean 1--2.5  keV and 2.5--10 keV, respectively.
 Those of NuSTAR refer to 3--10 keV and 10--60 keV.
 The ``Soft''  counts of  ASCA93  include the SNR contamination.
  \item[$^{\ddagger}$] 0.5--2 keV flux of 1E~1613, in $10^{-12}$ erg s$^{-1}$ cm$^{-1}$, from DL06 and \citet{Esposito19}.
  \item[$^{\#}$] The data quality measure, defined in subsection~\ref{subsec:obs_data_selection}.
  \end{itemize}
  \vspace*{-4mm}
\end{table*}
%%T%T%T%T%T%T%T%T%T%T%T%T%T%T%T%T%T%T%T%T%T%T%T%T%T%T%T%T%T%T%T%T%T%T%T%

\vspace*{-2mm}
%==========2.2========
\subsection{ASCA observation}
\label{subsec:obs_ASCA}
%==========2.2========

The ASCA93 observation  was  intended for  X-ray spectroscopy 
of RCW 103 using the Solid State Spectrometer (SIS).
1E~1613, which was  in the same fields of view of SIS and GIS,
was rediscovered by GPH97 on this occasion.
A serendipitous object, the 69 ms pulsar \AXS, 
was also in the same GIS field of view (\cite{Torii98}, 2000).

Since the SIS data have a poor time resolution in view of  Criterion (iii),
we use the GIS data,
which were processed in the standard way as in GPH97.
We extracted events from 1E~1613 and the 69 ms pulsar separately,
within $3'$ of respective image centroids.
The  arrival times of individual photons were  converted
to those at the Solar system barycenter,
and  events from the two detectors,  GIS2 and GIS3, were co-added.
Due to  the limited angular resolution of the GIS,
soft X-rays from the SNR heavily contaminate the  CCO events,
and to a lesser extent the pulsar data.
To avoid this effect,  we set $\EL \gtrsim 2.5$ keV.

The nominal upper energy bound of the GIS is  $\EU=10$ keV, 
but we raise it to  $\EU=12$ keV.
This is because the 10--12 keV photons are often useful
(e.g., \cite{Makishima23}, 2024b),
thanks to the high quantum efficiency of the GIS \citep{GIS1}.

Since 1E~1613  emits short bursts \citep{Rea16},
we produced 3--12 keV light curves from ASCA93,
using 10-s bins, and searched them for burst candidates.
However, none was found.

%==========2.3========
\subsection{XMM-Newton observations}
\label{subsec:obs_XMM}
%==========2.3========

From the XMM-Newton archive, 
we analyze the selected three data sets, XMM01, XMM05, and XMM16,
excluding the 2018 data for the reason 
described in subsection~\ref{subsec:obs_data_selection}.

The XMM01 data  in fact consist of two subsets, 
one acquired on September 3,
and the other on  September 3 to 4.
The EPIC-MOS data from neither subset are usable, 
because they were taken  in ``Full Frame" mode with 2.6 s  time resolution, 
violating Criterion (iii).
In the first subset,  EPIC-pn was in ``Timing" model,
with no imaging  information.
Based on Criterion (i), we discard this data subset.
In the second subset, EPIC-pn was  in ``Small Window" mode
with an adequate time resolution of 5.7 ms.
Thus,  our XMM01 data consist of EPIC-pn events in  the second subset.
On this occasion, 1E~1613  was about half as bright as  in ASCA93.

In the 2005 XMM-Newton observation
in which the 6.67 hr flux variation was clearly identified (DL06),
1E~1613 was rather faint,
but we still find $Q \sim 43$ thanks to a long exposure.
The EPIC data ware acquired  in the same modes 
as in the second 2021 subset.
We again utilize only the EPIC-pn data.

On 2016 June 22, the CCO suddenly brightened up,
emitting a short burst \citep{Rea16}.
About 2 months later,
when the source was still in an active state 
and nearly as bright as in ASCA93,
the XMM16 data were obtained in the same mode as in  XMM05.
Through this outburst, the 24 ks flux variation profile
changed from single-peaked to double-peaked ones \citep{Rea16, Esposito19}.
Therefore, when analyzing the EPIC-pn data from XMM16,
we must  keep in mind
that the source behavior could  somewhat differ 
from those on the earlier occasions,
possibly with increased timing noise as in some magnetars \citep{Younes17,Makishima21b}.

From each XMM-Newton data set, we produced an X-ay image,
and derived 0.2--15 keV keV events with grade $\lesssim 12$,
within $15''$ of 1E~1613.
Then, the barycentric correction was conducted.
These  steps utilized the SAS v21.0.0 package.
Unless otherwise stated,
we set $\EU=10$ keV,
above which there are few signal photons.
In these data, the event pile-up is estimated to be negligible.

In these observations,
the background was relatively calm, 
so we did not employ so-called SN filter.
To remove flaring particle backgrounds,
we produced 100-s bin light curves,
using 10--15 keV events, % from  the source region,
and discarded those  bins when the count rate
(typically 0.002 c s$^{-1}$) exceeds 0.05 c s$^{-1}$.
This resulted in a photon loss by $\approx 6\%$ in 
XMM01 and XMM05, and $<1\%$ for XMM16.

After removing the particle flares,
we searched 10-s binned light curves in 2--10 keV for short-burst candidates,
in the same way as for ASCA.
At about 6.31 ks from the start of XMM01,
we found a candidate bin which contains 32 counts
(against an average of  7.2 counts/bin), and removed that bin.
However,  no such candidates were found in XMM05 or XMM16.

%==========2.4========
\subsection{NuSTAR observations}
\label{subsec:obs_NuS}
%==========2.3========

As described in subsection~\ref{subsec:obs_data_selection},
we analyze two data sets of 1E~1613;
NuS16 and NuS17.
The object was very bright in the former,
whereas it became  4 times fainter in the latter \citep{Esposito19}.
The NuS16 data were acquired only 4 days after the activity onset,
and 54 days before XMM16. 
Therefore, the behavior of  1E~1613 in NuS16
may be more strongly subject to the issues pointed out for XMM16. 

For each data set, we co-added
the events from the two  Focal-Plane Modules,
and processed them (including the barycentric correction)
using HEASOFT version 6.35.1 and CALB 2025.4.15.
We excluded about 13\% of the NuS16 exposure
where the attitude solution suffers large errors.
%For NuS16 and NuS17,
On-source  photons were accumulated within $1'$ of 
the source position on the X-ray image.
The derived number of photons, below and above 10.0 keV,
are given in table~\ref{tbl:datasets}.
We searched a 10-s bin light curve in 3--70 keV 
for short bursts, but none was found  in either data set.

\vspace*{-3mm}
%%%%%% 3 %%%%%%
\section{Data analysis and results}
\label{sec:ana}
%%%%%%%%%%%%%

%=========== 3.1 =====================
\subsection{Basic strategy}
\label{subsec_ana_strategy}
%=========== 3.1 =====================

%--------------- 3.1.1 ---------------------
\subsubsection{Period search range}
\label{subsubsec:ana_strategy_searchrange}
%--------------- 3.1.1 ---------------------
Our central  objective is to search the six data sets
for a fast ($P \sim 1$ s) X-ray pulsation,  
assuming that it is not directly visible,
due to the PPM disturbance with the period of  $T=6.67$ hr = $24.0$ ks.
We try to remove the suspected PPM effects by two methods (subsection~\ref{subsec:ana_ASCA93}).
One  is newly developed for the present study, 
while the other, called ``demodulation analysis",
has been applied extensively to our magnetar studies (section~\ref{sec:intro}).

In \citet{Makishima24a},
the $P/T$ ratios of the seven magnetars 
distribute over  $(0.58 - 5.0)\times 10^{-4}$.
Then, if the present CCO with $T=24.0$ ks is in a similar condition,
its pulsation is expected to lurk in a range from $P\sim 1.4$ s to $\sim 12$ s.
Allowing for sufficient margins, 
we start our period  search, with ASCA93, 
over the range from  0.3 s to 30 s.
This interval is made  narrower later on.

%--------------- 3.1.2 ---------------------
\subsubsection{Periodograms (PGs) and $Z_m^2$ statistics}
\label{subsubsec:ana_strategy_Z2}
%--------------- 3.1.2 ---------------------

Our analysis tool is  periodogram (hereafter PG),
which displays  periodicity significance
against the  trial period $P$ 
with which the data are epoch-folded.
The periodicity significance is expressed 
by the $Z_m^2$ statistics \citep{Z2_94};
at each $P$, the Fourier power of the folded profile is
summed from the fundamental  to the $m$-th harmonic,
and normalized by the total photon number, 
to yield a quantity called  $Z_m^2$.
This  is similar to, but improves over, 
the more conventional chi-square evaluation \citep{Makishima23}.
Based on our  magnetar studies, we mostly use $m=4$, 
but sometimes $m=2$ or $m=3$.
For white noise data,  $Z_m^2$ follows a chi-square distribution 
with $2m$ degrees of freedom (d.o.f);
its standard deviation  is $\sqrt{4m}$ around the mean of $2m$.

To  calculate a PG,
usually the observing window function is folded
together with the photon counts,
and the former is used for exposure correction.
However,  the periods of interest in the present study are short 
enough for the exposure to be uniform to within a few percent
across the pulse phase \citep{Makishima21b}.
Therefore, the exposure correction is not performed.
The original photon counts are hence  preserved,
and the statistical fluctuations are solely Poissonian.

%--------------- 3.1.3 ---------------------
\subsubsection{Period  increment steps}
\label{subsubsec:ana_strategy_searchstep}
%--------------- 3.1.3 ---------------------
In calculating a PG,
we scan the period $P$ with an incremental step  $\Delta P$ which is given as
\begin{equation}
\Delta P =\frac{P^2}{\eta S} ~ \iff~ \frac{S}{P} -\frac{S}{P+\eta (\Delta P)} \approx 1,
\label{eq:dp}
\end{equation}
considering Fourier independence of adjacent periods.
Here, $S$ is the data span (gross exposure in table~\ref{tbl:datasets}),
and $\eta$ is an ``over sampling factor";
$\eta=1$ means the exact Fourier wavenumber relation,
and $\eta>1$ an over sampling.
When the period range is not too vast,
the epoch-folding analysis provides a better sensitivity than the Fourier transform scheme,
because of the freedom to select $\eta>1$,
and of the capability to incorporate the higher harmonics
by selecting $m>1$.

%--------------- 3.1.4 ---------------------
\subsubsection{Period change rate}
\label{subsubsec:ana_strategy_Pdot}
%--------------- 3.1.4 ---------------------

Since $\dot P$ of the putative pulsation is at first unknown,
the analysis of ASCA93 and XMM01
are first carried out assuming $\dot P=0$.
After an  estimate of $\dot P$ is found,
we iterate these analyses considering $\dot P$,
to derived the final results to be presented here.
However,  the results are nearly the same as the first-round outcomes.

\vspace*{-3mm}
%=========== 3.2====================
\subsection{The 69 ms pulsar}
\label{subsec:ana_strategy_69ms}
%=========== 3.2====================
Prior to the study of 1E~1613,
we  analyzed the  ASCA93 data  of the 69 ms pulsar,  \AXS,
with a few objectives in mind.
One is to validate our timing analysis framework
using this object that has a known periodicity,
and another is to prepare a counter example 
that is considered to be free from the PPM disturbance.
For these purposes, this  pulsar indeed provides ideal reference data,
which are subject to the same systematic effects 
(e.g,, data gaps and background variations)
as the CCO data.

The derived PG is shown in figure~\ref{fig:f1_PG_psr93+97}.
After \citet{Torii98}, we employed 
$\dot P=1.31 \times 10^{-13}$ s s$^{-1}$ and the 3--10 keV energy band;
the latter is to avoid the SNR contamination
which is still present at the position of this object.
We also chose $m=2$ based on the reported simple pulse profile.
The 69 ms pulsation is clearly seen,
and the measured period given in the figure is fully consistent 
with those  in \citet{Torii98}.
A series of side lobes seen on both sides of the peak
are due to the orbital period (5.6 ks) of ASCA.

The analysis of the pulsar data indicated
that  $\eta\approx 5$ is optimum in equation~({\ref{eq:dp}),
to reconcile the calculation time with the requirement
not to miss any PG peaks.
This constitute the third objective of analyzing the pulsar data.
We hence employ  $\eta\sim  5$ to $\eta\sim  10$
for  the  study of 1E~1613.

%%%%%%%%%%%% fig.1 %%%%%%%%%%%%%%%%%%%%%%%
\begin{figure}
\centerline{
\includegraphics[width=7.2cm]{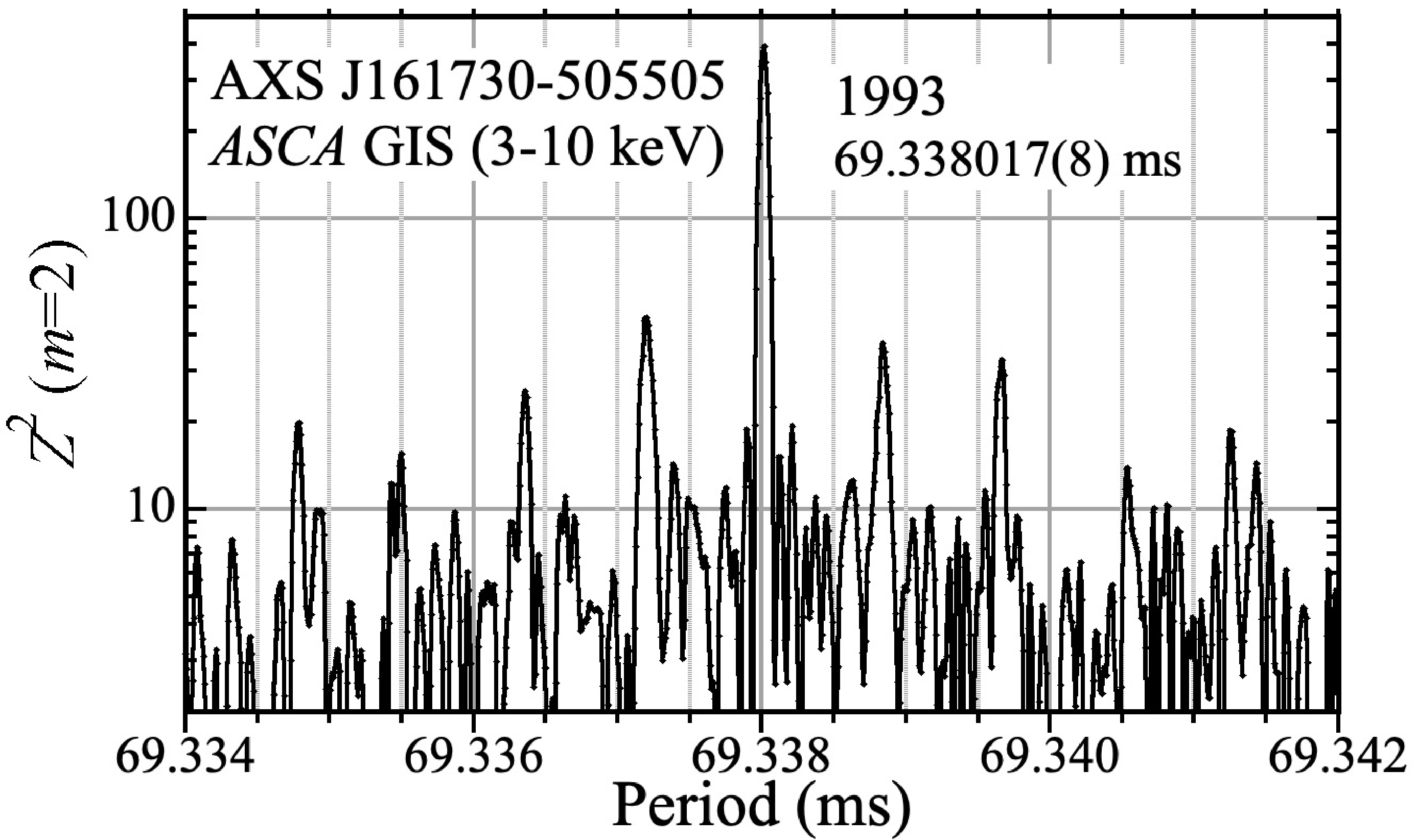}
}
\caption{A PG ($m=2$)  of  the 69 ms pulsar \AXS,
from the 3--10 keV ASCA93 data. The ordinate is logarithmic,
to show side lobes.
{Alt Text: A periodogram of the 69 millisecond pulsar from ASCA93.}
}
\label{fig:f1_PG_psr93+97}
\vspace*{-5mm}
\end{figure}
%%%%%%%%%%%%%%%%%%%%%%%%%%%%%%%%%%%

\vspace*{-2mm}
%=========== 3.3 =====================
\subsection{1E~1613 in the 1993 ASCA data  (ASCA93)}
\label{subsec:ana_ASCA93}
%=========== 3.3 =====================

%--------------- 3.3.1 -----------------------------
\subsubsection{Raw periodograms}
\label{subsubsec:ana_ASCA93_simplePG}
%--------------- 3.3.1 -----------------------------
Using the  CCO photons in  ASCA93,
we calculated a PG over  0.3 s to 30 s  (subsubsection~\ref{subsubsec:ana_strategy_searchrange}),
again in 3--10 keV and  with $m=2$ as a first attempt.
The derived ``raw" PG exhibited many peaks with similar heights
up to 35.3, but none were dominant.
%These peaks are statistically insignificant,
%because the chance probability for a peak with $Z_2^2>35.3$ 
%to appear at any period  in this PG is  estimated as $\Pch(>35.3)= 8\%$ 
%(see Appendix 1), which is not low.
Thus, in ASCA93, we find no evidence for regular pulsations 
of 1E~1613 in the relevant period interval.
This is consistent with various previous works (e.g., GPH97),
in which no fast periodicity was found
even though the attempts were often more thorough than ours.

%--------------- 3.3.2 -----------------------------
\subsubsection{Periodograms considering pulse-phase modulation}
\label{subsubsec:ana_ASCA93_phase-sorted}
%--------------- 3.3.2 -----------------------------

Even if a putative fast pulsation is  phase-disturbed by the PPM,
the pulse-phase alignment will be recovered 
if we use  those photons which fall in a limited phase of the cycle in $T$.
(This is similar to the case of a pulsar in a binary.)
We hence  divide all CCO photons in ASCA93
into $\Mseg$ (a positive intger) subsets. 
The $i$-th subset ($1 \leq i \leq \Mseg$)  comprises those photons
of which the arrival time $t$ satisfies $(i-1)/\Mseg \leq [t/T] < i/\Mseg$,
where $[t/T]$ means $t$ modulo $T$.
Then, we calculate PGs, one from each subset,
and sum up the $\Mseg$ PGs  into a single PG.
If $\Mseg$ is too small, the PPM effect would still remain,
whereas the sensitivity to a hidden periodicity would decrease if $\Mseg$ is too large.
This  method  is hereafter called ``phase-sorted PG".
As a drawback, any intrinsic peak in the  PG
will be accompanied by many side lobes,  
because each data subset  suffers artificial data gaps 
repeating  every $T$.

%%%%%%%%%%%% fig.2 %%%%%%%%%%%%%%%%%%%%%%%
\begin{figure}
\includegraphics[width=8.4cm]{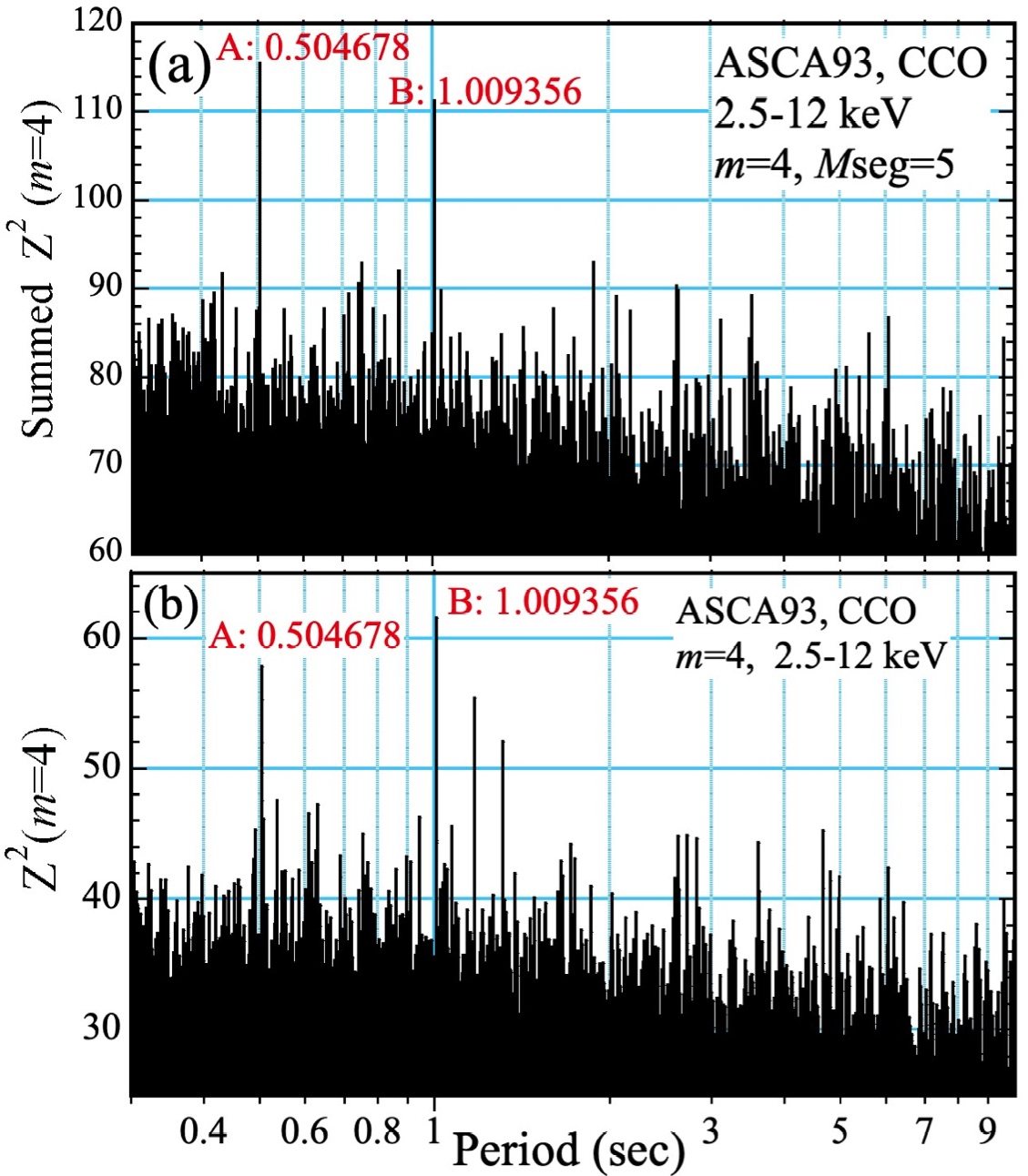}
\includegraphics[width=8.8cm]{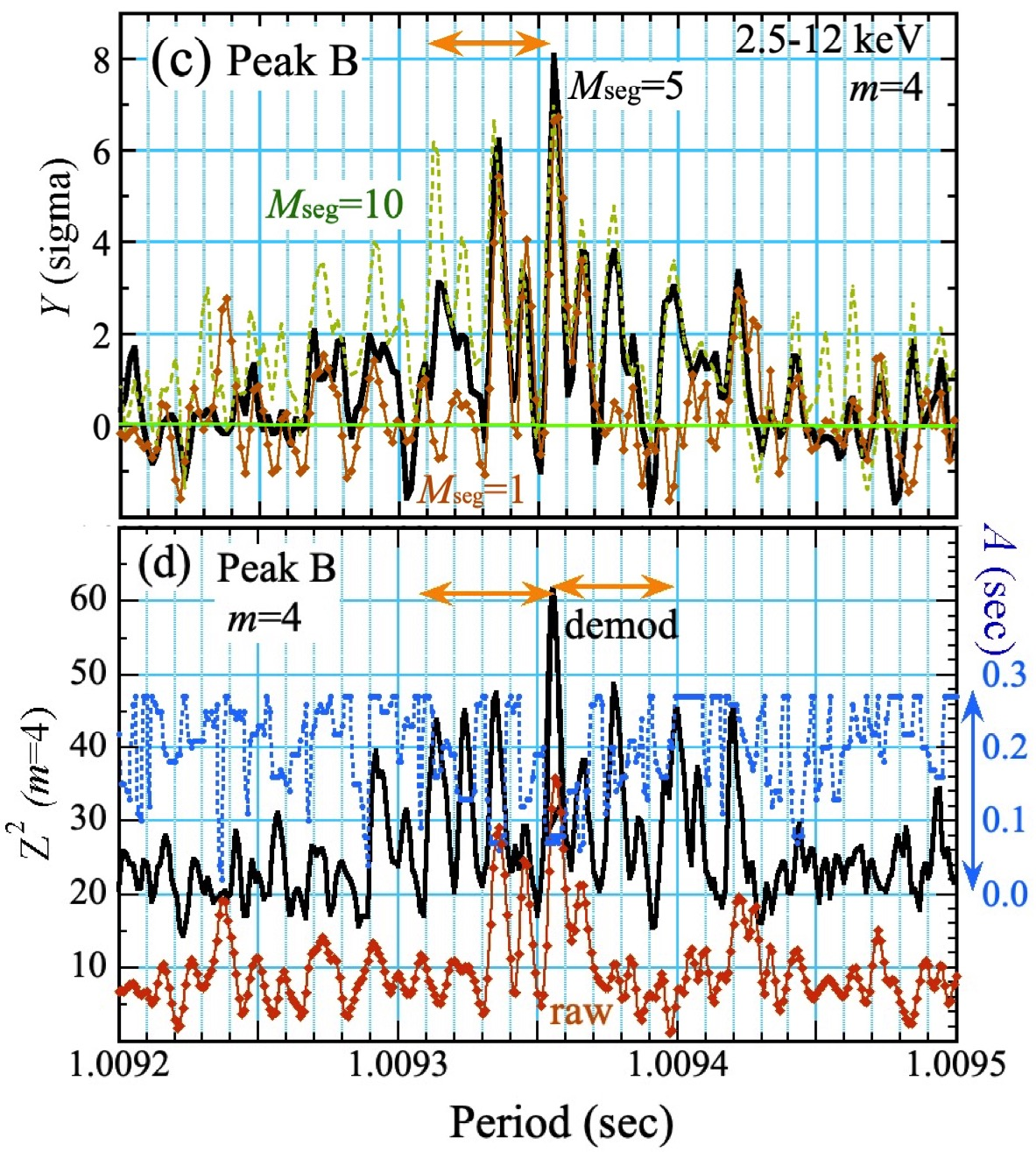}
\caption{(a) A phase-sorted  PG for 2.5--12 keV CCO photons in ASCA93,
calculated employing  $m=4$ and $\Mseg=5$.
The 10--30 s region is truncated.
(b)  Same,
but derived using the demodulation method assuming $T=24.0$ ks.
(c) Details of the PG in  (a) around Peak B.
Results with $\Mseg=1$ (brown with dots), $5$ (solid black),
and $10$ (dashed green) are overlaid.
The ordinate refers to equation~(\ref{eq:scaled_Z2}).
A horizontal arrow shows the expected fringe separation.
(d) Details of the PG in (c) around Peak B.
The amplitude $A$ is shown in dashed blue (right ordinate).
The dotted brown curve labeled ``raw" is the result before demodulation. 
{Alt Text: Four periodograms of 1E 1613 obtained with ASCA,
in photon energies of 2.5 to 12 kilo-electronvolts.
Top two cover the 0.3 to 30 second period range, 
whereas bottom two are their expanded views.}
}
\label{fig:f2_Pscan_ASCA93}
\end{figure}
%%%%%%%%%%%%%%%%%%%%%%%%%%%%%%%%%%%%

We applied this new analysis to the ASCA93 CCO data,
changing $\Mseg$ from 3 to 10,
and  $\EL$ from 3.0 keV down to 2.0 keV.
Also,  $\EU=10$ keV and $\EU=12$ keV were  tested.
These PGs consistently revealed  a pair of peaks,
which became  most prominent for $\Mseg=5$, $\EL=2.5$ keV, and $\EU=12$ keV.
The result under this condition is presented in figure~\ref{fig:f2_Pscan_ASCA93}a,
where the ordinate represents $\Sigma \zz$,
i.e., the $\zz$ values summed over the constituent five PGs.
The plot excludes the $10 < P \le 30$ s period range,
where no interesting feature was seen.

Figure~\ref{fig:f2_Pscan_ASCA93}a clearly reveals
the aforementioned pair of peaks,  denoted as A and B, 
of which the parameters are given as
\begin{subequations}
\begin{align}
{\rm A}:\;&\Sigma \zz=115.5,~P=0.5046777(4)\;{\rm s}
\label{eq:ASCA93_periodA}\\
{\rm B}: \;&\Sigma \zz=111.2,~P =1.0093558(8)\;{\rm s} \equiv P_{93}.
\label{eq:ASCA93_periodB}
\end{align}
\end{subequations}
The two periods  are  in the 1:2 harmonic ratio well within the errors.
Thus,  B can be regarded as the fundamental of an intrinsic periodicity in the data,
and A its second harmonic (half in period and twice in frequency).
However, the heights of the two peaks 
with $m=4$ are not independent,
because they are both contributed by the Fourier components 
with periods $P_{93}/2$ and $P_{93}/4$.

Figure~\ref{fig:f2_Pscan_ASCA93}c expands the PG in panel (a) around $P_{93}$
of equation~(\ref{eq:ASCA93_periodB}).
For easy comparison of the results with different  $\Mseg$, 
the ordinate was  converted to a new variable $Y$ defined as
\begin{equation}
Y \equiv \frac{\Sigma \zz - 8 \Mseg}{4 \sqrt{\Mseg}}~.
\label{eq:scaled_Z2}
\end{equation}
Since $\zz$ has the mean as $2m=8$
and the  standard deviation as $\sqrt{4m}=4$,
we expect $\Sigma \zz$, which sums $\Mseg$ independent variables,
to have a mean of $8 \Mseg$ and a standard deviation of $4 \sqrt{\Mseg}$.
Hence, this equation  expresses a procedure of subtracting the mean
and normalizing to the standard deviation.
 As a result, $Y$ represents the conventional ``sigma" value.

In figure~\ref{fig:f2_Pscan_ASCA93}c,
PGs with different $\Mseg$ commonly show a peak at $P_{93}$.
Even the $\Mseg=1$ result, identical to the raw PG,
reveals a  peak reaching  $Y=6.77$ (or $\zz=35.1$),
which is not much lower than
that ($Y=8.09$) for $\Mseg=5$.
Although this is in an apparent  contradiction to 
the pulse absence in \S~\ref{subsubsec:ana_ASCA93_simplePG},
this  paradox has a clear solution.
As  $\Mseg$ increases, the distribution of $\Sigma \zz$
starts deviating from chi-square functions,
and approaches a Gaussian due to the central limiting theorem.
Depending on $\Mseg$,
the same value of $Y$ translates to very different  probabilities;
for example, $Y=7$ means 
an upper probability of $1.76 \times 10^{-5}$ if $\Mseg=1$ ($\zz=36.0$),
but it decreases with $\Mseg$,
finally converging to the  Gaussian case of $1.28\times 10^{-12}$.
This provides a good lesson
that quoting  a probability in ``sigmas" is meaningless,  
unless the underlying probability distribution is rigorously specified.

To reconfirm the above argument,
the periods of  equations~(\ref{eq:ASCA93_periodA}) and (\ref{eq:ASCA93_periodB}) 
actually appear, in the raw PG ($\Mseg=1$) with $m=4$,
as local peaks with $\zz\approx 31$ and $\zz\approx 36$, respectively.
However, when plotted over wider period ranges,
they become exceeded by many noise peaks.
The phase-sorted analysis has made the two peaks far more outstanding,
by mitigating the PPM disturbance working on them,
and averaging out the noise peaks.

In figure~\ref{fig:f2_Pscan_ASCA93}c,
we observe clear side lobe fringes.
While those in the PGs with $\Mseg=5$ and $\Mseg=10$  
can be readily ascribed to the beat between the
artificial data gaps repeating with $T$
(see orange arrows) or $2T$,
those in the $\Mseg=1$ result (the raw PG) is somewhat puzzling, 
because the artificial data gaps are absent,
and the flux is not strongly modulated at  $T$.
Probably, the $\Mseg=1$ side lobes are produced 
because the pulse fraction is actually modulated mildly at 24.0 ks.

%--------------- 3.3.3 ---------------------
\subsubsection{Demodulation analysis}
\label{subsubsec:ana_ASCA93_demodulation}
%--------------- 3.3.3 ---------------------
The results in figure~\ref{fig:f2_Pscan_ASCA93}a
could be some artifacts specific to our new analysis tool.
We hence cross checked them
by a different tool, called ``demodulation analysis''.
It was first devised by \citet{Makishima14} to search Suzaku data 
of a magnetar for the PPM period $T$, given  the pulse period $P$.
By modifying the logic,
we here assume that an {\em unknown} fast pulsation in the  data  suffers 
the PPM  at a {\em known}\/ long period $T$ of  equation~(\ref{eq:6.67hr}).
Then,  the arrival time $t$ of each photon (including background) 
is modified to $t- \Delta t$, using a formula
\begin{equation}
\Delta t = A \sin\left(2 \pi t/T - \psi_0 \right),
\label{eq:demodulation}
\end{equation}
where $A \geq 0$ is an assumed modulation amplitude,
and $\psi_0$  (0 to $2 \pi$) is the phase of the first photon.
This  $\Delta t$ means a correction of the pulse arrival time 
for delays (or advances if $\Delta t<0$),
relative to the case of  precisely constant periodicity.
Then, a ``demodulated PG'' is computed 
as before over a certain period interval,
but at each $P$, we maximize $Z_m^2$
by scanning $A$ and $\psi_0$  over appropriate ranges.
We fix  $T=24.0$ ks, unless otherwise stated.
A significant increase of  $Z_m^2$ at a certain $P$
would be regarded as a pulse candidate.

Compared to the phase-sorted PG  
(subsubsection~\ref{subsubsec:ana_ASCA93_phase-sorted}),
two advantages of this method are 
that we can recover full pulse-phase coherence among all photons,
and the method has afforded the detections of free precession
from the seven magnetars altogether \citep{Makishima24b}.
On the other hand, its clear disadvantage is 
the required  much longer computational time.

Using the 2.5--12 keV energy interval 
which was found optimum in subsubsection~\ref{subsubsec:ana_ASCA93_phase-sorted},
together with $m=4$ and $\eta=5$, 
we computed a demodulated PG over the same period interval.
We fixed $T=24.0$ ks,
scanned $A$ from 0 to $\approx P/4$
(a plausible maximum; \cite{Makishima16})  typically in 10 steps,
and $\psi_0$  from 0  to $360^{\circ}$ with $10^{\circ}$ step.
The obtained PG is shown in figure~\ref{fig:f2_Pscan_ASCA93}b,
where  a  few outstanding peaks are seen.
Of them,  the highest two reproduce, within $\approx 1~\mu$s,
equations~(\ref{eq:ASCA93_periodA}) and (\ref{eq:ASCA93_periodB})
found in panel (a),
although their relative heights  have reversed.
The derived modulation parameters are summarized in table~\ref{tbl:demodulation}.
The two peaks demand nearly the same set of $(A, \psi_0)$,
suggesting them to have a common origin.
Thus, our results from the phase-sorted PGs have been
reconfirmed and reinforced.
For reference, a peak with $\zz \approx 40.1$ is observed also at $P_{93}/4$.

%%table 2 table 2 table 2 table 2 %%%%%%%%%%%%%%%%%%%%%%%%%%%%%%%%
\begin{table*}
\caption{Results of the demodulation analysis of \oneE\ using $m=4$.}
\label{tbl:demodulation}
\vspace*{-2mm}
\begin{center}
\begin{tabular}{lcllcrcccc}
\hline %-----------------------------------
Data          &  time 0$^{*}$  & Energy range & {~~~~~$P$ (s)} &  $\zz$  &  $A$ (s)  & $ \psi_0$
                                  & ${\Pchone}^{\dagger}$ &${\Pch}^{\dagger}$ &PF$^{\ddagger}$  \\
\hline %-----------------------------------
\hline %-----------------------------------
ASCA93 &49216.9211 & 2.5--12 keV & 1.0093558(8) =$P_{93}$&   61.5   &  0.08  & $40^\circ$ &$1.3\times 10^{-8}$ &0.54\%&0.16\\
%ASCA93 A &2.5--12  keV &  0.5046778(5)  &  57.9 & 0.07  & $50^\circ$& 49216.92107  & $1.3 \times 10^{-7}$   \\
%\hline %-----------------------------------

XMM01  &52155.8016 & 2.7--10 keV &$1.0096311(70) = P_{01}$&  44.2  & 0.24  & $40^\circ$  &$1.2 \times 10^{-5}$&0.92\% &0.14\\
%\hline %-----------------------------------
XMM05 &53605.3177  & 1.4--10  keV &$1.0097713(6) =P_{05}$ &  47.2   & 0.28    &$350^\circ$   &  $1.1 \times 10^{-5}$ &19\%&0.07\\ 
%-----------------------------------
XMM16 & 57619.7342  &6.6--12  keV & $1.0101518(12) =P_{16}^{\rm X}$ &47.7  &0.25 &$290^\circ$  &  $9 \times 10^{-6}$&11\%&0.25\\
%-----------------------------------
NuS16 &57563.6136&11--60  keV & $1.0101480(7) =P_{16}^{\rm N}$ &56.9$^{\|}$  &0.20 &$140^\circ$  &   ---$^{\S}$ & ---$^{\S}$&0.30\\
           &   & 3--6 keV      & $1.0101479(9) =P_{16}^{\rm N}$ &49.0$^{\|}$ &0.29 &$0^\circ$  &   ---$^{\S}$ & ---$^{\S}$&0.09\\    
%-----------------------------------
NuS17 & 57906.3899  &5.1-12 keV & $1.0101794(10)=P_{17}$&48.6$^{\|}$ &0.20 & $10^\circ$ & ---$^{\S}$ &---$^{\S}$ & 0.24\\
\hline %-----------------------------------
\end{tabular}
\end{center}
\begin{itemize}
\setlength{\itemsep}{0mm}
  \item[$^{*}$] MJD of the first photon in the data.
  \item[$^{\dagger}$] $\Pchone$ and $\Pch$ are pre- and post-trial probabilities, respectively.
The ASCA93 results refer to Peak B in the demodulated PG and the period search range of 0.3--30 s.
$\Pch$  of the other data sets are calculated against equation~(\ref{eq:P_search_range}).
  \item[$^{\ddagger}$] The pulse fraction defined in subsubsection~\ref{subsubsec:Pr_XMM01}.
 \item[$^{\S}$] Not evaluated due to technical difficulties (see text).
 \item[$^{\|}$]Derived using  only  a limited interval of the modulation phase.
 \end{itemize}
\end{table*}

An expanded view of figure~\ref{fig:f2_Pscan_ASCA93}b is given in panel (d).
It  is very similar to (c),
still showing the interference structure,
even though the data are now free from the 24 ks data gaps,
and the flux is not strongly modulated at this period.
Again, this behavior is possibly due to the 24 ks variation of the pulse fraction,
which we found to amount to a factor of $\sim 2$.
The dashed blue trace gives the values of $A$ that maximize $\zz$.
Also shown  is the PG before demodulation,
which is identical to the $\Mseg=1$ result in panel (c)
except the difference in the ordinate variable.

%--------------- 3.3.4 ---------------------
\subsubsection{Statistical significance of the $\sim 1.01$ s periodicity}
\label{subsubsec:ana_ASCA93_significance}
%--------------- 3.3.4 ---------------------

A vital question  is the statistical significance of the $\sim 1.01$ s periodicity.
We carried out its evaluation,
using the highest Peak B with $\zz=61.5$ in the demodulated PG 
of figure~\ref{fig:f2_Pscan_ASCA93}d.
As detailed in  Appendix 1, 
the probability for a peak with $\zz\geq 61.5$ 
to appear by chance fluctuations,
{\it at a given period}  in this PG,
was found to be $\Pchone(P_{93}) \approx 1.3 \times 10^{-8}$.
When the trial number in the period search is multiplied,
we obtain a post-trial probability of $\Pch(P_{93}) \approx 0.54\%$
(see Appendix 1).
This means a chance probability of finding $\zz\geq 61.5$ 
{\it at any period} from 0.3 s to 30 s.}
We regard this as a representative value for ASCA93,
and list it in table~\ref{tbl:demodulation}.
As a working hypothesis, hereafter we regard $P_{93}$
as the  spin period of  1E~1613 as of 1993 August.
For reference, a comparable value of $\Pch$ 
is derived for the Peak A in the phase-sorted PG.

%separately for the two analysis techniques.
%As to the phase-sorted PG in figure~\ref{fig:f2_Pscan_ASCA93}a,
%we find  post-trial probabilities of $\Pch({\rm A})\approx 2.1\%$ 
%for the second harmonic,
%and $\Pch({\rm B}) \approx 65\%$  for the fundamental,
%considering overall trial numbers.

%%%%%%%%%%%% fig.3 %%%%%%%%%%%%%%%%%%%%%%%
\begin{figure}
\includegraphics[width=7.85cm]{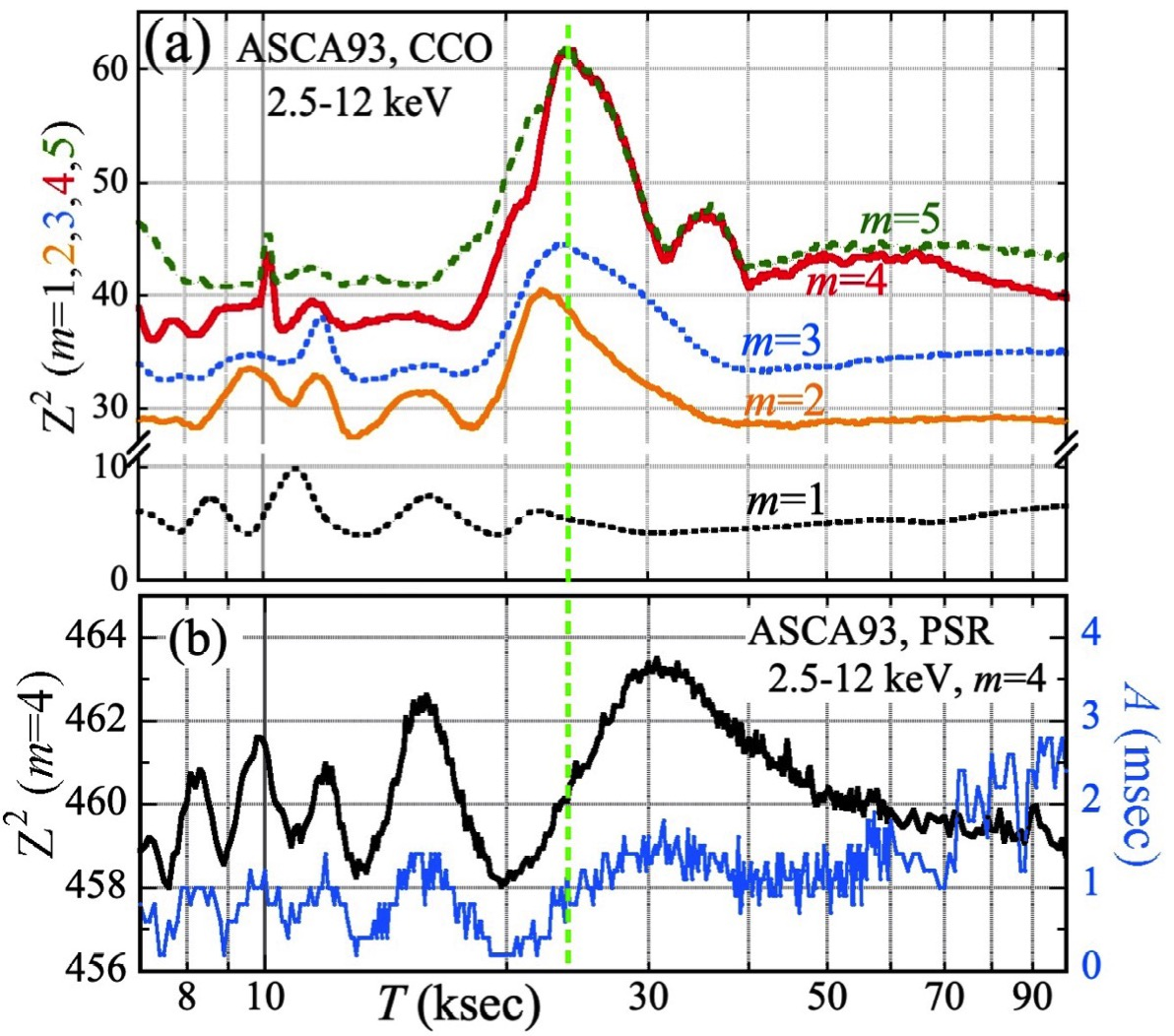}
\caption{(a) Demodulated $Z_m^2$ values from the ASCA93 CCO data in 2.5--12 keV,
calculated with different $m$ and shown against the assumed value of  $T$.
The pulse period is constrained to a vicinity of $P_{93}$.
(b) A control study using the 69 ms pulsar in ASCA93,
in the same energy range.
The optimum value of $A$ is shown in the lower trace (blue), using the right ordinate.
{Alt Text: Top panel shows the significance of the 1.01 second pulsation  with ASCA, 
as a function of the modulation period 
from 7 to 100 kilo-seconds.
Bottom panel is a control study using the 69 millisecond pulsar.} 
}
\label{fig:f3_Tscan_ASCA93}
\vspace*{-4mm}
\end{figure}
%%%%%%%%%%%%%%%%%%%%%%%%%%%%%%%%%%%

%--------------- 3.3.5 ---------------------
\subsubsection{An independent determination of $T$}
%--------------- 3.3.5 ---------------------
To reinforce the results derived so far, we carried out  yet another effort.
The present analysis is an attempt to  search a region
on the $(P,T)$ plane for the highest  $\zz$ peak.
The PGs in figure~\ref{fig:f2_Pscan_ASCA93} 
provide  cross sections of this peak along the $P$ axis,
in which  $T=24.0$ ks is pre-specified.
We may alternatively compute a cross section  along the $T$ dimension,
with $P\approx P_{93}$ fixed.
So, we computed $Z_m^2$ incorporating equation~(\ref{eq:demodulation}),
as a function of $T$ rather than $P$,
changing $T$ in the same manner as equation~(\ref{eq:dp}).
At each $T$, we maximized $Z_m^2$, 
allowing $P$  to float by $\pm 2~\mu$ s around $P_{93}$,
and scanning $A$ and $\psi_0$  in the same manner as before.

When applied to the 2.5--12 keV  CCO  data from ASCA93,
this analysis yielded figure~\ref{fig:f3_Tscan_ASCA93}a,
where  results with $m=1$ through $m=5$ are shown together.
For $m=3$ to 5, the pulse significance indeed becomes maximum at $T\approx 24$ ks,
or more specifically at 
\begin{equation}
T = 23.8^{+3.2}_{-1.5}~{\rm ks}~~(m=4).
%\label{eq:ASCA93_T}
\end{equation}
This  peak has a relative enhancement by
 $\Delta \zz \sim 20$ above  the surroundings,
and the probability for this increment to take place by chance is estimated as 
$\exp(-\Delta \zz/2) \approx 4.5 \times 10^{-5}$ \citep{Makishima23}.
After considering the effective number of trials in $T$, about 40, 
the probability still remains at $ \approx 0.2\%$.
The peak is hence  statistically  significant.
Solely via timing studies, we have thus arrived at an estimate of $T$ 
that is consistent with equation~(\ref{eq:6.67hr}),
even though the 24 ks  flux  variation of 1E~1613  is unclear in ASCA93.
%We also confirm that $m=4$ is the best choice.

%--------------- 3.3.6 ---------------------
%\subsubsection{A control study using the 69 ms pulsar}
%\label{subsubsec:ana_ASCA93_control}
%--------------- 3.3.6 ---------------------

The $\sim 1.01$ s periodicity 
could  be some instrumental artifacts of either  ASCA  or the GIS,
rather than specific to 1E~1613.
We hence applied the same analysis to the ASCA93 data 
of the ideal reference, the 69 ms pulsar,
and computed its demodulated $\zz$ as a function of $T$.
To reproduce the condition for 1E~1613,
the energy range was changed from the original 3--10 keV  to 2.5--12 keV,
and the harmonic number from $m=2$ (figure~\ref{fig:f1_PG_psr93+97}) to $m=4$.

The derived $T$-dependence of $\zz$ of the pulsar 
is presented in figure~\ref{fig:f3_Tscan_ASCA93}b.
Although some oscillatory behavior is seen,
 $\zz$ remains mostly within $\sim \pm 4.72$ (68\% error associated with a peak value)
of the average at $\sim 461$.
In particular, no enhancement is seen at $T=24$ ks.
Also, the  modulation amplitude remains
$A < 3$ ms, or $ <0.043P$.
We  conclude that the 69 ms pulsar signal is free from PPD  
at any period between $T=7.0$ and 100 ks, including  24.0 ks.

%%%%%%%%%%%% fig.4 %%%%%%%%%%%%%%%%%%%%%%%
\begin{figure}
\includegraphics[width=8.3cm]{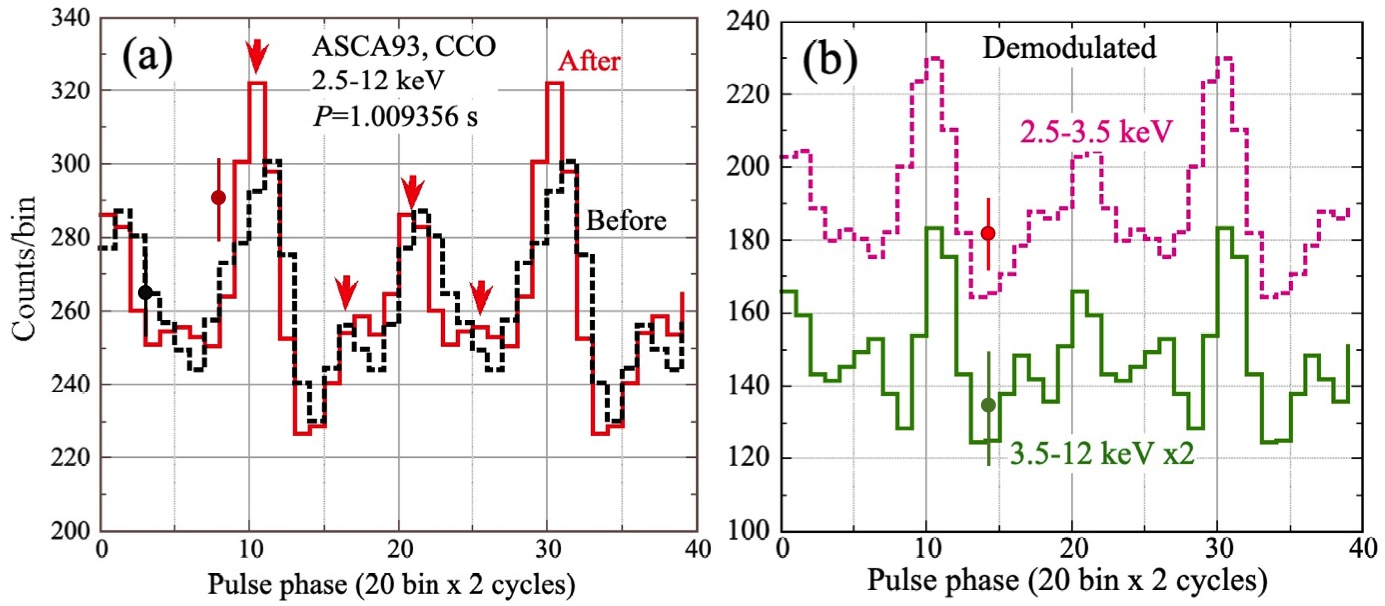}
\caption{Pulse profiles of 1E~1613 in ASCA93.
(a)  Results in 2.5--12 keV folded at $P_{93}$,
before (dashed black) and after (solid red) the demodulation using  $T=24.0$ ks.
(b) Demodulated pulse profiles in  the 2.5--3.0 keV and 3.5--12 keV energy ranges.
{Alt Text: Folded pulse profiles with ASCA.}
}
\label{fig:f4_Pr_ASCA93}
\vspace*{-5mm}
\end{figure}
%%%%%%%%%%%%%%%%%%%%%%%%%%%%%%%%%%%

%--------------- 3.3.6 -------------
\subsubsection{Pulse profiles}
\label{subsubsec:ana_ASCA93_Pr}
%--------------- 3.3.6 ------------
%
When evaluating a candidate periodicity,
folded pulse profiles provide  important information.
Figure~\ref{fig:f4_Pr_ASCA93}a shows
2.5--12 keV pulse profiles of 1E~1613 in ASCA93,
before (dashed black) and after (solid red) the demodulation,
which employs  equation~(\ref{eq:demodulation}) and the parameters in table~\ref{tbl:demodulation}.
Both are folded at $P_{93}$,
and smoothed with a running average which combines three adjacent bins
(e.g., \cite{Makishima21a}).
As a result, the error associated with  each data bin
is  0.61 times the Poissonian value.
The  profiles are both double-peaked,
with  a larger main peak and a smaller sub peak.

%Para 2 Four-peak structure
The demodulation is seen to make the main peak higher,
and  enhance fine structures in the profile.
As a result, the demodulated profile acquires a clear
4-peak structure, as indicated by four downward arrows (in red).
This reinforces the pulse reality,
because similar 4-peak structures are rather common to magnetars
(\cite{Makishima19}, 2021ab, 2024a),
even though its interpretation is still an open question.

%Para 3 Harmonic contribution
When the Fourier power of  the  demodulated profile 
in figure~\ref{fig:f4_Pr_ASCA93}a is summed up to the 4th harmonic,
the result gives the $P_{93}$ peak height,  $\zz=61.5$.
It is contributed by individual harmonic,
from the fundamental to the 4th harmonic,
by 6.7\%, 53.2\%,10.0\%, and 30.1\%, respectively.
Thus, the second harmonic is dominant,
in agreement with the double-peaked pulse profile,
and the strong 2nd harmonic signal emerging
in panels (a) and (b) of figure~\ref{fig:f2_Pscan_ASCA93}.
The 4th harmonic gives the  second strongest contribution.

%Para 4 Energy dependence
In figure~\ref{fig:f4_Pr_ASCA93}b, the pulse profiles were derived in
a softer (2.5--3.5 keV) and a harder (3.5--12 keV)  energy intervals,
both demodulated under the same condition as the red profile in panel (a).
They are very much alike,
except that the hard-band profile exhibits
a sharper main peak  and a more prominent 4-peak structure.
This energy dependence is similar to the case of 
the magnetar SGR~1900+14 \citep{Makishima21b}.

%%%%%%%%%%%% fig.5 %%%%%%%%%%%%%%%%%%%%%%%
\begin{figure}
\includegraphics[width=7.4cm]{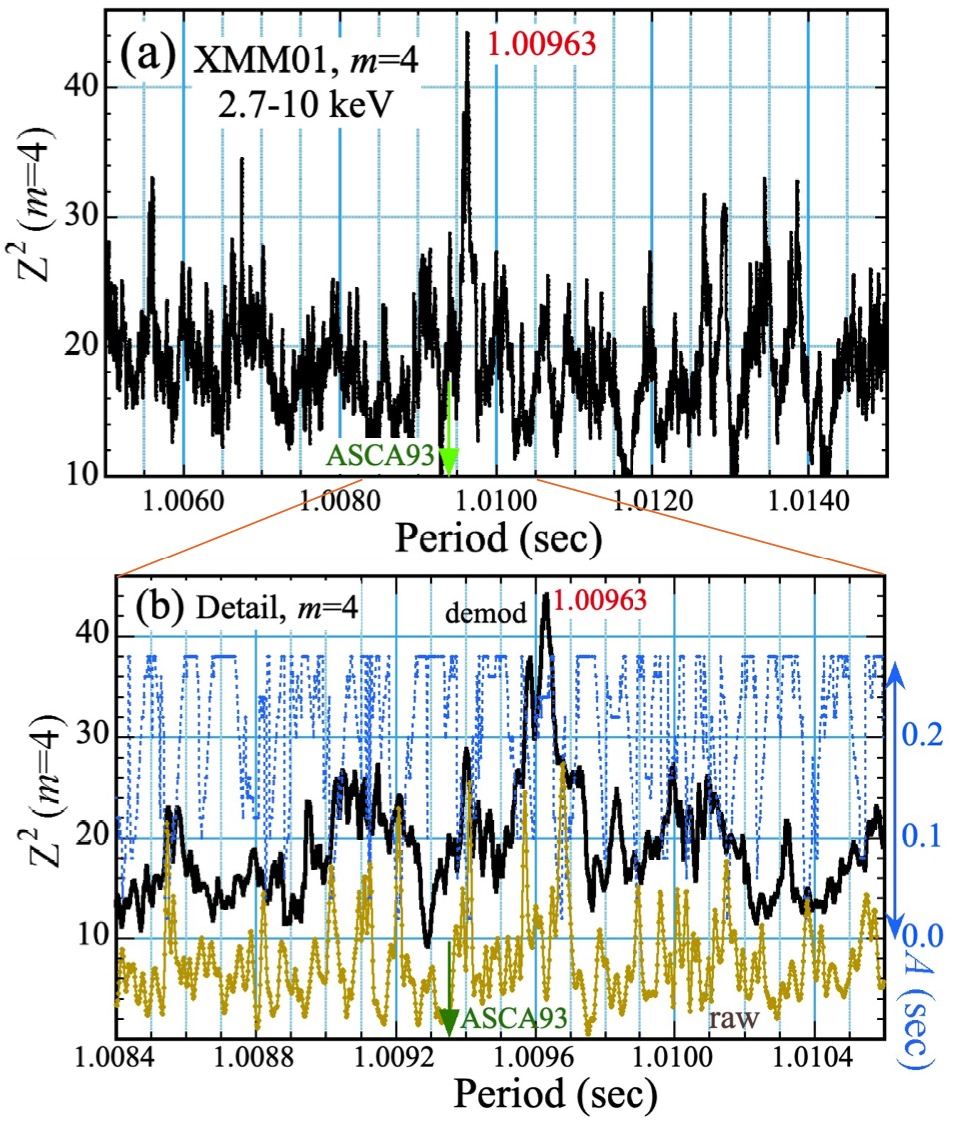}
\caption{Results on 1E~1613 from the 2.7--10 keV XMM01 data.
(a) A demodulated PG with $m=4$, 
derived in the same way as figure~\ref{fig:f2_Pscan_ASCA93}b
but over a more limited period range of equation~(\ref{eq:P_search_range}).
The  downward arrow indicates $P_{93}$.
(b) An expanded view (in solid black) of  panel (a)  around the peak, 
and  the optimum $A$ in dotted blue (right ordinate).
The  $m=4$  PG before the demodulation (labeled ``raw")
is superposed as an ocher line.
{Alt Text: Periodograms with XMM-Newton in 2001,
in 2.7 to 10 kilo-elecronvolts.}
}
\label{fig:f5_Pscan_XMM01}
\vspace*{-5mm}
\end{figure}
%%%%%%%%%%%%%%%%%%%%%%%%%%%%%%%%%%%

\vspace*{-2mm}
%=========== 3.4 =====================
\subsection{XMM-Newton data in 2001 (XMM01)}
\label{subsec:ana_XMM01}
%=========== 3.4 =====================
To reinforce the evidence from ASCA93
for the $\sim 1.01$ s pulsation of 1E~1613,
we proceed to the analysis of XMM01.
We then  need to redefine the period search range,
using some assumptions on $\dot P$.
As a simple way, we may assume
that the characteristic age of 1E~1613,
$\tau_{\rm c} \equiv P/2 \dot P$, 
should be comparable to the age of the host SNR, 
2.0-4.4 ky \citep{RCW_age2.0,RCW_age4.4}.
Selecting the shorter of the two,
we obtain a nominal spin-down rate as
$
 \dot P =7.9 \times 10^{-12} ~~{\rm s~s}^{-1}.
$
To be conservative, let us  allow a $\sim 2.5$ times larger value,
$\dot P \sim  2.0 \times 10^{-11} ~~{\rm s~s}^{-1}$, as an upper limit.
This implies a  spin down by $\Delta P \sim 5$ ms 
in 8 yrs from ASCA93 to XMM01.
Hence, we set the longest limit of the period search  at 1.015 s.
The shortest limit will be somewhat arbitrarily set at 1.005 s,
even though the case of $P< P_{93}$ is unphysical.
The selected period search range, 
hereafter called  a {\em fiducial interval},
becomes approximately symmetric around $P_{93}$ as
\begin{equation}
P=1.005-1.015~{\rm s}.
\label{eq:P_search_range}
\end{equation}
%

%----------------------------------
\subsubsection{PGs from XMM01}
\label{subsubsec:PG_XMM05}
%----------------------------------
Over the  fiducial period interval,
we calculated a raw PG from XMM01.
After ASCA93,  the 2.5--10 keV band was used,
together with  $m=4$.
For the reason described in subsection~\ref{subsec:obs_XMM},
$\EU$ was set at 10 keV  rather than 12 keV.
However, no significant peak exceeding $\zz \sim 35$ was found.

We hence incorporated the demodulation 
with equation~(\ref{eq:demodulation}),
because the XMM01 data are too short to analyze 
with the phase-sorted technique.
We again fixed $T=24.0$ ks.
Then, in the PG, a dominant peak with $\zz=41.3$  has
appeared at a period %
\begin{equation}
P= 1.0096311(70) ~{\rm s}~ \equiv P_{01},
%\label{eq:P01}
\end{equation}
which is  longer than $P_{93}$ by $\Delta P =0.275$ ms.
As shown in figure~\ref{fig:f5_Pscan_XMM01}a,
the peak became higher to $\zz=44.2$
(with $P$  unchanged),
when $E_L$ is raised from 2.5 keV to 2.7 keV.
Details around the peak is 
shown in figure~\ref{fig:f5_Pscan_XMM01}b,
together with some additional information,
while relevant parameters are given in table~\ref{tbl:demodulation}.
The peak width ($\sim 70~\mu$ s) is much larger
than that of $P_{93} (\sim 6~\mu$ s),
because of the short length of XMM01.

In Appendix 1, we examined the statistical significance of this periodicity, 
in the same way as for ASCA93.
Then, it was found to have
$\Pch(P_{01}) \approx 0.92\%$ (table~\ref{tbl:demodulation}),
when evaluated over the fiducial period interval,
and considering trials in $P$ and $\EL$ 
(but not in $m$ because we tried $m=4$ only).
Hence, the signal at $P_{01}$ is likely to be real as well,
and to represent the object's pulse period in 2001.
We are then allowed to connect $P_{93}$ with this $P_{01}$, 
to derive
\begin{equation}
\dot P = (1.09 \pm 0.04)\times 10^{-12}~{\rm s~s}^{-1}
\label{eq:spindown_9301}
\end{equation}
which is reasonable for a young pulsar (see subsection~\ref{subsec:discuss_nature}).

%%%%%%%%%%%% fig.6 %%%%%%%%%%%%%%%%%%%%%%%
\begin{figure}
\includegraphics[width=8.5cm]{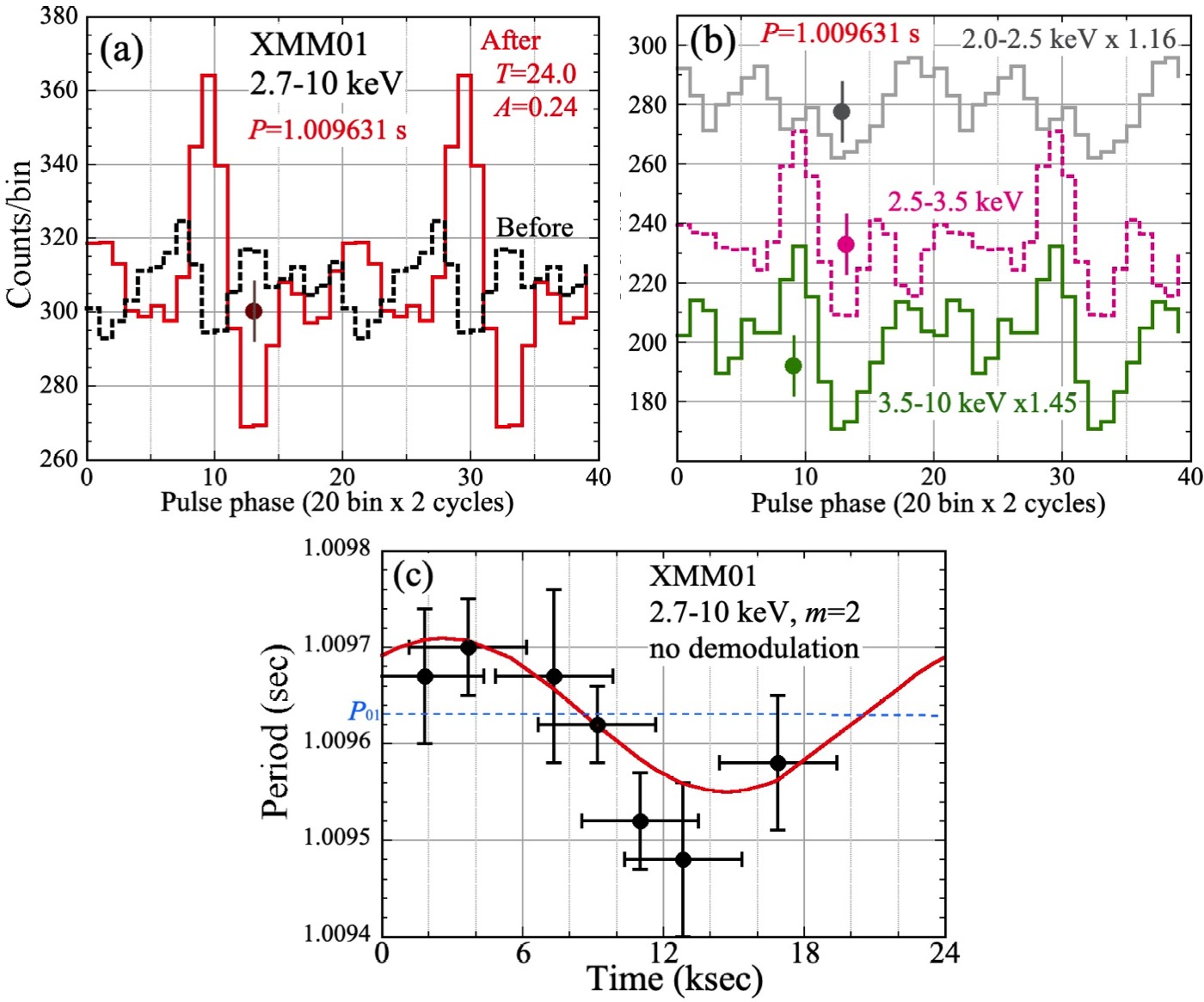}
\caption{
(a) XMM01 pulse profiles   in 2.7--10 keV, 
folded at $P_{01}$ and shown in the same way as figure~\ref{fig:f4_Pr_ASCA93}a.
(b) Pulse profiles at $P_{01}$ derived in three energy ranges.
(c) Instantaneous pulse periods in XMM01 determined with $m=2$ in 2.7--10 keV,
without demodulation,
shown against the elapsed time from the data start.
The solid red curve is a prediction by the optimum modulation parameters,
and the dashed horizontal line indicates $P_{01}$.
{Alt Text: Folded pulse profiles from XMM-Newton in 2001.
The bottom panel shows instantaneous pulse periods without demodulation.}
}
\label{fig:f6_Pr_XMM01}
\vspace*{-5mm}
\end{figure}
%%%%%%%%%%%%%%%%%%%%%%%%%%%%%%%%%%%
%
\vspace*{-2mm}
%----------------------------------
\subsubsection{Pulse profiles from XMM01}
\label{subsubsec:Pr_XMM01}
%---------------------------------

The demodulated 2.7--10 keV pulse profile of XMM01,
shown in solid red  in figure~\ref{fig:f6_Pr_XMM01}a,
is amazingly similar to that with ASCA93 (figure~\ref{fig:f4_Pr_ASCA93}a).
Also similar is the pulse fraction, 
defined as  $\tfrac{1}{2}{\rm (maximum- minimum)/average}$,
which is  0.16 for ASCA93 and 0.14 for XMM01
(table~\ref{tbl:demodulation}).
These support that  $P_{93}$ and $P_{01}$ represent the same phenomenon,
i.e., the fast rotation of the NS in 1E~1613.
As a difference, 
the raw profile in XMM01 is more strongly suppressed than that in ASCA93.
This is because the  modulation amplitude, $A \approx 0.08$ s in ASCA93,
increased to  $A\approx 0.24$ s in XMM01 (table~\ref{tbl:demodulation}),
which is close to the upper limit, $A \approx P/4$.
Such a change in $A$ is not rare among magnetars \citep{Makishima19}.

Figure~\ref{fig:f6_Pr_XMM01}b 
presents energy-sorted pulse profiles in XMM01,
all folded at $P_{01}$ using the same set of $(A, \psi_0)$.
The 2.5--3.5 and 3.5--10 keV results again look alike,
but  the 2.0--2.5 keV profile is considerably different.
This is also evidenced by the $P_{01}$ peak height  in the PG,
which decreases as $\EL$ is lowered from 2.5 keV downward.
Thus, some changes in the pulse properties 
are suggested at $<2.5$ keV.
This information was  unavailable with the 
SNR-contaminated ASCA data.

Utilizing the  high CCO flux  in XMM01, 
together with the large effective area of XMM-Newton,
we time-divided the data into overlapping ten segments.
From each of them, we calculated  a  2.7--10 keV PG,
{\em without demodulation} and using $m=2$.
Because the PPM effect is mild within each segment,
seven out of the ten PGs  showed clear peaks at $\sim P_{01}$.
We were hence able to determine the instantaneous pulse periods
as shown in figure~\ref{fig:f6_Pr_XMM01}c.
These  periods are consistent with  the solid curve in red,
which is predicted by $(A,\psi_0)$ in table~\ref{tbl:demodulation}.
The two turning points of this curve  are confirmed,
in the ``raw'' PG in figure~\ref{fig:f5_Pscan_XMM01}b,
as a pair of horn-like features on both sides of $P_{\rm 01}$.
These non-trivial  properties of $P_{01}$
strengthen its association with the 6.67 hr periodicity.

\vspace*{-3mm}
%=========== 3.5 ============
\subsection{XMM-Newton data in 2005 (XMM05)}
\label{subsec:ana_XMM05}
%=========== 3.5 ==============
Let us next analyze XMM05, hoping to find a periodicity 
that is consistent with $P_{93}$ and $P_{01}$.
In this observation, 1E~1613 was about 4 times fainter than in XMM01 (table~\ref{tbl:datasets}),
but this may be compensated for 
by the $\sim 4$ times longer exposure of XMM05.
On this occasion, 1E~1613  exhibited the 24.0 ks flux variation clearly,
with a rather sinusoidal waveform (DL06).

%
%----------------- 3.5.1 --------------------
\subsubsection{PGs from XMM05}
\label{subsubsec:ana_XMM05_PGs}
%----------------- 3.5.1 --------------------
%
If the fast pulsation of 1E~1613 has a  constant $\dot P$
as in equation~(\ref{eq:spindown_9301}),
the pulse period in XMM05 is predicted,
from  $P_{93}$ and $P_{01}$, as
\begin{equation}
{\rm XMM05} :P=1.009769(15)~{\rm s}.
\label{eq:period_prediction_XMM05} 
\end{equation}
The constant-$\dot P$ assumption, however,
may not  necessarily be warranted.
Therefore, we retained the fiducial search interval  
[equation~(\ref{eq:P_search_range})] used for XMM01,
and calculated an $m=4$ PG in 2.5--10 keV.
Since no outstanding periodicity was seen therein,  
we next incorporated the demodulation correction, 
keeping the same period and energy regions.
Considering the long data span, 
$T$ was allowed to vary from 23.8 to 24.2 ks.
The demodulated PG showed several peaks up to $\zz \sim 45$,
but none of them was dominant,
and none fell in or close to the error range of equation~(\ref{eq:period_prediction_XMM05}).

\begin{figure}
\vspace*{-3mm}
\includegraphics[width=8.cm]{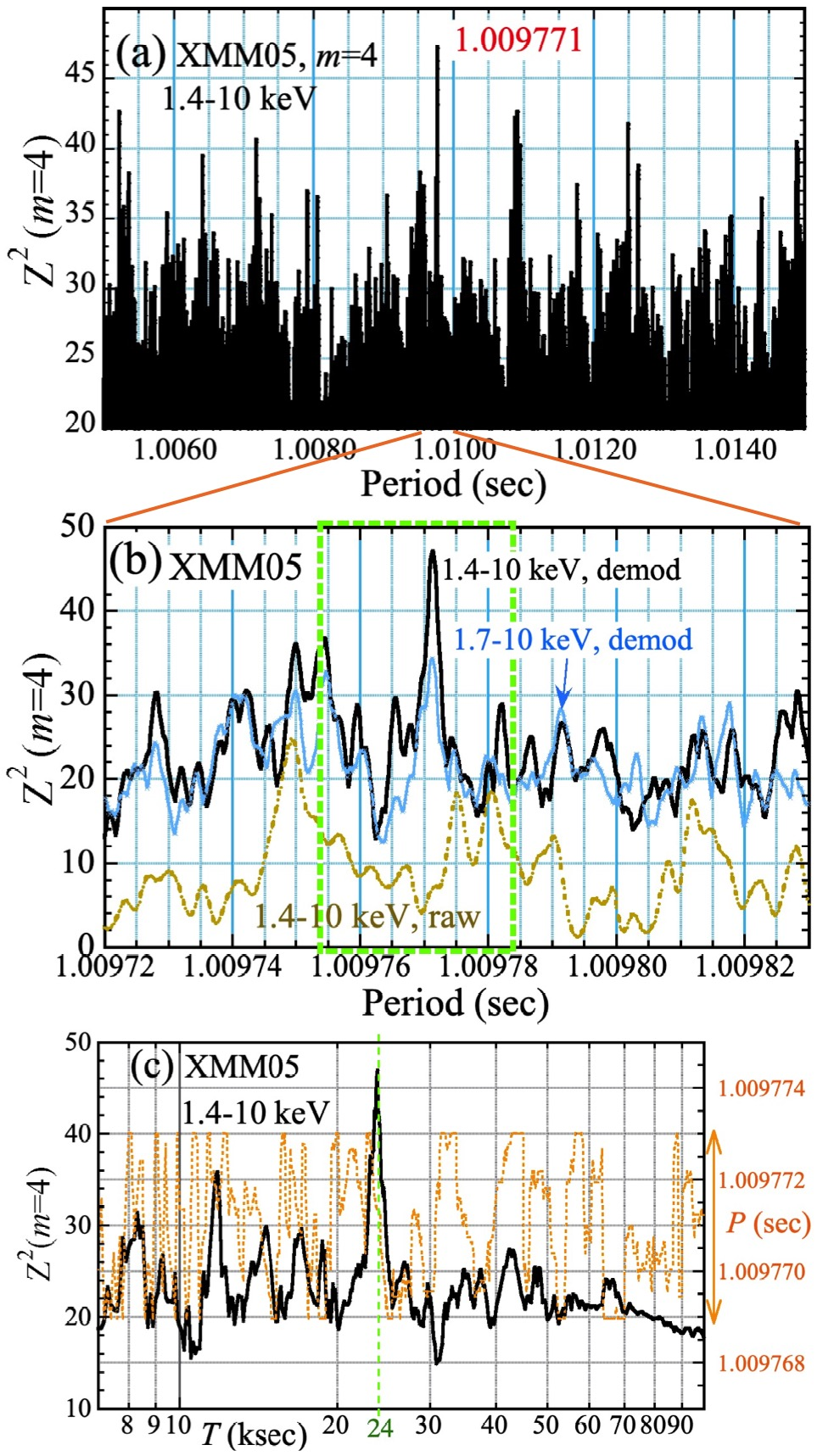}
\caption{Results from XMM05.
(a) A demodulated $m=4$ PG, in 1.4--10 keV.
(b) Details of (a), together with a result in 1.7--10 keV (thin cyan),
and the 1.4--10 keV ``raw" PG (in ocher).
A green box indicates the prediction of equation~(\ref{eq:period_prediction_XMM05}).
(c) Values of $\zz$ in 1.4--10 keV, shown against $T$.
A dashed orange curve is the optimum $P$.
%(allowed to float by $\pm 2~\mu$ s).
{Alt Text: 
Periodograms with XMM-Newton in 2005,
in photon energies of 1.4 to 10 kilo-electronvolt.
Bottom panel shows the pulse significance against the modulation period.}
}
\label{fig:f7_PTscan_XMM05}
\vspace*{-3mm}
\end{figure}

We repeated the demodulation, keeping $\EU=10$ keV,
but this time changing $\EL$ both above and below 2.5 keV.
(Although the pulsation in XMM01 was suppressed for $\EL< 2.5$ keV,
the pulse properties may  have changed as the flux decreased.)
As $\EL$ is lowered from 1.6 keV downwards,
a dominant peak has emerged in the PG,
and its height became maximum at $\EL=1.4$ keV.
The demodulated $m=4$ PG under this condition
is presented in figure~\ref{fig:f7_PTscan_XMM05}a,
where the periodicity appears as the highest peak at
\vspace*{-1mm}
\begin{equation}
P = 1.0097713(6)~{\rm s} \equiv P_{05}.
%\label{eq:P05}
\end{equation}
Because this $P_{05}$ falls right on the prediction of equation~(\ref{eq:period_prediction_XMM05}),
we may identify it with the source pulsation.
This also suggests that $\dot P$ is rather constant,
and updates equation~(\ref{eq:spindown_9301})  to 
\begin{equation}
\dot P = 1.097(4) \times 10^{-12}~{\rm s~s}^{-1}.
\label{eq:spindown_930105}
\end{equation}

Figure~\ref{fig:f7_PTscan_XMM05}b gives details 
of figure~\ref{fig:f7_PTscan_XMM05}a around the peak.
Shown together are results from two other conditions;
one with the same $\EL=1.4$ keV but without demodulation (labeled ``raw"),
and the other with the demodulation but using $\EL=1.7$ keV.
Thus, the pulse detection from 1E~1613 requires
not only the demodulation correction with equation~(\ref{eq:demodulation}),
but also a selection of an appropriate  $\EL$, 
possibly depending on the data set.
While the decrease of $\zz$ for $\EL\gtrsim 1.6$ keV
is due likely to a  decrease of the signal photons,
that toward $\EL<1.4$ keV could be intrinsic.

%%%%%%%%%%%% fig.7 %%%%%%%%%%%%%%%%%%%%%%%
%%%%%%%%%%%% fig. 7%%%%%%%%%%%%%%%%%%%%%%%

We also calculated the $\zz$ value as a function of $T$,
and show the result in figure~\ref{fig:f7_PTscan_XMM05}c,
in a similar way to figure~\ref{fig:f3_Tscan_ASCA93}a.
The $T=24.0$ ks peak is reproduced very clearly,
at $23.9 \pm 0.3$ ks,
and more sharply because of the larger amplitude,
$A \approx 0.28$ in XMM05, than $A \approx 0.08$ in ASCA93.
Also, the optimum $P$ is seen to converge to $P_{05}$;
the $\zz$ peak is hence confined to a well-defined region on the $(P,T)$ plane.

The chance probability of the $P_{05}$ peak in 
figure~\ref{fig:f7_PTscan_XMM05}a is rather modest,
$\Pch(P_{05})\approx19\%$ (Appendix 1),
if we count all  period trials in
the interval of figure~\ref{fig:f7_PTscan_XMM05}a,
together with those in $\EL$.
However, it decreases to $\Pch(P_{05})\approx0.2\%$
when $\Ntrial^{\rm P}$ is calculated over 
the period range of figure~\ref{fig:f7_PTscan_XMM05}b.
Since this range fully accommodates the 
prediction from $P_{93}$ and $P_{01}$,
we consider that the XMM05 data have provided
another detection of the 1.01 s pulse period.

%------------------ 3.5.2 --------------------
\subsubsection{Pulse properties in XMM05}
\label{subsubsec:ana_XMM05_Pr}
\begin{figure}
\hspace*{-5mm}
\includegraphics[width=9.5cm]{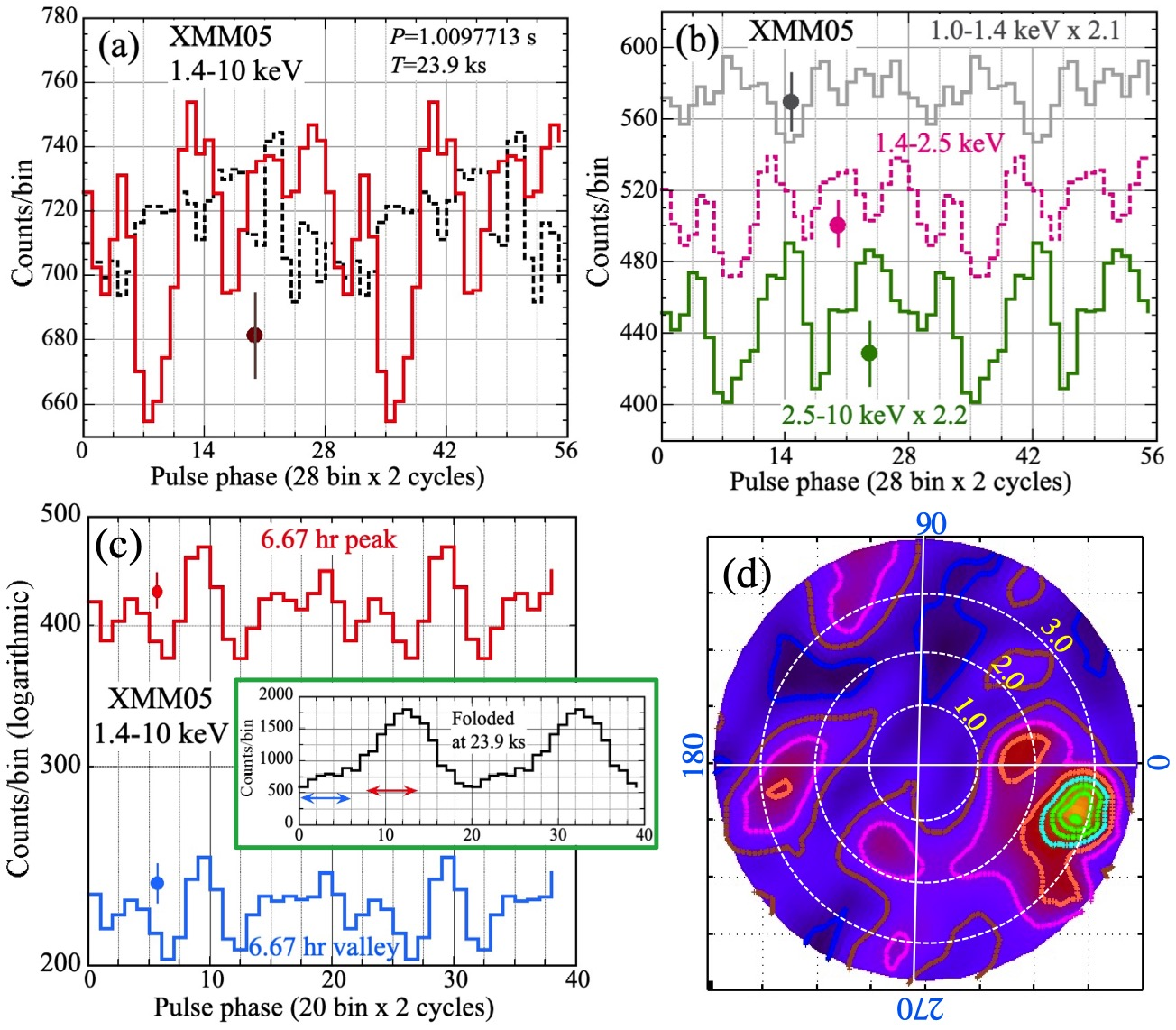}
\caption{
(a) Pulse profiles from XMM05 in 1.4--10 keV, folded at $P_{05}$ into 28 bins.
(b) Same, but derived in three separate energies.
(c) Same as (a), but the profiles are accumulated 
over the peak (top in red) and valley (bottom in blue) phases of the 6.67 hr cycle.
The 1.4--10 keV counts folded at 23.9 ks are shown in the inset.
(d) Values of $\zz$ in XMM05, shown on the $(A, \psi_0)$ polar coordinates.
They are derived with demodulation
using  $T= 23.9$ ks and $P=P_{\rm 05}$.
{Alt text: Some properties of the pulsation of 1E~1613,
in the 2005 XMM-Newton data.}
}
\label{fig:f8_Pr_XMM05}
\vspace*{-4mm}
\end{figure}

%------------------ 3.5.2 --------------------
When folded at $P_{05}$, the 1.4--10 keV XMM05 data
yielded pulse profiles shown in figure~\ref{fig:f8_Pr_XMM05}a.
The demodulated profile reveals the 4-peak structure very clearly.
(To resolve the fine structure, here one pulse cycle is divided into 28 bins,
instead of 20).
This profile is contributed by the fundamental to the 4th harmonic, 
by 19.7\%, 18.4\%, 8.7\%, and 53.2\%,  respectively.
Thus, in contrast to the dominance of the 2nd harmonic 
in the profile of ASCA93 (and of XMM01), 
now the 4th harmonic becomes dominant.

Figure~\ref{fig:f8_Pr_XMM05}b shows energy dependence
of the demodulated pulse profile.
The 4-peak feature, seen in 1.4--2.5 keV,
changes into a 3-peak structure in 2.5--10 keV,
whereas it  splits into finer peaks below 1.4 keV;
this may explain the decrease of the peak $\zz$ for $\EL<1.4$ keV.

The XMM05 results provide not only an additional support to the pulse reality,
but also  several pieces of valuable information.
The inset to figure~\ref{fig:f8_Pr_XMM05}c
is the 1.4--10 keV photon counts folded at 23.9 ks (maximizing the PG peak), 
which reconfirms the reported flux modulation (DL06).
When the  photons are accumulated
over the peak and valley phases of the cycle as indicated by two arrows, 
we obtain the demodulated pulse profiles shown at the top (in red) 
and  bottom  (in blue) of  figure~\ref{fig:f8_Pr_XMM05}c, respectively.
Here, the ordinate is logarithmic to enable a direct comparison of the two profiles.
Between the two modulation phases,
neither the pulse profile nor the pulse fraction differs significantly.

As in the figure~\ref{fig:f8_Pr_XMM05}c inset,
the 6.67 hr flux variation in XMM05 happens to
be close to the minimum at $t=0$.
Then, at this epoch, how is the pulse-timing  lag $\Delta t$
in equation~(\ref{eq:demodulation}) behaving? 
For this purpose,  figure~\ref{fig:f8_Pr_XMM05}d shows 
the values of $\zz$ as a  contour plot on a $(A, \psi_0)$ polar plane,
where $P = P_{05}$ and $T = 23.9$ ks are  fixed.
Thus, the demodulation solution is constrained to $A \approx 0.28$ s
and $\psi_0 \approx 350^\circ$.
The pulse-phase lag is hence close to zero at $t=0$,
and is increasing with time.
This  result fits very naturally into the possible emission geometry
presented in subsection~\ref{subsec:discuss_geometry}.

%%%%%%%%%%%% fig.8 %%%%%%%%%%%%%%%%%%%%%%%
%%%%%%%%%%%% fig. 8%%%%%%%%%%%%%%%%%%%%%%%

%==========3.6=================
\subsection{XMM-Newton data in 2016 (XMM16)}
\label{subsec:ana_XMM16}
%==========4.1=================

%%%%%%%%%%%% fig.9 %%%%%%%%%%%%%%%%%%%%%%%
\begin{figure}
%\vspace*{-5mm}
\centerline{
\includegraphics[width=8.5cm]{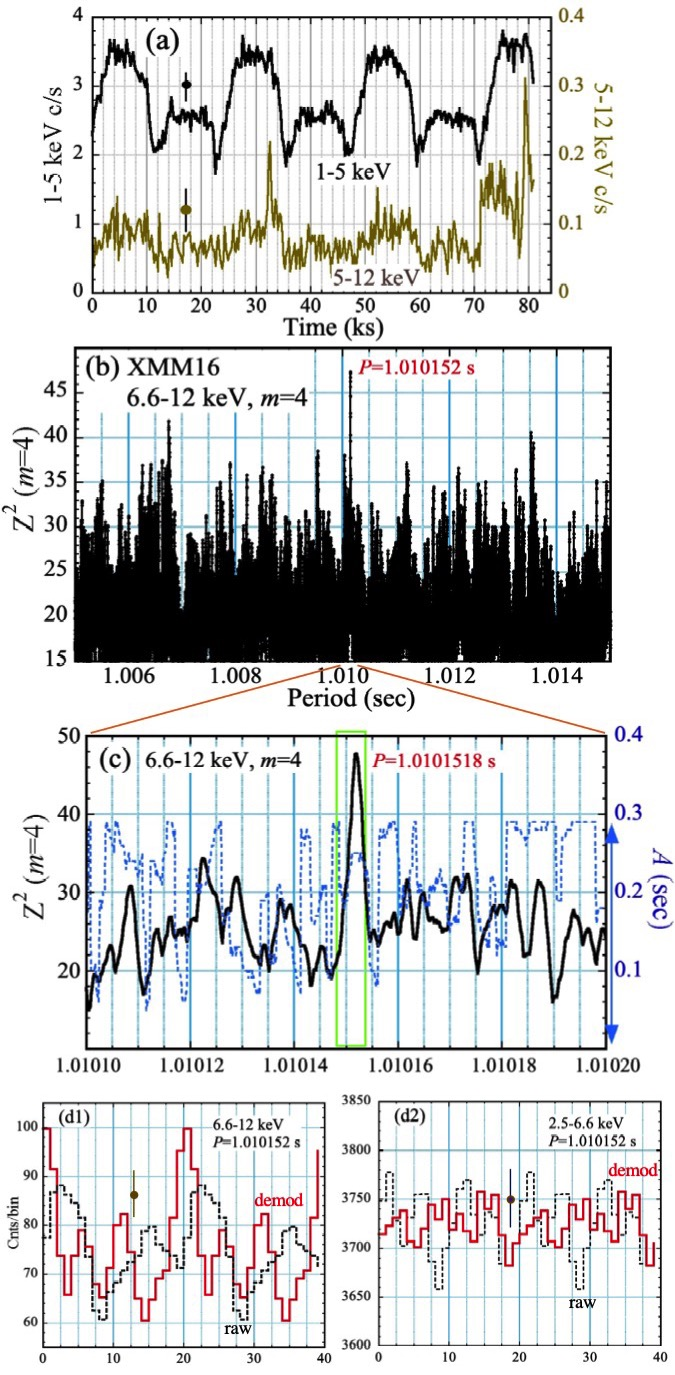}
}
\caption{
Results from XMM16.
(a) Light curves in 1--5 keV (left ordinate) and 5--12 keV (right ordinate), 
with 250 s binning.
(b)  A demodulated 6.6--12 keV PG with $m=4$.
(c)  Details of (b), where the green box indicates equation~(\ref{eq:period_prediction_XMM16}).
(d1) Pulse profiles in 6.6--12 keV folded at $P_{16}^{\rm X}$,
before (dashed black) and after (solid red) the demodulation.
(d2) Same as (d1), but in 2.5--6.6 keV.
The demodulation uses the same parameters as for (d1).
{Alt Text: Results on 1E 1613 from the XMM-Newton data in 2016.
Top panel shows two-band light curves with 250 second binning. 
Next two graphs are periodograms, and the bottom two 
give folded pulse profiles in two energy ranges.}
}
\label{fig:f9_XMM16}
%\vspace*{-2mm}
\end{figure}
%%%%%%%%%%%% fig.9 %%%%%%%%%%%%%%%%%%%%%%%

The primary aim of analyzing XMM16  is to find
another piece of evidence for the 1.01 s pulsation.
A successful detection would be followed by  two more aims;
to refine the knowledge on $\dot P$ (e.g., its constancy) 
utilizing the time span from XMM05,
and to look for possible effects of  the 2016 June episode.
We must be cautious
that the pulse properties may change across $\sim 5$ keV,
as suggested by the two-band light curves 
in figure~\ref{fig:f9_XMM16}a,
which reconfirm figure~4 of \citet{Esposito19}.
To enhance the information above $\sim 5$ keV,
we raise $\EU$ from 10 keV to 12 keV.

Assuming that  $\dot P$ of the source is constant 
at equation~(\ref{eq:spindown_930105}),
the pulse period  in XMM16 is predicted as
%
%\vspace*{-1mm}
\begin{equation}
{\rm XMM16} : P=1.010151(2)~{\rm s}
\label{eq:period_prediction_XMM16}
\end{equation}
where the uncertainty is already dominated by those in $P$ rather than $\dot P$.
However, considering again  possible changes in $\dot P$,
over 11 yrs an/or across the brightening episode,
we retained the fiducial period-search range of equation~(\ref{eq:P_search_range}),
and computed simple PGs in various energy ranges  above and below 5 keV.
No noticeable periodicity was found,
just as in the preceding three data sets.

We then applied the demodulation to the data, 
using $m=4$ and $\EU=12$ keV.
The search failed for  $\EL \lesssim 6$ keV, 
as each PG showed multiple peaks with similar heights
(typically $\zz \lesssim 38$).
In contrast, PGs  with $\EL \gtrsim6$ keV revealed a  unique candidate,
which became most prominent ($\zz \approx 47.7$) at  $\EL =6.6$ keV.
%(although the photon number decreased to $\approx 1500$).
The result is shown in panels (b) and (c) of figure~\ref{fig:f9_XMM16},
where the candidate stands out at
%\vspace*{-1mm}
\begin{equation}
P = 1.0101518(12) ~{\rm s} \equiv P_{16}^{\rm X}.
%\label{eq:P16X}
\end{equation}
%\vspace*{-1mm}
%
As  described in Appendix 1,
this peak is estimated to have
$\Pchone(P_{16}^{\rm X})\approx 9\times 10^{-6}$,
and  $\Pch(P_{16}^{\rm X}) \approx 11\%$,
when  we consider trials in $P$ over equation~(\ref{eq:P_search_range}),
plus those in $\EL$.
Although this result is modest, 
the peak remains the highest over the entire interval of equation~(\ref{eq:P_search_range}),
and yet falling right on the prediction of  equation~(\ref{eq:period_prediction_XMM16}).
As a result, $\Pch(P_{16}^{\rm X})$ decreases to $0.1\%$,
just as we found with XMM05,
when $\Ntrial^{\rm P}$ is calculated over the period range of 
figure~\ref{fig:f9_XMM16}c.
It further decreases to 0.04\%,
if calculated against the predicted period uncertainty of 
equation~(\ref{eq:period_prediction_XMM16}).
Therefore, $P_{16}^{\rm X}$ is again likely to be a real pulse period.
In addition, the constancy of  $\dot P$ at 
equation~(\ref{eq:spindown_930105}) has been reinforced.

Pulse profiles from XMM16, in 6.6--12 keV and 2.5--6.6 keV,
are given in panels (d1) and (d2) of figure~\ref{fig:f9_XMM16}, respectively.
The demodulated 6.6--12 keV profile is similar to those in ASCA93 and XMM01,
reinforcing the reality of $P_{16}^{\rm X}$.
In contrast, the low-energy folded profiles in (d2) are much featureless,
regardless of the demodulation correction;
the pulse fraction, if any,  is at most $\sim 1/30$ of that in (d1).
These low-energy properties did not change even 
when allowing $(A, \psi_0)$ to take different  values from those in 6.6--12  keV.
These results reconfirm the apparent  failure in detecting the pulsation below 6.6 keV.

Then, what is taking place in the softer energies?
The pulses must be present there as well,
because the emission should not become suddenly isotropic at $< 6$ keV.
At the same time, the PPM must also be operating,
because otherwise the pulse would be directly visible.
A likely scenario to explain these conditions is
presented in subsubsection~\ref{subsubsec:discuss_EUEL}.

As anticipated in advance
(subsection~\ref{subsec:obs_XMM}),
the pulse detection in XMM16 was less straightforward 
than in the preceding three data sets,
mainly due to complex pulse behavior in  $<6$ keV
where most of the photons are present.
This may possibly due, in turn, to the source activation 2 months before.
Nevertheless, the $\sim 1.01$ s pulsation has been reconfirmed in XMM16 as well,
and $\dot P$ has been found to be quite constant
for 23 yrs from 1993 to 2016, 
with no precursory changes preceding the activation episode.

\vspace*{-3mm}
%==========3.7=================
\subsection{NuSTAR data in 2016 (NuS16)}
\label{subsec:ana_NuS16}
%==========4.1=================

\begin{figure}
\vspace*{-8mm}
\centerline{
\includegraphics[width=8cm]{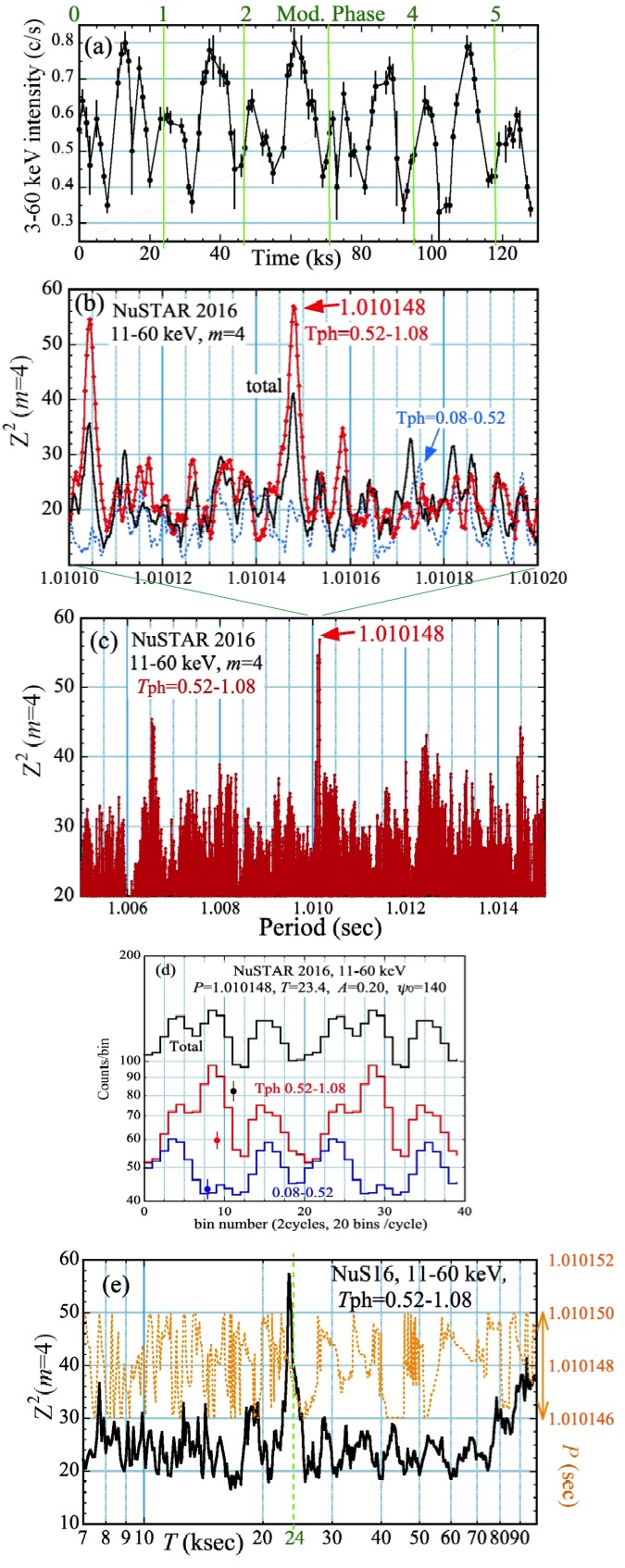}
}
\caption{
Results from NuS16, mainly above 10 keV.
(a)  A background-inclusive 3--60 keV light curve from NuS16,
with 1 ks binning. 
The modulation phase $t/T$, 
using $T=23.4$ ks and the origin at the data start,
is indicated in the top abscissa.
(b)  A demodulated 11--60 keV PG with $m=4$ (solid black curve),
presented over a narrow period range. 
The PGs for $[t/T]=0.52-1.08$ and $0.08-0.52$
are given by a solid red curve with dots
and a dashed cyan curve, respectively.
(c)  The same as the red curve in (b), 
but the period range  is expanded to 
that of equation~(\ref{eq:P_search_range}).
(d) Pulse profiles in the three modulation-phase 
intervals (see text).
(e) The $\zz$ values shown as a function of $T$,
calculated under the same condition as the red curve in (b),
but limiting $P$ to within $\pm 2~\mu$s of  $P=P_{16}^{\rm N}$.
{Alt Text: Results from the 11-60 kilo-electronovolt  data of NuSTAR in 2016.
Shown are a light curve, two periodograms, three pulse profiles, 
and the pulse significance against the assumed modulation period.}
}
\label{fig:f10_NuS16H}
\vspace*{-1mm}
\end{figure}

In an attempt to obtain the pulse information
for the first time in hard X-rays above $\sim 10$ keV,
we analyze the NuS16 data,
kepping in mind the following reservations.
(i) As described in subsection~\ref{subsec:obs_data_selection},
the data were taken only 3 d after its activation on 2016 July 22.
Therefore, an enhanced timing noise may be anticipated.
(ii) Since the data were taken only 54 days before XMM16,
equation~(\ref{eq:spindown_930105}) 
accurately predicts the pulse period  to appear at 
$
P=P_{16}^{\rm X}-5.1~\mu{\rm s} \approx 1.010147~{\rm s},
$
unless short-term changes in $P$ or $\dot P$ took place.
For the first-cut pulsation search,
we hence select  a narrower interval, $P=1.0101-1.0102$ s.
(iii) As described by \citet{Rea16} and reconfirmed in figure~\ref{fig:f10_NuS16H}a,
the 6.67 hr intensity variation on this occasion had 
a double-peaked profile,  just as in XMM16.
(iv) Like magnetars, the wide-band spectrum of 1E~1613 
generally  exhibits two-component nature \citep{Rea16}, 
wherein a thermal-like soft component and a power-law 
like hard component  cross over at $\sim 10$ keV.
We hence analyze the data above and below 10 keV separately.

%%%%%%%%%%%% fig.10 %%%%%%%%%%%%%%%%%%%%%%%
%%%%%%%%%%%% fig.10 %%%%%%%%%%%%%%%%%%%%%%

%------------------ 3.7.1 --------------------
\subsubsection{Hard X-ray results from NuS16}
\label{subsubsec:ana_NuS16_hard}
%------------------ 3.7.1 -------------------

Simple PGs, either above or below 10 keV, did not yield convincing results.
We thus  proceed to the demodulation analysis,
first in hard X-ays because the pulsation in XMM16 was 
less disturbed in harder energies.
When using $\EL \gtrsim 10$ keV and $\EU =45-70$ keV,
the demodulated PGs consistently revealed a noticeable peak,
at a period in agreement with the prediction.
As shown in figure~\ref{fig:f10_NuS16H}b by a solid black line,
it reached the maximum of $\zz \approx 41$,
for $\EL=11$ keV and $\EU=60$ keV.
The secondary peak at $P\approx 1.010105$ s
is its prograde beat between the $T=24$ ks periodicity.

Although the hard X-ray data have thus provided  
evidence  for the expected pulsation, 
the peak is not high enough,
and is overwhelmed by other noise peaks 
when the PG is calculated over the fiducial interval 
of equation~(\ref{eq:P_search_range}).
Presumably, the NuS16 data  take during the high activity
violate the basic assumption behind equation~(\ref{eq:demodulation}),
that  all features in the pulse profile shift in parallel, 
and in a sinusoidal manner,
throughout the modulation phase $[t/T]$.

Given the above possibility, we repeated the demodulation 
using only  those photons that fall in a limited modulation phase
\footnote{This is similar to the phase-sorted PG 
(subsubsection~\ref{subsubsec:ana_ASCA93_phase-sorted}),
but differs in that we focus on a single modulation-phase interval,
and apply demodulation.},
between $[t/T]_1$ and $[t/T]_2$
($0 \leq [t/T]_1< [t/T]_2 \leq 2.0$)
referring to panel (a).
As  shown in figure~\ref{fig:f10_NuS16H}b by a red line with dots,
the peak in the demodulated 11--60 keV PG 
became much higher, reaching $\zz=56.9$,
when $[t/T]_1=0.52$ and $[t/T]_2=1.08$ are chosen.
%(The secondary peak also became higher.)
The main peak position has remained the same, and is given as
\begin{equation}
P=1.0101480 (7) \equiv P_{16}^{\rm N},
%\label{eq:P16N}
\end{equation}
which agrees well with the prediction from $P_{16}^{\rm X}$.
The peak disappears (dashed cyan curve in figure~\ref{fig:f10_NuS16H}b)
when the complementary modulation phase, $0.08-0.52$, is used.
Details of the modulation-phase dependence are given in Appendix 2.

In figure~\ref{fig:f10_NuS16H}c,  we calculated the demodulated PG, 
using the same energy range and the same modulation phase (0.52--1.08),
but reviving the  period range of equation~(\ref{eq:P_search_range}).
There, the $P_{16}^{\rm N}$ peak still remains the highest,
even though the number of period trials have 
increased by two orders of magnitude.
This may provide some flavor as to the peak reality,
although we skip calculating its statistical significance,
because the effective trial number in choosing the modulation phase
is difficult to estimate.

Folded pulse profiles in 11--60 keV are shown in figure~\ref{fig:f10_NuS16H}d,
for the conditions of  the  three  PGs in panel (b).
They are all demodulated with  $A=0.2$ s and $\psi_0=140^\circ$.
The middle one (in red), using $[t/T]=0.52-1.08$,
yields $\zz=56.9$, and exhibits a clear 3-peak structure.
Among the three subpeaks, the hightest one at a pulse-phase 
bin 8 disappears in the bottom (blue) profile
which uses the complementary modulation phase, resulting in $\zz \approx 21$.
The top profile (in black), with $\zz \approx 41$,  is for the total modulation phase.
We now understand the reason why the demodulation was most successful
when using $[t/T]=0.52-1.08$; outside this modulation phase, 
the brightest emission component becomes invisible,
due, e.g., to self-eclipse, or deflection of the X-ray beam off our line-of-sight.
This strong dependence on the modulation phase 
makes a  contrast with figure~\ref{fig:f8_Pr_XMM05}c for XMM05,
where two distinct  modulation phases gave similar pulse profiles.
Further information on this issue is provided in Appendix 2.

As the last piece of hard-band information from NuS16,
figure~\ref{fig:f10_NuS16H}e presents
how the pulse $\zz$ derived via demodulation  depends on the assumed $T$,
just as in figure~\ref{fig:f3_Tscan_ASCA93}a and figure~\ref{fig:f7_PTscan_XMM05}c.
The solution indeed stands out, and gives a constraint as
\begin{equation}
T = 23.4 \pm 0.2~{\rm ks}.
\label{eq:T_NuS16}
\end{equation}
This appears somewhat smaller than those from the other data sets;
this issue is examined later in  subsectio ~\ref{subsec:discuss_additional}.

\vspace*{-3mm}
%------------------ 3.7.2 --------------------
\subsubsection{Soft X-ray results from NuS16}
\label{subsubsec:ana_NuS16_soft}
%------------------ 3.7.2 --------------------

The pulsation search was performed also
in two  soft X-ray bands of NuS16, 3--6 keV and 6--10 keV,
which contain much larger number of  background-inclusive photons,
31479 and 6431 respectively,
than the 11--60 keV band (2367 photons).
By changing $[t/T]_1$ and $[t/T]_2$,
we searched for evidence of the pulsation.

%%%%%%%%%%%% fig.11 %%%%%%%%%%%%%%%%%%%%%%%
\begin{figure}
\vspace*{-4mm}
%\centerline{
\includegraphics[width=7.8cm]{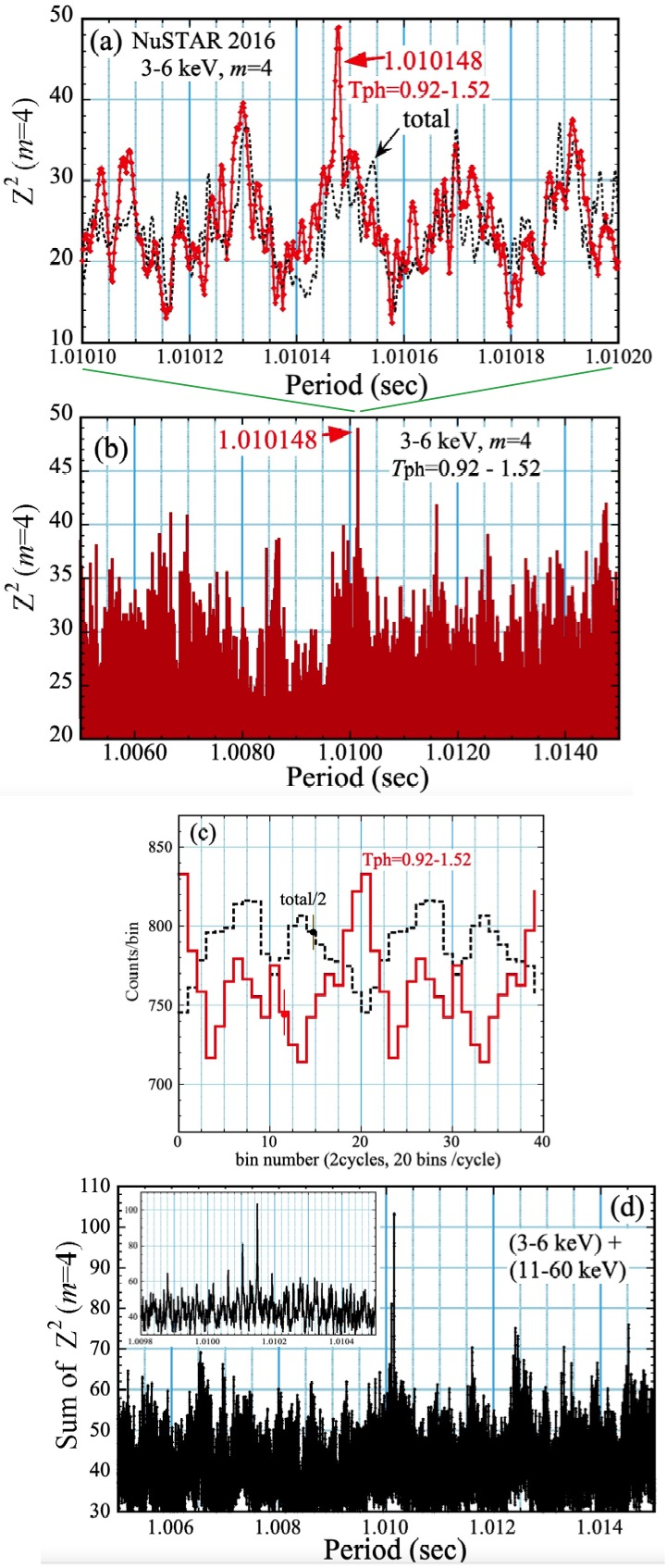}
\caption{  
Results from NuS16 in the 3--6 keV band.
(a)  Demodulated 3--6 keV PGs with $m=4$,
presented over a narrow period range. 
The PG for $[t/T]=0.92-1.52$
is  given by a solid red line with dots,
and that for $[t/T]=0-1$
by a dashed black curve.
(b) The same as the red curve with dots in (a), 
but the period range is expanded to 
that of equation~(\ref{eq:P_search_range}).
(c) Folded pulse profiles for $[t/T]=0.92-1.52$
(solid red) and $[t/T]=0-1$ (dashed black; halved).
(d) A sum of two PGs, one in panel (b) 
and the other in figure~\ref{fig:f10_NuS16H}c.
The inset expands a region around the peak.
{Alt Text: Same as the previous figure,
but using the 3--6 kilo-electronovolt  photons.}
}
\label{fig:f11_NuS16S}
\vspace*{-4mm}
\end{figure}
%%%%%%%%%%%% fig.11 %%%%%%%%%%%%%%%%%%%%%%

In the 3--6 keV demodulation analysis,
a choice of $[t/T]=0.92-1.52$
(see Appendix 2 for details)
yielded figure~\ref{fig:f11_NuS16S}a,
where the pulsation has been recovered
exactly at $P_{16}^{\rm N}$ with $\zz=49.0$,
with several side lobes on both sides.
While $T$ of equation~(\ref{eq:T_NuS16}) is reconfirmed,
the preferred modulation phase 
is nearly complementary to  
what was favored  by the hard X-ray photons.
The  optimum modulation parameters 
($A$ and $\psi_0$; table~\ref{tbl:demodulation},) 
also differ from those in 11--60 keV.
As shown in figure~\ref{fig:f11_NuS16S}b, 
the peak remains the highest,
even when the period range is expanded 
to equation~(\ref{eq:P_search_range}).

Folded pulse profiles in 3--6 keV are presented in 
figure~\ref{fig:f11_NuS16S}c,
for $[t/T]= 0.92-1.52$ (solid line in red),
and for $[t/T] = 0-1$ (total, in dashed black), 
with the latter scaled by 1/2.
Both have been demodulated using the same 
parameters in table~\ref{tbl:demodulation}.
The former profile  is similar to those in ASCA93, XMM01, and XMM16,
but the pulse fraction is $\sim 0.09$,
which is much lower than that in 11-60 keV ($\sim 0.3$).
The dominant peak, seen in the profile with the limited $[t/T]$,
turns into a valley in the other one.

%To assess the consistency between the 11--60 keV PG in figure~\ref{fig:f10_NuS16H}c 
%and the 3--6 keV one in figure~\ref{fig:f11_NuS16S}b,
%we have summed up the two PGs,

To assess the 11--60 keV  vs. 3--6 keV consistency,
we have incoherently summed up the PGs in
figure~\ref{fig:f10_NuS16H}c and  figure~\ref{fig:f11_NuS16S}b,
ignoring their differences in the pulse profile, $A$, $\psi_0$, and the $[t/T]$ interval.
The result is presented in figure~\ref{fig:f11_NuS16S}d, 
together with its details in the inset.
Thus, the hard and soft peaks at $P_{\rm 16}^{\rm N}$
nicely add up to form an outstanding peak,
accompanied by a series of side lobes on both sides
due to the beat with $T=24$ ks.
This result  provides  visual evidence
for the pulsation in NuS16.

We also analyzed the 6--10 keV photons from NuS16,
by changing $[t/T]_1$ and $[t/T]_2$
in a similar way.
However, no evidence for the pulsation 
at $\sim P_{\rm 16}^{\rm N}$ was found (Appendix 2).
In this intermediate energy band,
the hard and soft emission components may
have comparable contributions.
Because they are visible over distinct modulation phases,
and have different modulation parameters (particularly $\psi_0$),
the demodulation might become ineffective.

\vspace*{-2mm}
%==========3.8=================
\subsection{NuSTAR data in 2017 (NuS17)}
\label{subsec:ana_NuS17}
\begin{figure}

%\centerline{
\includegraphics[width=7.8cm]{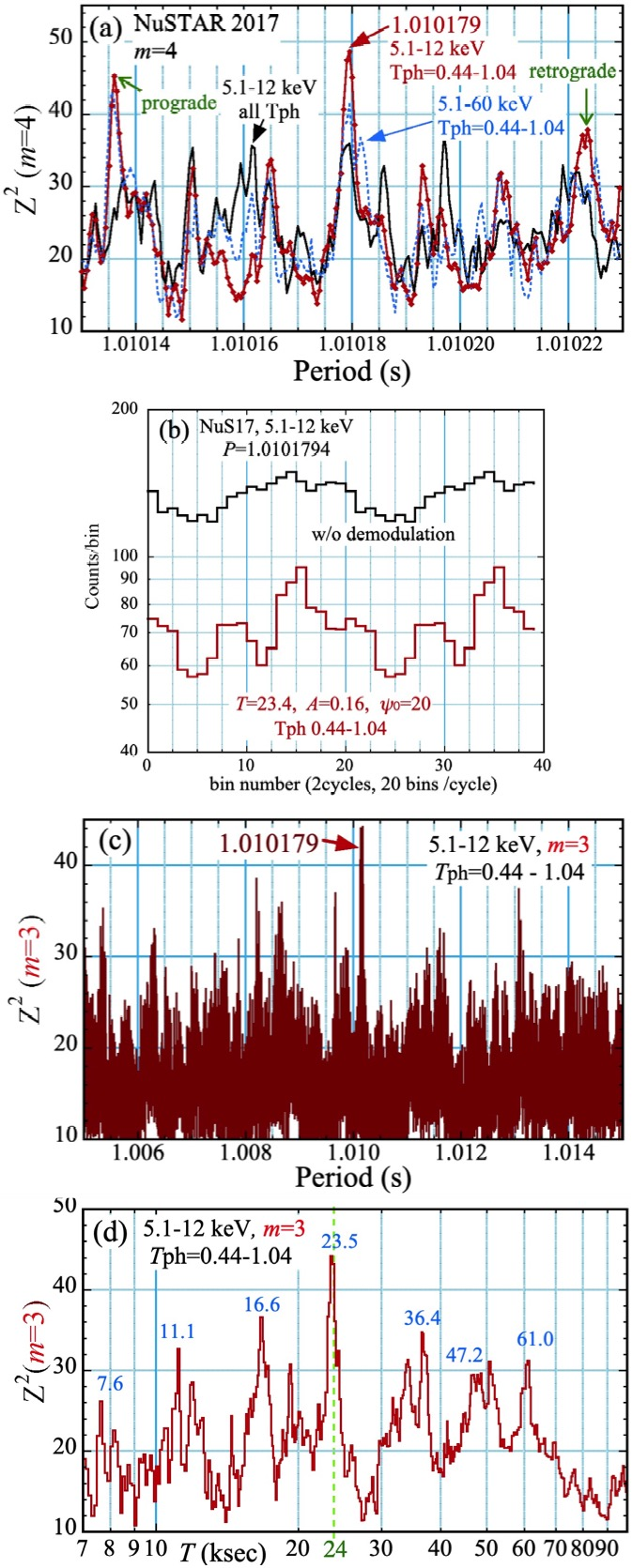}
\vspace{2mm}
\caption{
Results from NuS17.
(a) Three PGs with $m=4$, shown over a narrow period range.
The demodulated PG, using all photons in 5.1--12 keV,
is shown in solid black.
A constraint of $[t/T]=0.44-1.04$ turns it into
the one in solid brown line with dots.
It further changes into the dashed blue one, 
when the energy range is expanded to 5.1--60 keV.
(b) Pulse profiles in 5.1--12 keV, folded at $P_{\rm 17}$.
The upper one is before the demodulation,
using all modulation phases.
The lower one is after demodulation,
and limiting $[t/T]$ to $0.44-1.04$.
(c) A demodulated $m=3$ PG in 5.1--12 keV,
with $[t/T]=0.44-1.04$,
shown over the fiducial period range.
(d) The behavior of $Z_3^2$ as a function of $T$,
using the same condition as (c).
Values of $T$ are given for some peaks.
{Alt Text: Results from NuS17,
mainly using 5.1--12 kilo-electronvolt photons.}
}
\label{fig:f12_NuS17}
%\vspace*{-2mm}
\end{figure}

%==========4.1=================
We finally  analyze the NuS17 data, 
more concisely than NuS16 
because the source became $\sim 4$ times fainter.
The demodulation analysis was conducted
using the full modulation phase,
by changing $\EL$ and $\EU$.
Since the pulsation is expected  at $P=1.01018$ s
from equation~(\ref{eq:spindown_930105}),
the search was performed, as for NuS16,
in a narrow period  range  around the prediction, 
The solid black line in figure~\ref{fig:f12_NuS17}a  is a result
using $\EL=5.1$ keV and $\EU=12$ keV,
where a minor peak appeared at a period close to the prediction.

%%%%%%%%%%%% fig.12 %%%%%%%%%%%%%%%%%%%%%%%
%%%%%%%%%%%% fig.12 %%%%%%%%%%%%%%%%%%%%%%

When we limit the modulation phase to $[t/T]=0.44-1.04$,
the photon number was halved,
but the above peak grew higher to $\zz \approx 48.6$,
and became rather dominant,
as shown in the same figure by a brown line with dots.
Its period is determined as
\begin{equation}
P=1.0101792 (10) \equiv P_{17}
\label{eq:P17}
\end{equation}
which agrees well with the prediction.

Since the employed 5.1--12 keV  interval is much narrower
than the NuSTAR bandpass,
we tried to expand it.
When $\EL$ is lowered to $<5.1$ keV,
the $P_{17}$ peak quickly diminished,
suggesting a pulse suppression below 5.1 keV.
As  $\EU$ is raised from 12 keV upwards,
the peak also decreased, but more gradually.
As given by a dashed blue line in figure~\ref{fig:f12_NuS17}a,
the peak is still visible for $\EU=60$ keV.
As in NuS16, the pulsation in NuS17 is likely to be  present in $>12$ keV, 
and the gradual $\zz$ decrease could be 
a result of an increased  background.
Hereafter, we retain the 5.1--12 keV interval.

Figure~\ref{fig:f12_NuS17}b presents two pulse profiles
in this energy range.
One is before the demodulation,
while the other is after applying the demodulation 
and limiting the modulation phase.
The latter profile is much more structured than the former;
again, appropriate timing corrections and optimization of the energy range
are inevitable for the pulse detection.
The demodulated profile exhibits
3 sub-peaks per cycle, instead of 4.
This is evidenced by a rather small difference
between $\zz=48.6$ and $Z_3^2=44.3$ at $P_{17}$.

Switching to $m=3$,
we repeated the period search,
but over the fiducial period interval of equation~(\ref{eq:P_search_range}).
As shown in figure~\ref{fig:f12_NuS17}c,
the $P_{17}$ peak has been confirmed to be the highest therein.
This statement still remains valid when $m=4$ is used,
but the peak has a lower contrast to the surroundings.
Thus, the NuS17 data are considered to have 
afforded another pulse detection from 1E~J1613,
even though the source was rather dim at that time.

Figure~\ref{fig:f12_NuS17}d gives the behavior of $Z_3^2$,
under the same condition as panel (c), but as a function of $T$,
with $P$ allowed to float by $\pm 2~\mu$ s around $P_{17}$.
This yields the best estimate of $T$ as
\begin{equation}
T = 23. 5 \pm 0.2~{\rm ks}
\label{eq:T_NuS17}
\end{equation}
which appears also deviated from 24.0 
as the NuS16 result in equation~(\ref{eq:T_NuS16}).
There, $T$-values of several other peaks are also given.
Some may have approximate integer ratios to 23.5 ks;
e.g., 7.6 ks ($\approx 1/3$), 11.1 ks ($\approx 1/2$), and 47.2 ks ($\approx \times 2$).

\vspace*{-2mm}
%%%%%% 4 %%%%%%
\section{Discussion}
\label{sec:dicuss}
%%%%%%%%%%%%%

%==========4.1=================
\subsection{Summary of the timing results}
\label{subsec:discuss_summary}
%==========4.1=================
Based on our  magnetar studies,
as well as the prediction by HH02,
we  assumed that equation~(\ref{eq:6.67hr}) 
is  the beat period of a freely precessing NS,
and  its rotation period must be much shorter, 
but  is invisible due to the PPM.
By cancelling the PPM disturbance,
and appropriately selecting the energy range,
we tried to detect the predicted fast ($\sim 1$ s) pulsation
of 1E~1613 in the six X-ray data sets,
ASCA93, XMM01, XMM05, XMM16,  NuS16, and NuS17.
The attempt has resulted in a success
through the following steps.

Assuming the PPM period of 24.0 ks,
we began the pulse search with the ASCA93 data, 
over the 0.3--30 s period range.
The two analysis methods consistently revealed,
in the 2.5--12 keV range,
an outstanding  periodicity at $P_{93}=1.009356$ s,
together with its strong second-harmonic signal.
Employing the demodulation, we next searched the 2.5--10 keV (later 2.7--10 keV) 
XMM01 data for the corresponding signal,  over the fiducial 1.005--1.015 s interval.
The  2.7--10 keV PG revealed the highest  peak at $P_{01}=1.00963$ s,
which overwhelmed other peaks by $\Delta \zz \gtrsim 8$.
In the same way, the XMM05  data were analyzed  
over the fiducial period interval,
but with the energy range expanded  to 1.4--10 keV.
Then, at $P_{05}=1.009771$ s,
we detected the highest peak,
which exceeded other peaks  by $\Delta \zz \gtrsim 5$.
Because $P_{93}$, $P_{01}$, and $P_{05}$
have modest statistical significance individually, 
and align well under a constant $\dot P$, 
they may be regarded as the pulse periods of 1E~1613.

We finally analyzed the XMM16, NuS16, and NuS17 data,
all acquired after the 2016 June outburst.
From XMM16, the strongest periodicity 
over the fiducial period interval was found at  
$P_{16}^{\rm X}=1.010152$ s,
even though we had to discard photons below 6.6 keV.
The NuS16 data afforded the pulse period of
$P_{16}^{\rm N}=1.010148$ s, both in 11--60 and 3--6 keV,
even though we had to use limited modulation-phase intervals
which differ between the two energy bands.
Through a similar demodulation analysis 
of the 5.1--12 keV photons from NuS17,
using a limited modulation-phase interval,
the pulse period was found at  $P_{17}=1.010179$ s.

\begin{figure}
\centerline{
\includegraphics[width=7.7cm]{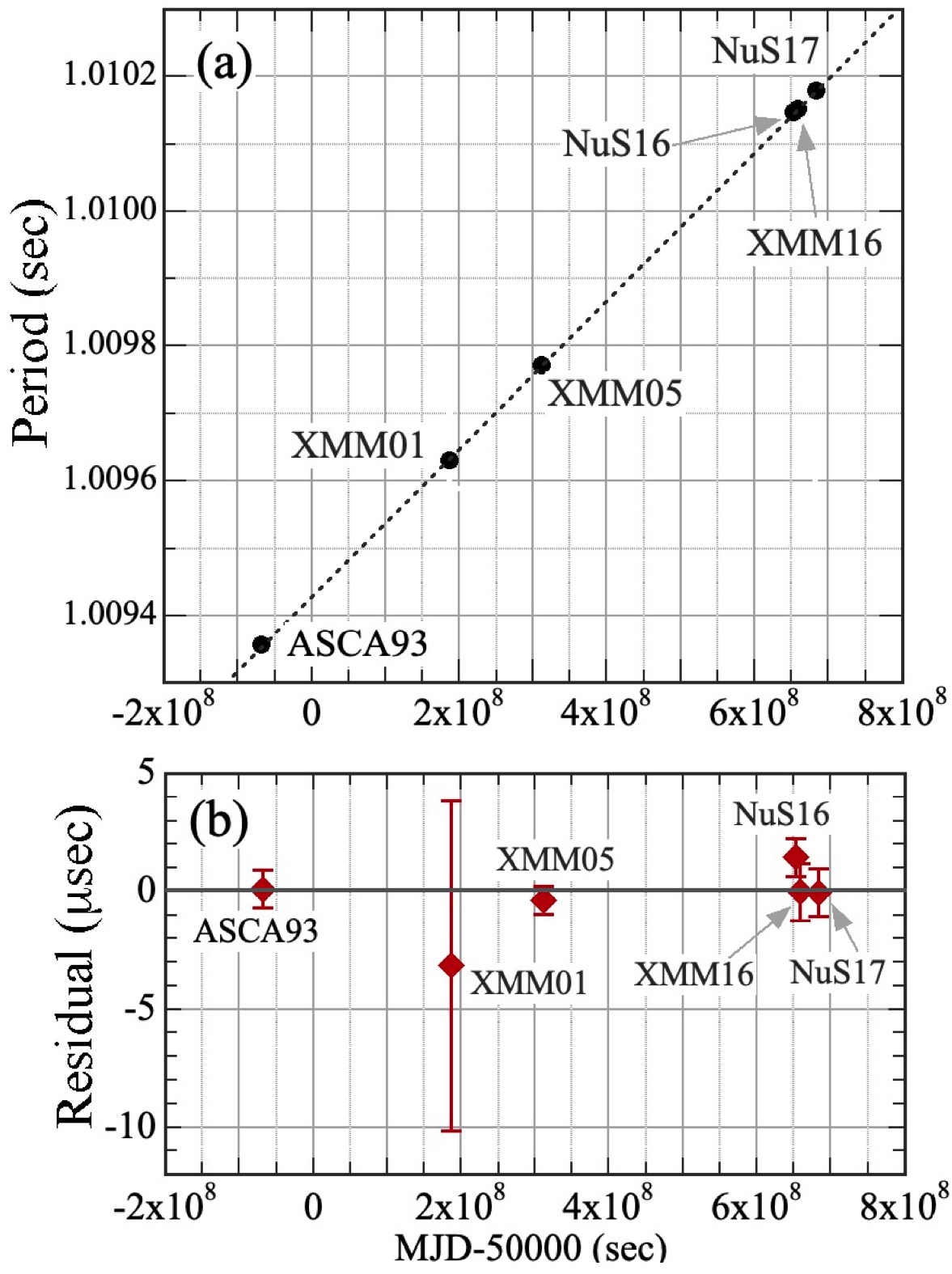}
}
\caption{(a) The six pulse periods measured from 1E~1613,
shown against the time in seconds from MJD 50000.0.
Measurement errors are much smaller than the plotting symbols.
The fit with equation~(\ref{eq:linear_spindown}) is
presented in a dashed  line.
(b) The period residuals in $\mu$s from the linear fit in (a).
{Alt Text: (a) The pulse periods measured in the six observations, 
plotted against the time.
(b) Residuals from the linear fit.
}
}
\vspace*{-5mm}
\label{fig:f13_linear_spindown}
\end{figure}

Figure~\ref{fig:f13_linear_spindown}a summarizes 
the measured six periods as a function of time.
We emphasize that these periods were all identified
as the highest peak of the PG over the fiducial range (1.005--1.015 s), 
except for the ASCA93 result 
which dominates the much wider interval of 0.3--30 s.
Also plotted is a linear fit,
which updates equation~(\ref{eq:spindown_930105}) and  is expressed as
\begin{equation}
P(t) = 1.0094300(2) + 1.097(2) \times 10^{-12} t~~{\rm s},
\label{eq:linear_spindown}
\end{equation}
where $t$ is the time measured from MJD 50000.0 in sec.
Thus, the data points line up precisely on the linear fit,
because the fit residuals, shown in panel (b),
are generally within measurement errors.
No noticeable period excursions are seen,
in the latest three points acquired immediately after 
the onset of a high activity.

%%%%%%%%%%%% fig.13 %%%%%%%%%%%%%%%%%%%%%%%
%%%%%%%%%%%% fig.12 %%%%%%%%%%%%%%%%%%%%%%%

%==========4.2===========
\subsection{Overll pulse significance}
\label{subsec:discuss_significance}
%==========4.5==========

\subsubsection{Statistical significance}

Needless to say, the most critical issue with the present work
is whether the $\sim 1.01$ s pulsation is real or false.
As detailed in Appendix 1 and summarized in table~\ref{tbl:demodulation},
the highest-$\zz$ period identified in each data set
has a chance occurrence probability of 
roughly 1 to 10 percent, except in NuS16 and NuS17
where we skipped the probability estimation
(because of the difficulty in modeling the modulation-phase selection).
Although these values may not be convincing enough individually,
they are contingent to the selection of the period search range,
and could be considerably tighter,
as touched on at the end of subsubsection~\ref{subsubsec:ana_XMM05_PGs} for XMM05,
and in subsection~\ref{subsec:ana_XMM16} for XMM16.
Furthermore, we can utilize an obvious advantage of analyzing multiple data sets;
namely, we can evaluate the overall pulse significance by combining the individual results.
Below, we perform this attempt in two slightly different ways.

\smallskip
First, let us consider a hypothesis:\\
\hspace*{4mm}  H: {\em A real pulse period of 1E~1613 is present at $\sim 1.01$ s.}\\
We regard the ASCA93 result as the prior knowledge on H,
and examine how H is reinforced by the addition of the XMM01 result
which is expressed by a symbol ${\rm X}_{01}$.
The posterior probability %$P(\rm{\bar H|X_{01}})$, 
for H to still remain  false, after $\rm{X}_{01}$ occurred,
is  given  as
\begin{equation}
P(\rm{\bar H|X_{01}}) =\frac{P(\rm{X_{01}|\bar H})  P(\rm{\bar H})}
{P(\rm{X_{01}|\bar H})P(\rm {\bar H}) + P(\rm{X_{01}|H}) P(\rm H)}
\end{equation}
by the  Bayes' formula.
Here,  $P(\rm{\bar H})=\Pch(P_{93})=0.54\% \equiv \delta$ 
and $P({\rm H})\equiv 1-\delta$ 
are  prior probabilities for the ASC detection of pulsation
to be false and true, respectively,
and $P(\rm{X_{01}|\bar H})=\Pch(P_{01})=0.92\%\equiv \beta $ is 
the probability that $P_{01}$ is detected by chance 
even when H is false (i.e., no intrinsic periodicity).
Finally, $P(\rm{X_{01}|H})\equiv \gamma$ is a likelihood with 
which we expect to detect the periodicity in XMM01, 
when the ASCA detection is assumed true. 
Although $\gamma$ is rather uncertain,
it should not deviate significantly from unity, because ASCA93 and XMM01 
have similar number of photons.
Assuming $\beta \delta \ll 1$ and ${\cal O} (\gamma) =1$,
we find
\begin{equation}
P(\rm{\bar H|X_{01}}) =\frac{ \beta \delta}{\beta \delta  +\gamma (1-\delta)}
\approx \beta \delta /\gamma = 5.0\times 10^{-5}/\gamma.
\end{equation}
Thus, even when a conservative value as, e.g., $\gamma \gtrsim 0.5$, is assumed,
we obtain $P(\rm{\bar H|X_{01}}) \lesssim 1.0 \times10^{-4}$.
By repeating the argument by further taking into account the XMM05 and XMM16 results,
this value will further decrease to $\lesssim 10^{-5}$.

We may alternatively consider in the following way.
Among the four $\Pch$ values in table~\ref{tbl:demodulation},
those of MM01, XMM05, and XMM16 were calculated 
against the fiducial interval [equation~(\ref{eq:P_search_range})].
In contrast, that of ASCA93 referred to the
much wider  0.3--30 s range [equation~(\ref{eq:ASCA93_trials})].
Now that the six periods from the six data sets have all been 
identified as the highest peak over the fiducial period range, 
we are allowed to recalculate the period trial number for ASCA93, 
this time using the fiducial interval.
This yields $N_{\rm trial}^{\rm P} \approx 1243$
instead of equation~(\ref{eq:ASCA93_trials}).
Thus, considering feedback information from the other five data sets,
the false probability of ASCA93 may be revised as 
$\Pch (P_{93}) \approx 1.6 \times 10^{-5}$,
which is comparable to the Bayesian result.

From these two evaluations, an overall false probability may be quoted as $< 1\times 10^{-5}$.
We can now conclude that the $\sim 1.01$ s pulsation of 1E~1613 is  real
{\em at least statistically}.

\vspace*{-2mm}
\subsubsection{Qualitative supports}

Although the statistical significance of the 1.01 s periodicity 
has been assessed as above, this is not necessarily sufficient,
because we also need to rule out systematic artifacts.
This includes the case that the true pulse period resides
somewhere which is slightly deviated from 1.01 s,
and we are observing its side lobes.
Since there is no simple way of quantifying the probability of various systematic effects,
below we list qualitative pieces of evidence
that are thought to  strengthen the reality of the periodicity
against artifacts (both systematical and statistical).
\vspace*{1mm}
\begin{enumerate}
\setlength{\itemsep}{1mm}
\item  
In figure~\ref{fig:f13_linear_spindown},
the measured five periods very accurately line up on a constant-$\dot P$ trend. 
This would be difficult to explain 
when some, if not all, of these periods were  due to  some artifacts 
(e.g.,  statistical fluctuations or some side lobes).
\item  
While the ASCA and NuSTAR data are subject to 
periodic data gaps synchronized with the spacecraft orbits,
those of XMM-Newton to sporadic ones due to the flare removal. 
Their consistency hence argues against  
the pulsation being instrumental artifacts
that are related to  data gaps.
\item  
In ASCA93,  our start point,  the harmonic  pulse periods
were revealed consistently by the two algorithms,
the phase-sorted analysis and the demodulated PGs.
Therefore, the pulses are unlikely to be an algorithm-specific artifact.
\item  
In the ASCA93 data, the CCO photons were 
affected by the PPM at $T=24.0$ ks (figure~\ref{fig:f3_Tscan_ASCA93}a),
whereas the pulsar data were free from such effects
over $T=7.0-100$ ks (figure~\ref{fig:f3_Tscan_ASCA93}b).
\item Except in XMM05 when the source was rather faint,
the demodulated pulse profiles below $\sim 10$ keV are very  similar.
They consist of a dominant main peak and a weaker sub-peak,
on which the 4-peak structure is often superposed.
\item Except in XMM16,
the 6.67 hr period has been identified
via timing studies alone,
although the NuS16 and NuS17 data favor 
somewhat shorter periods
[equations~(\ref{eq:T_NuS16}) 
           and (\ref{eq:T_NuS17})].
\item
In figure~\ref{fig:f6_Pr_XMM01}c, the  pulse periods in XMM01,
determined over  short intervals without demodulation,
vary just as predicted by the precession scenario
(subsection~\ref{subsec:discuss_geometry}).
\item
Both in the 11--60 keV and 3--6 keV  bands,
the pulses in NuS16 are clearly seen only for $\sim 50\%$ of the modulation phase
(figure~\ref{fig:f10_NuS16H}, figure~\ref{fig:f11_NuS16S}).
If $P_{16}^{\rm N}$ arose due to some artifact
at least in one of these energy intervals,
this sort of correlated behavior would have scarcely taken place.
\end{enumerate}

\subsubsection{A composite PG}

\begin{figure}
\centerline{
\includegraphics[width=8.8cm]{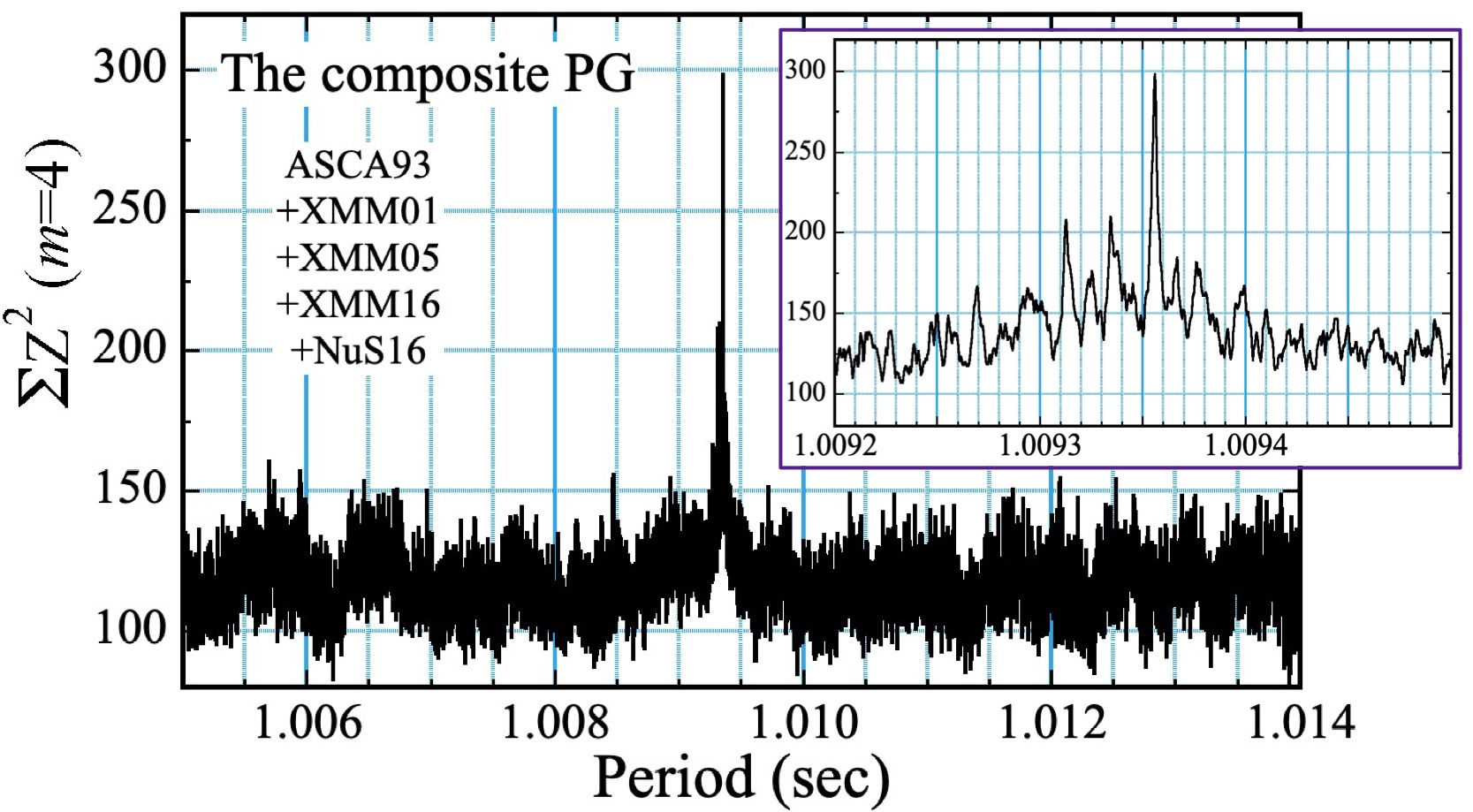}
}
\caption{The final composite PG, derived by summing up
five PGs from the five data sets except NuS17 (see text).
The period of each constituent PG is converted to that of ASCA93,
referring  to equation~(\ref{eq:linear_spindown}).
The inset shows details around the peak,
on the same scale as panels (c) and (d) of figure~\ref{fig:f2_Pscan_ASCA93}.
{Alt Text: The final periodogram, obtained by summing up
five individual periodograms.}
}
\label{fig:f14_composite_PG}
\vspace*{-3mm}
\end{figure}

As a visualization of the overall pulse significance,
figure~\ref{fig:f14_composite_PG} shows our final composite PG,
obtained as an incoherent sum of five demodulated PGs,
one from each data set
(except Nu17 which favors $m=3$); 
figure~\ref{fig:f2_Pscan_ASCA93}b 
(but limiting the period range to the fiducial interval) from ASCA93,
figure~\ref{fig:f5_Pscan_XMM01}a from XMM01,
figure~\ref{fig:f7_PTscan_XMM05}a from XMM05,
figure~\ref{fig:f9_XMM16}b from XMM16,
and
figure~\ref{fig:f11_NuS16S}d from NuS16.
As the only one freedom in summing up the five PGs,
we converted the period in each PG 
to that to be recorded at the epoch of ASCA93,
referring to equation~(\ref{eq:linear_spindown}).
Consequently,
the highest peaks of the five constituent  PGs have all 
added up into an extremely prominent single peak,
at a period which accurately coincide with 
equation~(\ref{eq:ASCA93_periodB}).
The peak height,  $\sum \zz =298.4$,
is lower, only by $\Delta \zz=5.0$, 
than the sum of the five peak values
of the constituent PGs.
Details around the peak, given in the inset,
reconfirm the series of side lobes 
that were observed in figure~\ref{fig:f2_Pscan_ASCA93}d.

%%%%%%%%%%%% fig.14 %%%%%%%%%%%%%%%%%%%%%%%
%%%%%%%%%%%% fig.14 %%%%%%%%%%%%%%%%%%%%%%%

\vspace*{-2mm}
%==========4.3===========
\subsection{Possible emission geometry}
\label{subsec:discuss_geometry}
%==========4.3==========
Now that we have identified the $\sim 1.01$ s period 
with the spin period of 1E~1613,
we must find an emission geometry 
that makes the pulses directly invisible,
and produce, at the same time, 
the  large-amplitude X-ray modulation at equation~(\ref{eq:6.67hr}).
The former condition is not difficult,
as the hard X-ray pulsation of magnetars is often undetectable,
unless we conduct the demodulation with the knowledge of the 
soft X-ray pulse period (e.g., \cite{Makishima21b}).
In contrast, the latter condition, seldom seen among magnetars
(\cite{Makishima16}, 2019), is not trivial.

Already, a successful  model to explain the above 
two requirements has been  presented by HH02,
who  assumed  a slightly aspherical NS,
and surface emission from a hot spot 
that is  displaced from the stellar symmetry axis.
The model, however,  may need some modification,
because the assumed simple surface emission would not explain
the observed pulse profiles which are rich in fine structures.
Below, we consider a revised (though very primitive)  free-precession scenario, 
which invokes asymmetric beaming of the emission,
instead of the source's   positional displacement.
The terminology which we have adopted in section~\ref{sec:intro} is retained.

We assume that the NS is prolately deformed by the toroidal field 
of  $B_{\rm t} \sim 10^{16}$ G, to an asphericity of 
$\epsilon  \equiv (I_1 - I_3)/I_3\sim 10^{-4}$ (e.g., HH02, \cite{Makishima24a}),
where  $I_j~ (j=1,2,3)$ are the principal moments of inertia
in the coordinates $(\hat{x}_1,\hat{x}_2,\hat{x}_3)$ fixed to the star.
Its symmetry axis $\hat{x}_3$ (to be identified with the magnetic axis)
precesses around the  angular momentum vector $\vec L$,
with a  period $\Ppr = 2 \pi I_1/| \vec L| = (1+\epsilon) \Prot$,
where $\Prot= 2 \pi I_3/| \vec L| $  is the rotation period around $\hat{x}_3$.
The precession and rotation periods differ slightly,
and their beat appears  at a long  period given as
\begin{equation}
T_{\rm sl} = \frac{\Ppr}{\epsilon \cos \alpha} 
= \frac{1}{\cos \alpha} \left(P_{\rm rot}^{-1} -{\Ppr}^{-1} \right)^{-1}~.
\label{eq:slip_period}
\end{equation}
Here, $\alpha$ is a constant ``wobbling angle" between $\hat{x}_3$ and $\vec L$.
If the emission is symmetric around $\hat{x}_3$,
a strictly periodic pulsation is observed at $\Ppr$ (not $\Prot$),
but the PPM sets in if the emission breaks the symmetry.

We identify $\Ppr$  with the $\sim 1.01$ s pulse period,
and $T_{\rm sl}$ with $T$ in equation~(\ref{eq:6.67hr}).
These yield  $\epsilon \cos \alpha = \Ppr/T_{\rm sl} =0.42\times 10^{-4}$,
which is typical of magnetars \citep{Makishima24a}.

\begin{figure}
\hspace*{-5mm}
\centerline{
\includegraphics[width=9.cm]{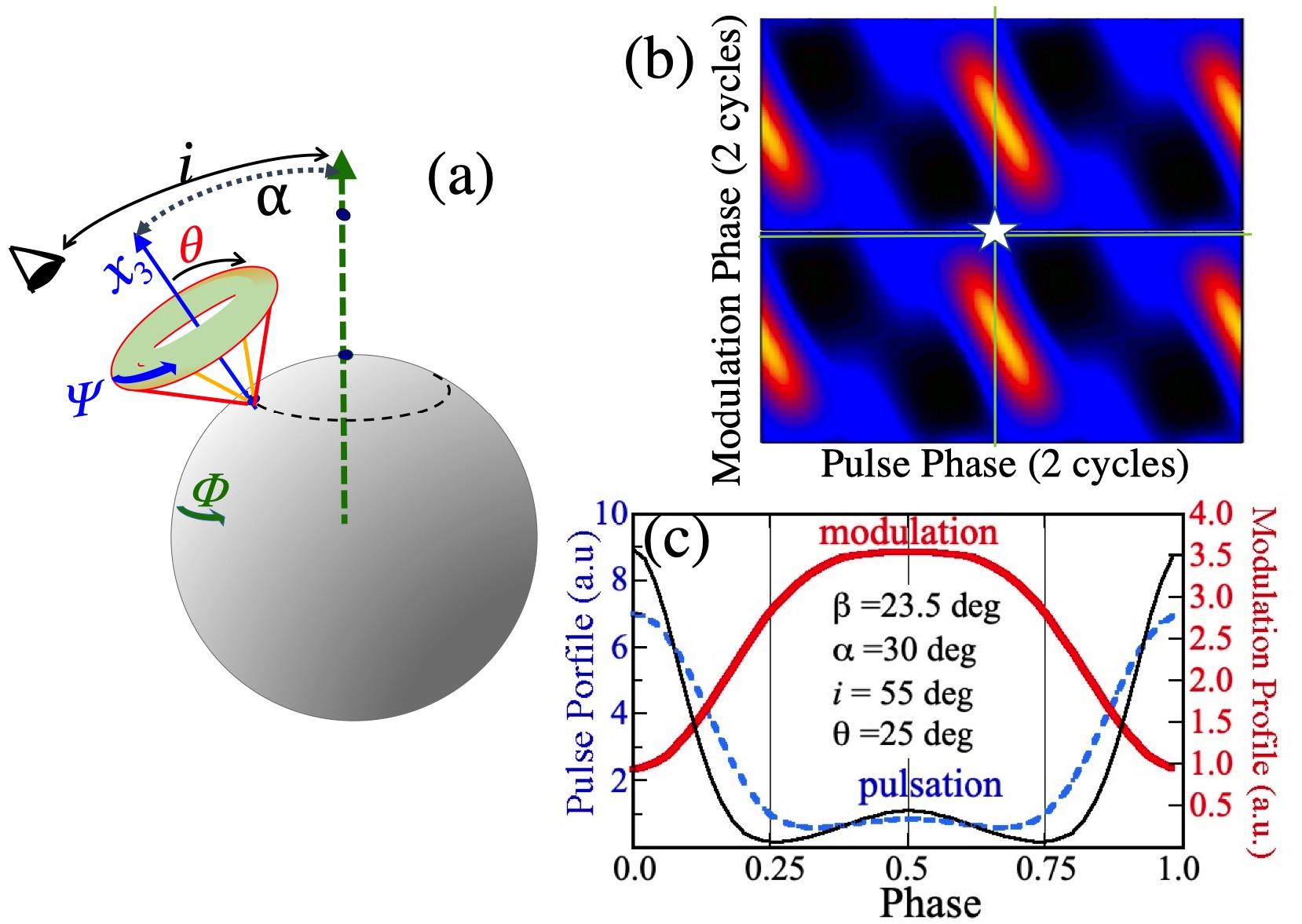}
}
\caption{(a) A possible geometry, after \citet{Makishima21a}.
(b) Prediction of the model, 
shown as a color-coded photon intensity map
on a plane of pulse phase vs. modulation phase.
(c) Projections of panel (b) onto the pulse-phase (thick solid red)
 and modulation-phase (dashed blue) axes.
 The thin black line shows the pulse profile without the PPM.
 {Alt Text: A possible emission geometry,
 together with its predictions.}
}
\label{fig:f15_emission_model}
\vspace*{-5mm}
\end{figure}

Adopting the toy model in \citet{Makishima21a},
we assume that the emission comes from one  magnetic pole,
in a conical beam pattern as illustrated in figure~\ref{fig:f15_emission_model}a.
The cone  is hollow,  symmetric  around $\hat{x}_3$
with a half-opening angle $\theta$, 
and emits  beams  along its generatrices,
each with a  divergence angle $\pm 23.^\circ 5$.
To break the symmetry around $\hat{x}_3$, 
the beam  brightness is assumed to vary as $ \propto 1 + a \cos \Psi$.
Here, $a ~(0 \le a \le 1)$ describes the degree of asymmetry,
and $\Psi$ (0 to $ 2 \pi$)  is the cone azimuth 
measured in the $(\hat{x}_1,\hat{x}_2,\hat{x}_3)$ frame.
(For a remote observer,
$\Psi$ appears as the modulation phase, $[t/T]$).
Then, $(\Phi, \alpha, \Psi)$ define the Euler angles
from the observer frame to the stellar frame,
where $\Phi$ is the pulse phase.
As described in subsubsection~\ref{subsubsec:drifting},
this model is analogous to
that used to explain some behavior of radio pulsars.

Assuming $\alpha=30^\circ$, $i=55^\circ$
(so the emitting pole is not self-eclipsing),
$\theta=25^\circ$, and $a=1$,
we numerically calculated the expected pulse profile.
The result is  presented in figure~\ref{fig:f15_emission_model}b
on a pulse-phase vs. modulation-phase plane,
with the photon intensity color coded.
Figure~\ref{fig:f15_emission_model}c gives the pulse profile (dashed blue) 
and the modulation profile (solid red), obtained by projecting panel (b)
onto the pulse-phase and modulation-phase axes, respectively.

The result reproduces three observed properties.
First, the pulse phase is modulated 
with an amplitude close to $\pm 0.25$ cycles.
(When fine structures are present,
the pulse visibility would further decrease.)
Next, the modulation profile swings from 1.0 to 3.5, 
by a factor $3.5$ just as observed 
in the figure~\ref{fig:f8_Pr_XMM05}c inset.
Finally, when the 6.67 hr flux modulation is minimum (a white star),
the pulse timing delay is $\Delta t \sim 0$  and is increasing with  the modulation phase, 
as  found in subsection~\ref{subsec:ana_XMM05}.

%%%%%%%%%%%% fig.15 %%%%%%%%%%%%%%%%%%%%%%%
%%%%%%%%%%%% fig.15 %%%%%%%%%%%%%%%%%%%%%%%

Since the adopted  angles,
$\alpha=30^\circ$,  $i=55^\circ$, and $\theta=25^\circ$, 
are ordinary, the results appear to apply to any  magnetar.
However, one secret recipe is
that $i = \alpha + \theta$ holds.
As a result, the brightest part of the cone directly 
points to us when $\Phi=\Psi=0$,
producing the bright spots in figure~\ref{fig:f15_emission_model}b.
This may explain the peculiarity of this object.
While  $\alpha$ and $i$ should be constant,
$\theta$ and $a$ can change with time, 
depending, e.g.,  on the  luminosity.
Then, the above condition can sometimes be violated, 
and the swing of the modulation profile will decrease.
The absence of clear 6.67 hr intensity variations
in the ASCA data  may be thus explained.

Of course, this model  is still subject to many inadequacies.
It considers neither the general-relativistic light bending effects,
nor emission from the other pole.
No attempt has been made to reproduce the 4-peak structure seen in the pulse profiles. 
It is unclear if the double-peaked modulation waveform 
of XMM16 in figure~\ref{fig:f9_XMM16}a 
and of NuS16 in figure~\ref{fig:f10_NuS16H}a
can be explained, 
even though the flat-topped peak is relatively well reproduce.
The pulse profile is smeared to some extent  compared with the case of no PPM, 
but is still clearly visible without demodulation.
Also, it is not obvious, either, 
whether the $[t/T]$ independence of the pulsation in 
XMM05 (figure~\ref{fig:f8_Pr_XMM05}c)
and strong $[t/T]$ dependence in NuS16 
(figure~\ref{fig:f10_NuS16H}, figure~\ref{fig:f11_NuS16S})
and nuS17 (figure~\ref{fig:f12_NuS17})
can be consistently modeled.
Instead of trying to improve the model,
we may quote a phrase from \citet{Makishima21a};
{\em figure~\ref{fig:f15_emission_model} is meant to show 
that some emission geometry can roughly explain the  observation, 
rather than to claim that the selected model or the geometrical parameters are correct.}

\vspace*{-3mm}
%==========4.4===========
\subsection{The nature of the object}
\label{subsec:discuss_nature}
%==========4.4==========
Assuming that 1E~1613 has a $\sim 1.01$ s spin period
which changes with a rate given by equation~(\ref{eq:linear_spindown}),
and identifying  $T=24.0$ ks with the beat period,
we can derive basic properties  of 1E~1613 as a young pulsar.
First, the characteristic age is readily calculated as
\begin{equation}
\tau_{\rm c} \equiv P/2 \dot P = (14.7 \pm 0.1) ~{\rm kyr}.
%\label{eq:tauc}
\end{equation}
When compared with the estimated age of  RCW 103,
$\tau_{\rm SNR}= 2.0-4.4$ kyr
\citep{RCW_age2.0,RCW_age4.4},
we find $\tau_{\rm c}/\tau_{\rm SNR} =3.3-7.3$.
This ratio is comparable to those found with 
other magnetar-SNR pairs by \citet{Nakano15},
who showed that the $\tau_{\rm c}$ vs. $\tau_{\rm SNR}$ discrepancy 
my be an artifact caused by the decay of dipole magnetic fields.
Or else, invoking rather long natal pulse periods
can also solve the discrepancy.

Next, the spin-down luminosity is calculated as
\begin{equation}
L_{\rm rot} =  (2 \pi)^2 I_{45} \dot P P^{-3}\approx 4.2 \times 10^{34}~{\rm erg~s}^{-1}
%\label{eq:sdluminosity}
\end{equation}
where $I_{45}$ is the moment of inertia of a NS in $10^{45}$ g cm$^{2}$.
Considering the typical conversion efficiency from rotational to radiative
energies in pulsars, $10^{-4}$ to $10^{-2}$ \citep{Possenti02},
and the X-ray luminosity of 1E~1613
which varies over $10^{33}-10^{35}$ erg s$^{-1}$ (DL06),
this $L_{\rm rot}$ can sustain, at best,  the lowest observed  luminosity.
Therefore, 1E~1613 must be magnetically powered, 
at least in outbursts.

Assuming $\alpha \sim 30^\circ$ from the above model,
the dipole magnetic field of this object is given as 
\citep{M2rad}
\begin{equation}
B_{\rm d} \sim 2.2\times 10^{19} (\sin^{-1}\alpha) \sqrt{P\,\dot P} =4.6 \times 10^{13}~{\rm G}.
%\label{eq:Bd}
\end{equation}
This is typical of  ``high field pulsars" rather than  magnetars,
but it possibly reaches the critical field.

As to $B_{\rm t}$, we can derive an estimate as
\begin{equation}
B_{\rm t} \sim  \sqrt{(10^4 P/T \cos\alpha)} \times 10^{16} =7 \times 10^{15}~{\rm G}
%\label{eq:Bt}
\end{equation}
after \citet{Makishima24a}.
This is  only  slightly lower than those of typical magnetars.
Then, 1E~1613 is implied to have $B_{\rm t}/B_{\rm d} \sim 152$.
This ratio is somewhat higher  than the values of 40 to 100,
which are predicted by the $\tau_{\rm c}$ vs.  $B_{\rm t}/B_{\rm d}$ scaling  
found by \citet{Makishima24a} among  the seven magnetars.

The present work for the first time provided quantitative support to the view,
that 1E~1613 is a magnetically-powered NS (section~\ref{sec:intro}).
However, we also notice several differences from typical magnetars \citep{McGill},
such as  the shorter $P$ (though not the shortest) and the weaker $B_{\rm d}$. 
More importantly, 
the timing residuals of 1E~1613 (figure~\ref{fig:f13_linear_spindown}b)
is much smaller than those of magnetars
(e.g., $\sim 1\%$; \cite{Younes17}).
In some aspects, 1E~1613 might resemble, 
particularly in quiescence, rotation-powered pulsars, 
even though its radio quietness remains to be explained.

%==========4.5===========
\subsection{Some additional remarks}
\label{subsec:discuss_additional}
%==========4.5==========

%\vspace*{-2mm}
%----------------------4.5.1--------------------
\subsubsection{Luminosity dependent effects}
%-----------------------4.5.1--------------------

Through the analysis, we noticed some luminosity-dependent  source properties.
Namely, a luminosity decline my be accompanied by an increase in  $A$,
a decrease in the pulse fraction (table~\ref{tbl:demodulation}),
a possible decrease of $\EL$ above which equation~(\ref{eq:demodulation}) works,
and changes of double-peaked profiles into 4- (or 3-) peaked ones.
Interpretations of these effects, however,
are not attempted here.

\vspace*{-7mm}
%----------------------4.5.2--------------------
\subsubsection{Meanings of  $\EU$ and $\EL$}
\label{subsubsec:discuss_EUEL}
%-----------------------4.5.2--------------------
In the present work, the pulse detection required us to optimize,
in a data-dependent way, 
the upper ($\EU$) and lower ($\EL$) energy thresholds (table~\ref{tbl:demodulation}).
Of these, the meaning of the employed $\EU$ is rather simple to interpret,
because $\EU=12$ keV of ASCA93, 
$\EU=10/12$ keV of XMM-Newton, and $\EU=60$ keV of NuS16,
are all close to the instrumental upper-limit energy.
Also,  as evidenced by figure~\ref{fig:f12_NuS17}a,
$\EU=12$ keV of NuS17 is a relatively loose boundary,
above which the background graually becomes dominant.
Thus,  the pulsation of 1E~1613  is likely to be
always detectable in harder energies,
as long as the signal statistics allow, 
and the PPM correction is properly performed.

The meaning of $\EL$, in contrast,  is more difficult to understand
for a few reasons.
Although $\EL$  is rather sharply specified by the individual data sets,
it scatters considerably among observations  (table~\ref{tbl:demodulation}).
Putting aside ASCA93,
$\EL$ was at 2.7 keV (XMM01), 1.4 keV (XMM05), 
6.6 keV (XMM16), 11 keV (NuS16), and 5.1 keV (NuS17).
It is not yet clear what causes $\EL$ to vary,
as the suggested correlation  between $\EL$ and the luminosity
is not tight enough.
Last but not least, we cannot associate $\EL$
with particular features or components
in the  spectra of 1E~1613  (DL06; \cite{Esposito19,Rea16}),
partly because the spectrum is rather featureless.
At present, what we can suggest is the following candidates.
(1) In XMM05, the spectral normalization varies with 6.67 hr,
but this behavior disappears in $\lesssim 1.4$ keV  (DL06).
(2) In NuS16, the hard power-law component and the hotter blackbody cross over,
at $\sim 9$ keV \citep{Rea16} which is close to $\EL=11$ keV.
(3) In XMM16, two blackbodies cross over at
about $\EL =6.6$ keV \citep{Esposito19}.

In magnetars, the timing vs. spectrum relation is much clearer
\citep{Makishima14, Makishima24b}.
Their hard X-ray emission (seen in $\gtrsim 10$ keV) is thought 
to break the symmetry around the magnetic axis,
whereas their soft X-ray emission is axially symmetric.
As a result, their hard X-ray pulses are  PPM affected,
wheres their soft X-ray pulses are directly detectable without PPM.
In contrast,  the pulsation of 1E~1613 below $\EL$ is not visible, 
even applying the demodulation.
Therefore, as mentioned in subsection~\ref{subsec:ana_XMM16},
the softer X-rays from 1E~1613 may still be affected by the PPM
that cannot be adequately rectified by the simple 
form of equation~(\ref{eq:demodulation}).
For example,  in softer energies where the beam divergence must be larger,
the emission from 1E~1613 may start to be contributed by the other pole, 
or from other regions of the same pole.
If this component is beamed in a different direction,
it would require a different $(A, \psi_0)$ pair,
and make the demodulation difficult.

To better understand these intriguing issues,
and clarify the fundamental relation of 1E~1613  to magnetars,
it may help if we examine the distribution of photons 
on the plane of pulse phase vs. modulation phase.
Alternatively,  we may produce the spectra of 1E~1613,
sorted by the pulse phase and/or the modulation phase,
although the photon statistics might become severe.

Considering these complex properties of 1E~1613,
we were extremely lucky to have started with ASCA93.
At that time,  the object was luminous with  a high pulse fraction,
and yet the simple PPM worked down to $\EL = 2.5$ keV,
which was preset with little ambiguity
by the SNR contamination.
If we had started  from any other data sets,
we may not have achieved the present results.

\vspace*{-1mm}
%--------------------4.5.3--------------------
\subsubsection{Is $T$ time variable?}
%--------------------4.5.3--------------------

Generally, the PPM period $T$ was consistent with equation~(\ref{eq:6.67hr}),
but the NuS16 data [equation~(\ref{eq:T_NuS16})],
and possibly the NuS17 data too [equation~(\ref{eq:T_NuS17})],
suggest a marginally shorter period.
(The XMM16 data were inconclusive on this point.)
In view of equation~(\ref{eq:slip_period}),
any intrinsic change in $T$, if  significant,
should be attributed to those in $P$,  $\epsilon$, or $\alpha$,
presumably synchronized with the 2016 June activity.
Of them, the change in $P$ can be ruled out
as a cause of the possible decrease in $T$,
because $\dot P$ has the opposite sign, and is way too small.
An increase in $\epsilon$ could make $T$ shorter,
but we expect  $\epsilon$ to gradually decrease,
as the magnetic energy is spent in radiation.
The most likely scenario is a sudden decrease 
in the wobbling angle $\alpha$ by $\sim 3^\circ$
across the June 2016 outburst,
possibly coupled with the modulation profile change.

\vspace*{-2mm}
%--------------------4.5.3--------------------
\subsubsection{Analogy to the subpulse drifting phenomenon}
\label{subsubsec:drifting}
%--------------------4.5.3--------------------
The geometrical model  in figure~\ref{fig:f15_emission_model}a
has an interesting analogy to that used to
describe the ``subpulse drifting" phenomenon
often observed from radio pulsars (e.g., \cite{DrigtingSubpulse05}).
In both cases, the system is assumed to have 
two rotational modes around different axes.
One is around the angular-momentum vector with a period $P$,
and produces the periodic pulsation.
The other  is around the magnetic axis, with a period $P'$,
and modifies the pulsation.
In the present case, $P'=P/(1+\epsilon)$
is so close to $P$ that the beat between $P$ and $P'$ 
appears as a PPM with the long modulation period 
$T=P/\epsilon \sim 10^{4} P$.
As to the drifting subpulse phenomenon,
$P'$ is typically 10--100  times $P$,
and corresponds to the period $P_3$
in which the subpulse completes its drift across $P$.
However, even putting aside 
that $P_3$  is much shorter than $T$,
the two cases have a fundamental difference.
Namely, $P'$ of 1E~1613 (and magnetars)  arises 
as rigid-body dynamics of  NSs under axial deformation,
whereas $P'$ of radio pulsars is due to 
some magneto-plasma effects in the magnetosphere
of spherical NSs.

\vspace*{-4mm}
%==========4.6===========
\subsection{Future tasks}
\label{subsec:discuss_tasks}
%==========4.6==========
The present result will become a start point of a novel paradigm for \oneE,
to be followed by a series of questions.
First of all,
can we reconfirm the pulsation in other archival data sets?
In addition to the Chandra data,
these include the 1997 ASCA data and those from XMM-Newton in 2018,
which were excluded based on Criteria (v) and (iii), respectively.
By thus increasing the number of pulse detections,
we may better assess if  the decrease in $T$ 
suggested by the NuS16  and NsS17 data is real or not.

A second set of questions are related to the puzzling behavior of $\EL$.
Namely, can we identify any spectral component 
that carries the PPM-affected pulsation,
and explain the epoch-dependent changes of 
its lower threshold at $\EL$?
Conversely, does this object harbor a ``magnetar-like" soft component
that is pulsating without PPM?
Or else, by improving the demodulation procedure,
can we detect the pulsation below 
$\EL$ from at least some data sets?

The geometrical mode, of course, must be improved, to answer
such inquiries as:
What is a more rigorous geometry
to explain the overall properties of 1E~1613,
and to what extent is that geometry special?
Do the model parameters depend systematically
on the source luminosity, and/or the source activity
as in 2016 June?
How to understand the overall $[t/T]$ dependence of 
the pulsation beyond the arguments in Appendix 2?
How can the model explain the change in the modulation profile across the 2016 activity,
from the single-peaked to double-peaked ones?

Finally, as discussed  in \citet{Makishima24a},
does the present scenario apply also to some other CCOs,
and/or so-called long-period radio transients (e.g., \cite{ULPP}),
including, e.g., ASKAP J1832$-$0911 \citep{ASKAP}?
In the overall zoo of NSs,
how can we locate 1E~1613, and clarify its relation to magnetars?
These issues are all left as future tasks.

\vspace*{-1mm}
%%%%%%%%%%%%5%%%%%%
\section{Conclusion}
\label{sec:conclusion}
%%%%%%%%%%%%5%%%%%%
We conducted timing studies of  \oneE, 
the Central Compact Object of RCW 103,
using six archival data sets: one from ASCA take in 1993,
three from XMM-Newton (2001, 2005, and 2016),
and two from NuSTAR (2016 and 2017).
We obtained evidence for a $\sim1.01$ s pulsation, 
with a combined chance probability of  $<1\times 10^{-5}$,
because  the six data sets all gave $\sim 1.01$ s pulse periods 
that line up, with an accuracy of  $<10\mu$ s,
on  a constant spin-down trend of
$\dot P = 1.097(2) \times 10^{-12}~{\rm s~s}^{-1}$.
The pulsation, however, was not directly detectable 
in either data set with a simple periodogram analysis.
Instead, it became detectable only after removing 
pulse-phase modulations occurring at the period of 6.67 hr, 
with which the object's X-ray intensity is known to vary.
The spin  period of this NS is hence concluded to be $\sim 1.01$ s,
which is in reality the free precession period.
The  6.67 hr periodicity arises as the beat between this period,
and the period of stellar rotation around the symmetry axes,
which differ only by $0.4 \times 10^{-4}$ in fraction.
This small difference is thought to arise from the NS deformation, 
due to  toroidal magnetic fields reaching 
$\sim 7\times 10^{15}$ G  (HH02).
The object is  implied to have a characteristic age of 14.7 kyr,
a spin-down luminosity of $4.2\times 10^{34}$ erg s$^{-1}$,
and dipole magnetic fields  of $\sim 4.6 \times10^{13}$ G.
It is basically a magnetically powered NS,
and resembles a magnetar,
but in some aspects it behaves like a rotation-powered NS.
%with no detectable deviation across the 2016 June activity episode. 

%%%%%%%%%%%%%%%%%%%%%%%%%
\section*{Appendix 1: Statistical significance of  the periodicity}
%%%%%%%%%%%%%%%%%%%%%%%%%
Statistical significance of a peak of height $Z_m^2 = X$,
found in a PG at a period $P_0$, can be estimated in the following two steps.
We first quantify the statistical distribution of $Z_m^2$ in the data,
and referring to it, derive a  probability $\Pchone(>X)$ for a value of $Z_m^2 \geq X$ 
to appear by chance fluctuations {\em in a single-trial},
assuming that the data are noise dominated.
Then, the final post-trial probability is obtained  as $\Pch(>X) = \Ntrial \Pchone(>X)$.
Here, $\Ntrial$  is the effective number of overall trials contained in the  PG,
which, in the present case,  is factorized as
$\Ntrial= \Ntrial^{\rm P}\Ntrial^{\rm E}$,
where $\Ntrial^{\rm P}$ is the trial number in the period, 
and $\Ntrial^{\rm E}$ is that in selecting the optimum energy range
by changing $\EL$ and $\EU$.
Although the maximum harmonic number $m$ can also be varied,
we do not need to consider its trial number, 
because we keep using $m=4$
whenever the pulse significance is concerned.
(For NuS17, we  switch from $m=4$ to $m=3$, 
but do not calculate $\Pch$.)

Among these trial numbers,
the dominant factor $\Ntrial^{\rm P}$  may be identified
with the number of independent Fourier waves contained in the PG,
as
\vspace*{-1mm}
\begin{equation}
\Ntrial^{\rm P} = \eta_{\rm hh}(S/P_1 -S/P_2) .
\label{eq:Ntrials}
\end{equation}
%\vspace*{-1mm}
%
Here, $S$ is again the data span,
while $P_1$ and $P_2$ are the minimum and maximum periods, 
hence $S/P_1$  and $S/P_2$ are the maximum and minimum 
wave numbers, respectively.
The factor $\eta_{\rm hh}$ is same as  $\eta$  in equation~(\ref{eq:dp}),
but in this case represents the effect of higher harmonics;
if the $n$-th harmonic ($n=1, 2, ...$) dominates,
the  period $P$ would represent a wave number  
$\eta_{\rm hh} S/P$ rather than $S/P$.

Except in limited cases (e.g., subsubsection~\ref{subsubsec:ana_ASCA93_simplePG})
where the distribution of $Z_m^2$ is analytically known,
the first step of this evaluation is difficult.
For example, the calculation of a demodulated PG involves, at each $P$,
many search steps in $A$, $\psi_0$, and sometimes $T$,
of which the  trial numbers are hard to estimate.
An obvious solution is to employ Monte-Carlo simulations;
we produce many fake data sets
that are free from any fast periodicity,
and analyze them in the same way as the actual data,
to construct a statistical distribution of $\zz$ at $P \approx P_0$.
However, to derive a reliable result,
each fake data set must reproduce such effects 
as the data gaps (particularly for ASCA and NuSTAR),
background variations, and the intrinsic source variation
including both the 6.67 ks periodicity 
and sporadic changes (e.g., figure~\ref{fig:f9_XMM16}a).

To avoid this difficulty,
the data themselves may be used instead \citep{Makishima23}.
Namely, we calculate a demodulated ``control" PG, over a period range
which is close to $P_0$, 
but excludes $P_0$ itself as well as its major side lobes.
Here, an implicit assumption is 
that the interval thus selected is noise dominated,
and the chance probability for a noise peak with a certain height 
to appear at any period in the interval is the same as that at $P_0$.
To ensure Fourier independence between the adjacent periods, 
$\eta\sim 0.3$ in equation~(\ref{eq:dp}) is employed.
Then, at each $P$, we maximize $\zz$,
by scanning  $A$ and $\psi_0$ (as well as $T$ in some cases),
exactly in the same way as in the actual  pulsation search.
The derived values of maximum $Z_m^2$, one from each period,
are collected over a single control PG (or a few of them with different intervals),
and used to produce a distribution of $\zz$.
Previous examples are given in in Appendix B of \citet{Makishima23}.

%%%%%%%%%%%% fig.16 %%%%%%%%%%%%%%%%%%%%%%%
\begin{figure}
\hspace*{-2mm}
\centerline{
\includegraphics[width=9.cm]{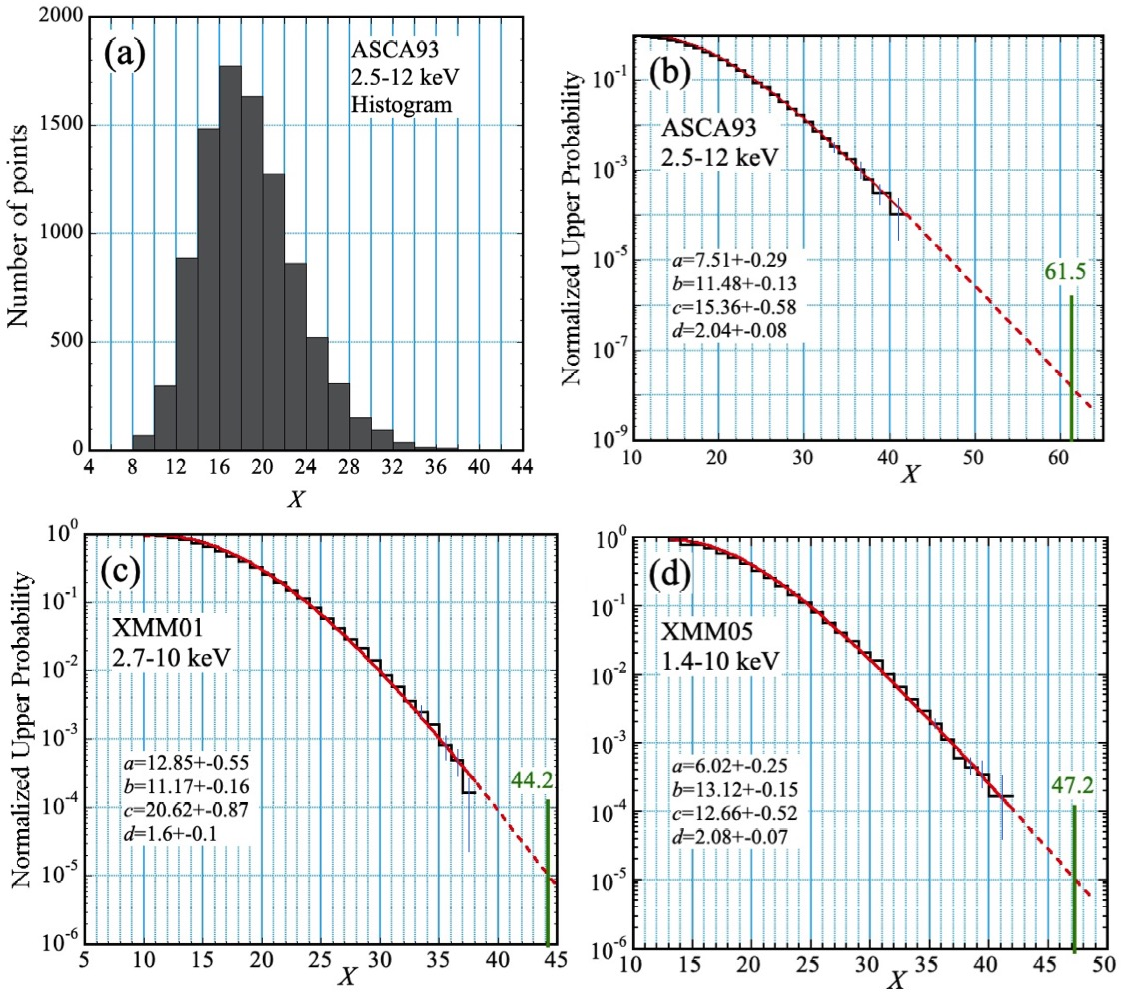}
}
\caption{
(a) A histogram of 9445 samplings of $\zz$,
obtained via a control study of the 2.5--12 keV ASCA93 data.
The abscissa gives $X\equiv \zz$. 
(b) An upper probability distribution of $X$, 
obtained by integrating the histogram in (a) 
from the maximum point downwards. 
A fit with equation~(\ref{eq:CNTL}) is shown in a red curve,
and the best-fit parameter values are given in the panel.
(c) The same as (b), but using the 2.7--10 keV photons from XMM01.
(d) A similar result for XMM05 in 1.4--10 keV.
 {Alt Text: Results of the control studies for ASCA93, XMM01, and XMM05.}
}
\label{fig:f16_CNTL}
\vspace*{-2mm}
\end{figure}
%%%%%%%%%%%% fig.15 %%%%%%%%%%%%%%%%%%%%%%%

\vspace*{2mm}
\noindent
{\bf ASCA93:} 
We applied the above procedure to the 2.5--12 keV ASCA GIS data,
to evaluate the statistical significance of Peak B
in figure~\ref{fig:f2_Pscan_ASCA93}b,
 because our standard tool is the demodulation
rather than the phase-sorted technique.
Two control  PGs with $m=4$ were calculated
over two period ranges, one on either side of $P_{93}$;
$0.80-1.00$ s and  $1.02-1.22$ s,
with a period step of $4.0 \times 10^{-5}$ s 
and $4.5 \times 10^{-5}$ s, respectively.
These steps correspond to $\eta = 0.3-0.45$ in equation~(\ref{eq:dp}).
Since the two control PGs gave consistent distributions of $\zz$,
we co-added them, and arranged 
a  total  9445 samplings of $\zz$ into a histogram
presented in figure~\ref{fig:f16_CNTL}a.
The distribution is asymmetric, because small values of $X$ are
discarded when selecting the maximum $\zz$ at each $P$.

By integrating the histogram,
from the maximum point at $\zz=43.3$ downwards,
we obtained figure~\ref{fig:f16_CNTL}b,
which gives the upper probability integral \citep{Makishima19}
as a function of $X \equiv \zz$.
It  gives the chance  probability for a $\zz$ value
exceeding a given $X$ to appear by chance, in a single trial in $P$.
We fitted this distribution with an empirical function \citep{Makishima23},
\begin{equation}
y = \exp \left\{a - d^{-1}\sqrt{ (X - b)^2 + c^2} \right\},
\label{eq:CNTL}
\end{equation}
where $y$ stands for the upper probability,
while $a, b, c$ and $d$ are parameters.
We expect $b \approx 2.0$ as an asymptotic analytic form \citep{Makishima23}.
The fitted function is superposed on figure~\ref{fig:f16_CNTL}b as a red curve,
where the derived parameters are also given.
By extrapolating the fit, 
Peak B with $\zz=61.5$ was found to have
$\Pchone(P_{93})\approx 1.3 \times 10^{-8}$
with  a typical error by a factor 1.5.

Next,  equation~(\ref{eq:Ntrials}) gives, in the present case, 
%
%\vspace*{-2mm}
\begin{equation}
N_{\rm trial}^{\rm P} = 2 (S/0.3 -S/30) = 4.18 \times 10^5.
\label{eq:ASCA93_trials}
\end{equation}
Here, we assumed $\eta_{\rm hh}=2$, 
because the folded pulse profile of ASCA93  is dominated by the 2nd harmonic
(subsubsection~\ref{subsubsec:ana_ASCA93_Pr}).
At $P \approx 1.01$ s, the implied period step is $\Delta P \approx 8.0~\mu$ s;
since this is comparable to the full-width at half-maximum 
of Peak B ($\approx 6~\mu$ s) in figure~\ref{fig:f2_Pscan_ASCA93}d, 
the above estimate of $N_{\rm trial}^{P}$ is considered reasonable.
When this is multiplied to  $\Pchone$ derived above,
$\Pch(P_{93}) \approx 0.54\%$ is obtained.
There is no need to consider $N_{\rm trial}^{E}$,
because we tested the 2.5--12 keV band only.
We do not need to count the trial number in $m$, either,
because the $P_{93}$ information was derived
solely from the single  PG using the $m=4$ demodulation.

%---------------------------------
\vspace*{2mm} \noindent
{\bf XMM01:} 
Following the procedure as for ASCA93,
we calculated two control PGs,
using the 2.7--10 keV XMM01 data.
The selected period intervals are
0.7--1.0 s and 1.03--1.33 s,
both with period steps of $1\times 10^{-4}$.
The difference from ASCA93 is due to the shorter data span,
which requires more sparse period samplings,
to make the adjacent ones Fourier independent.
Then, a histogram of total 6000 samplings were constructed,
and was converted to the upper probability distribution
which is presented in figure~\ref{fig:f16_CNTL}c.
Extrapolating a fit with equation~(\ref{eq:CNTL}),
the  peak at $P_{01}$ with $\zz=44.2$ was  found 
to have $\Pchone(P_{01}) \approx 1.2 \times 10^{-5}$,
with a similar uncertainty as in ASCA93.

To calculate $\Ntrial^{\rm P}$ of  XMM01 from equation~(\ref{eq:Ntrials}),
we employed $S=19.6$ ks and the period search range 
of equation~(\ref{eq:P_search_range}), and set  $\eta_{\rm hh}=2$ 
because the pulse profile is very similar to that in ASCA93.
These yield $N_{\rm trial}^{\rm P} = 384$.
Also, the choice between $\EL=2.5$ keV and 2.7 keV gives 
$N_{\rm trial}^{\rm E}=2$.
The trial number  in $m$ is unity, as we tested no other values than $m=4$.
We hence obtain  $\Pch(P_{01}) =\Pchone(P_{01})\; N_{\rm trial}^{\rm P}
N_{\rm trial}^{\rm E} \approx 0.92\%$.

\vspace*{2mm} \noindent
{\bf XMM05:} 
The statistical significance of the 1.4--10 keVpeak at $P_{05}$ in figure~\ref{fig:f7_PTscan_XMM05}a, 
with  $\zz =47.2$, was examined again in the same way as the preceding two data sets.
Then, as presented in figure~\ref{fig:f16_CNTL}d,
the peak  was found to have $\Pchone (P_{05})\approx 1.1 \times 10^{-5}$.
On the other hand, 
equation~(\ref{eq:P_search_range}) and $S=87.9$ ks
specify $N_{\rm trial}^{\rm P} \approx 3446$,
where $\eta _{\rm hh}=4$ was employed in equation~(\ref{eq:Ntrials})
because the XMM01 pulse profile is dominated by the 4th Fourier harmonic
(subsection~\ref{subsubsec:ana_XMM05_Pr}).
Multiplying the above $\Pchone (P_{05})$ by this $N_{\rm trial}^{\rm P}$,
and further  by $\Ntrial^{\rm E}\sim 5$ in the $\EL$ scan steps
(considering the energy resolution),
we obtain $\Pch(P_{05})\approx 19\%$.

\vspace*{2mm} \noindent
{\bf XMM16:}  
In the same way as above,
we produced a curve of upper probability,
which is very similar (though not shown)
to those from the preceding three data sets.
An extrapolation of the fitted function by 
equation~(\ref{eq:CNTL}) indicates 
that  the $P_{16}^{\rm N}$ peak in the 6.6--12 keV range
has $\Pchone\approx 9\times 10^{-6}$.
For the period range of equation~(\ref{eq:P_search_range}),
and assuming $\eta_{\rm hh}=3$ in view of the relatively dominant
3rd harmonic in figure~\ref{fig:f9_XMM16}(d1),
we find  $\Ntrial^{\rm P}\approx 2382$.
Further assuming $N_{\rm trial}^{\rm E}\approx 5$ 
in optimizing $\EL$, we find $\Pch=11\%$.

\vspace*{-4mm}
%%%%%%%%%%%%%%%%%%%%%%%%%
\section*{Appendix 2: Modulation-phase dependence of the pulse significance}
%%%%%%%%%%%%%%%%%%%%%%%%%

%%%%%%%%%%%% fig.17 %%%%%%%%%%%%%%%%%%%%%%%
\begin{figure*}
\centerline{
\includegraphics[width=16cm]{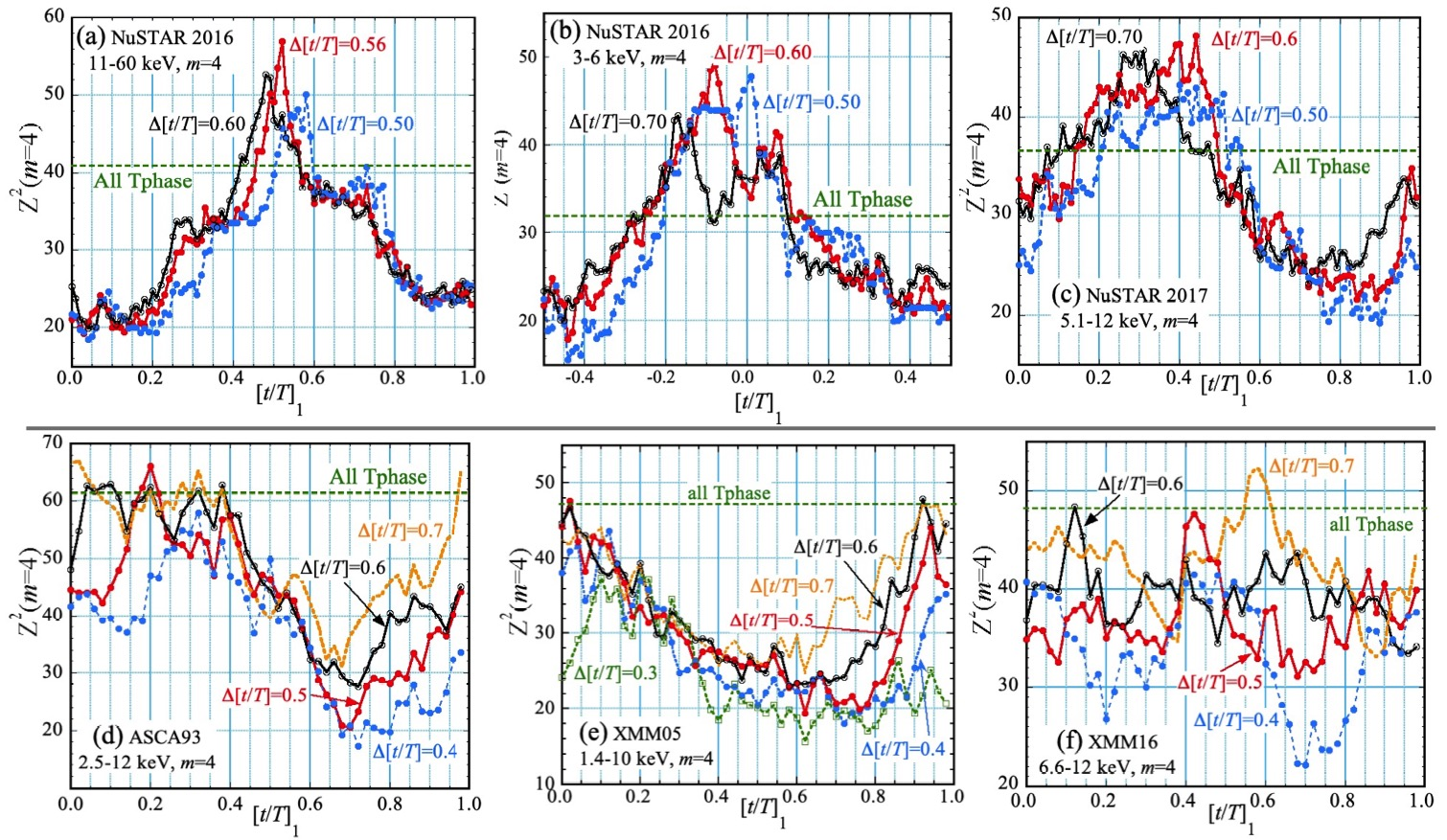}
}
\caption{
The $\zz$ value of the pulsation in each data set,
shown as a function of the starting phase $[t/T]_1$ of the 24 ks modulation (see text).
The effect of changing the modulation-phase width, $\Delta [t/T]$,
is represented by different curves (with different symbols/colors).
The dashed horizontal line  indicates the $\zz$ value 
obtained without limiting the modulation phase.
The employed data set and energy range are given in the title of each panel.
 {Alt Text: Effects of limiting the modulation phase on the pulse significance.}
}
\label{fig:f17_modp1_scan_summary}
\end{figure*}
%%%%%%%%%%%% fig.17 %%%%%%%%%%%%%%%%%%%%%%%

In the NuSTAR data analysis, we found
that the pulse-peak $\zz$ can increase,
in spite of a loss of photons,
when we use a particular $\sim 50\%$  interval of the 24 ks cycle.
We systematically studied this effect for all data sets,
except XMM01 which is shorter than 24 ks,
and summarize the results in figure~\ref{fig:f17_modp1_scan_summary}.
There, we calculated the $\zz$ values of the pulsation in each data set,
as a function of $[t/T]_1$ (the starting phase)
which is regarded as the primary variable.
Different curves specify the interval width, 
$\Delta [t/T] \equiv [t/T]_2 -[t/T]_1$,
which serves as the secondary variable.
In each panel, the dashed horizontal line in green 
indicates $\zz$ derived  from 
the entire modulation phase ($\Delta [t/T]=1$).

\subsubsection*{NuSTAR data sets}
In panel (a), the result from the 11--60 keV NuS16 data
reveals a clear peak at $[t/T]_1 \approx 0.52$ 
and $\Delta [t/T]\approx 0.56$ (namely, $[t/T]_2\approx 1.08$),
where $\zz$ exceeds the full-phase result by $\Delta \zz \gtrsim 15$.
This increase is considered  significant,
because its chance  probability is estimated
as  $\lesssim \exp(-\Delta \zz/2) =0.06\%$.
The 3--6 keV NuS16 data in panel (b) exhibit very similar behavior,
except that the optimum starting phase, 
$[t/T]_1\approx 0.92$, is reversed from  that in 11-60 keV.
As in panel (c), similar behavior was also observed 
from the 5.1--12 keV  NuS17 data.
Thus, in these three cases,
the pulse $\zz$ increased significantly,
when we limit the data to a particular modulation phase
which has a 50\% to 60\% duty ratio.
However, the optimum $[t/T]_1$ shows no correlation 
with  $\psi_0$ in equation~(\ref{eq:demodulation}).

The 6--10 keV result from NuS16 (not shown) differed significantly.
The $\zz$ values showed a mild scatter over the rage of 15 to 25
(which includes the full-phase value of 20.1),
without  systematic dependence on  $[t/T]_1$ or $\Delta [t/T]$.
Therefore, the pulsation was not confirmed
even using the phase-limiting analysis.
Nevertheless, these $\zz$ values are larger than that ($\sim 2m = 8$)
for purely Poissonian data.
Therefore, the count rate does vary intrinsically,
depending on the pulse phase and modulation phase.

\vspace*{-2mm}
\subsubsection*{The other  data sets}
We applied the same analysis also to the ASCA93, XMM05, and XMM16 data,
and produced panels (d), (e), and (f), respectively.
Although the behavior differs to some extent among them,
what is common is that $\zz$ generally stays below 
the dashed horizontal line.
Although some data points exceed this full-phase value by $\lesssim 7$,
they can be considered as statistical fluctuations,
because their estimated chance probability is higher than a few percent.
Therefore, the optimum solution for these data sets is
to employ the entire modulation phase.

The behavior of XMM05 in (e) implies
that the pulse significance remains relatively high
even when we limit the modulation phase to  0.0--0.5,
but this corresponds to the {\em minimum} phase of the 24 ks flux variation
as seen in figure~\ref{fig:f8_Pr_XMM05}c.
This apparent contradiction must be solved in our future studies.

We further examined the XMM16 and NuS17 data,
by limiting the modulation phase in various ways,
hoping to detect the pulsation over 
an energy range that is much wider than used so far.
However, this attempt was unsuccessful.

\vspace*{-2mm}
\subsubsection*{A tentative interpretation}
From figure~\ref{fig:f17_modp1_scan_summary},
we may  derive the following tentative interpretation.
Generally, the pulse profile (amplitude, shape, and peak phase)
of this object varies  to a certain extent,
depending on the modulation phase.
On some occasions, the variation is so mild
that the profiles from different phases 
look similar with good mutual coherence.
If so, the highest $\zz$ will be obtained 
by summing up the profile over the entire phase.
This condition is thought to apply to panels (c), (d), and (e).
On other occasions, the profile differs considerably  among the phases, 
just as  in figure~\ref{fig:f10_NuS16H}b,
and the pulse visibility would degrade by mixing up different phases.
In this case, we can maximize $\zz$ by selecting only those phases
where the pulse profile remains coherent with a high visibility.
Panels (a) through (c) are thought to correspond to this condition.

We are tempted to consider that the strong phase dependence 
was a result of the source activation in 2016 June.
Although this view is suggested by the NuS16 and NuS17 results,
the behavior of XMM16 argues against it.
Then, the modulation-phase dependence may vary from time to time, 
due to some mechanisms that are unrelated to the source activity.

To further explore this intriguing phenomenon,
two more attempts may be considered.
As mentioned in subsubsection~\ref{subsubsec:discuss_EUEL},
one is to derive a series of pulse profiles from 
different modulation phases of the same data,
and compare them quantitatively.
The other is to return to  the free-precessing magnetars,
and search their data for the same effect.
However, both these attempts will be our future tasks.
In particular, the former 
would require accurate background subtraction 
and exposure correction,
which we skipped in the present work.

%\vspace*{-1mm}
\section*{ORCID ID}
\noindent
Kazuo Makishima https://orcid.org/0000-0002-1040-8769

\vspace*{-1mm}
%%%%%%%%%%%%%%%%%%%%%%%%%%%%%%%%%%%%%
\section*{Acknowledgements}
%%%%%%%%%%%%%%%%%%%%%%%%%%%%%%%%%%%%%
The authors thank Y. Furuta, for his help in early data analysis.
The present work was  supported in part by the Japan Society for the Promotion of Science,
grant-in-aid (KAKENHI), no.18K03694 and 24H01612.
The  ASCA archive is provided by the JAXA/C-SODA.
%Thanks are also due to Wise Babel Ltd,
%for improving the English text.

\end{document}